\documentclass[onecolumn,pre,floats,aps,amsmath,amssymb,nofootinbib]{revtex4-2}
\usepackage{graphicx}
\usepackage{bm,braket}
\usepackage{verbatim}
\usepackage{microtype}
\usepackage{amsmath}
\usepackage{siunitx}
\usepackage{physics}
\usepackage{amssymb}
\usepackage{placeins}
\usepackage{slashed}
\usepackage{hyperref}
\newcommand{\bea}{\begin{eqnarray}}
\newcommand{\eea}{\end{eqnarray}}
\newcommand{\dfour}[1]{d^{\hspace{1.5pt}4}{#1}}
\newcommand{\dthree}[1]{d^{\hspace{1.5pt}3}{#1}}

\newcommand{\bs}{\boldsymbol}

\newcommand{\vecp}[1]{\vec{#1}\hspace{1.2pt}}
\newcommand{\nn}{\nonumber}

\newcommand{\vecpp}[1]{\vec{#1}\hspace{2.5pt}}

\newcommand{\ocharming}{O_{1,2}^{(c)}}
\newcommand{\msbar}{$\overline{\mathrm{MS}}$~}
\newcommand{\MSbar}{\overline{\mathrm{MS}}}
\newcommand{\RS}{\mathrm{SMom}}

\begin{document}
\vskip 0.5 cm 
\title{Theoretical framework for lattice QCD computations of {\boldmath$B\to K \ell^+ \ell^-$} and {\boldmath$\bar{B}_s\to \ell^+\ell^- \gamma$} decays rates, including contributions from charming penguin diagrams} 
\author{Roberto Frezzotti}
\affiliation{Dipartimento di Fisica and INFN, Universit\`a di Roma ``Tor Vergata", Via della Ricerca Scientifica 1, I-00133 Roma, Italy}
\author{Nazario Tantalo}
\affiliation{Dipartimento di Fisica and INFN, Universit\`a di Roma ``Tor Vergata", Via della Ricerca Scientifica 1, I-00133 Roma, Italy}
\author{Giuseppe Gagliardi}
\affiliation{Dipartimento di Fisica, Universit\`a  Roma Tre and INFN, Sezione di Roma Tre, Via della Vasca Navale 84, I-00146 Rome, Italy}
\author{Vittorio Lubicz}
\affiliation{Dipartimento di Fisica, Universit\`a  Roma Tre and INFN, Sezione di Roma Tre, Via della Vasca Navale 84, I-00146 Rome, Italy}
\author{Guido \surname{Martinelli}}
\affiliation{INFN, Sezione di Roma, P.le A. Moro 2, I-00185 Roma, Italy}
\affiliation{Dipartimento di Fisica, Universit\`a di Roma La Sapienza, P.le A. Moro 2, I-00185 Roma, Italy}
\author{Chris T.~Sachrajda}
\affiliation{Department of Physics and Astronomy, University of Southampton,\\ Southampton SO17 1BJ, UK}
\author{Francesco Sanfilippo}
\affiliation{Istituto Nazionale di Fisica Nucleare, Sezione di Roma Tre,\\ Via della Vasca Navale 84, I-00146 Rome, Italy}
\author{Luca Silvestrini}
\affiliation{INFN, Sezione di Roma, P.le A. Moro 2, I-00185 Roma, Italy}
\author{Silvano  Simula}  
\affiliation{Istituto Nazionale di Fisica Nucleare, Sezione di Roma Tre,\\ Via della Vasca Navale 84, I-00146 Rome, Italy}

\vskip 0.5 cm

\begin{abstract}
We develop a strategy for computing the $B\to K\ell^+\ell^-$ and $\bar{B}_s\to\gamma\ell^+\ell^-$ decay amplitudes using lattice QCD (where $\ell^\pm$ are charged leptons). 
We focus on those terms which contain complex contributions to the amplitude, due to on-shell intermediate states propagating between the weak operator and electromagnetic current(s). 
Such terms, which are generally estimated using model calculations and represent significant uncertainties in the phenomenological predictions for these decays, cannot be computed using standard lattice QCD techniques.  It has recently been shown that such contributions can be computed using spectral-density methods and our proposed strategy, which we discuss in detail, is built on this approach. 
The complex contributions 
include the ``charming penguins" (matrix elements of the current-current operators $O_1^{(c)}$ and $O_2^{(c)}$ defined in Eq.\,(\ref{eq:O12def}) below), in which the charm-quark loop can propagate long distances, particularly close to the region of charmonium resonances. They also include the contributions from the chromomagnetic operator ($O_8$ in standard notation, defined in Eq.\,(\ref{eq:O78def}) below). 
We discuss the renormalization of the ultra-violet divergences, and in particular those which arise due to ``contact" terms, and explain how those which appear as inverse powers of the lattice spacing can be subtracted non-perturbatively.
We apply the spectral density methods in an instructive exploratory computation of the charming penguin diagram in $B\to K\ell^+\ell^-$ decays in which the virtual photon is emitted from the charm-quark loop (the diagram in Fig.\,\ref{fig:CPBK123}(a) below) and discuss the prospects and strategies for the reliable determination of the amplitudes in future dedicated computations; computations which are however, beyond the scope of the present paper. 
\end{abstract}

\maketitle

 \section{Introduction}\label{sec:intro}
 
In the Standard  Model,  Flavor Changing Neutral Current (FCNC)  $B_{(s)}$ decays  are strongly  suppressed and thus represent a window for discovering New Physics (NP) effects.
At the quark level   $b\to s (d) $  transitions give rise to a  variety of interesting processes that have been  measured experimentally and compared to theoretical predictions, e.g   $B\to K^*  \gamma$ or 
$B \to K^{(*)} \ell^+ \ell^-$ and $\bar{B}_s \to \gamma  \ell^+ \ell^-$\, \cite{CLEO:1993nic,LHCb:2012quo,Belle:2014sac,LHCb:2014vnw,LHCb:2016oeh,LHCb:2019vks,CDF:2011tds,BaBar:2008fao,Belle:2009zue,LHCb:2013zuf,LHCb:2013ghj,CMS:2013mkz,LHCb:2015svh,CMS:2015bcy,Belle:2016fev,CMS:2017rzx,ATLAS:2018gqc,LHCb:2020lmf, CMS:2024atz, BaBar:2007lky, LHCb:2021awg, LHCb:2024uff} . The main limitation in the comparison between experimental measurements and theoretical predictions is that the evaluation of the hadronic matrix elements generally relies  on phenomenological models for which it is very difficult to estimate the uncertainty.  In this paper we show that, as the result of a number of recent theoretical developments \cite{Hansen:2019idp,Frezzotti:2023nun}, a full class of semileptonic FCNC decays, specifically those with 0 or 1 stable hadron in the final state and hence including several of the ones mentioned above, can be computed quantitatively from first principles using Lattice QCD, paving the way towards truly quantitative, and in the future precise, tests of the theory.  

\begin{figure}[t]
\begin{center}
\includegraphics[width=0.4\hsize]{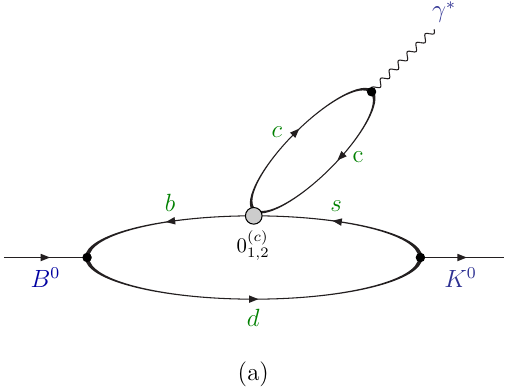}\qquad\quad
\includegraphics[width=0.4\hsize]{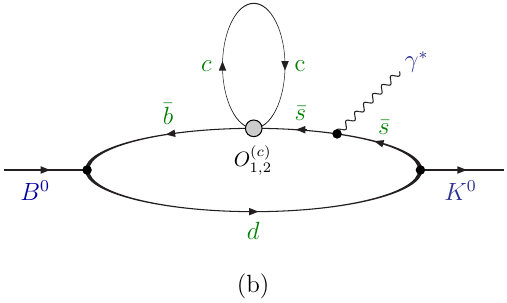}\\[0.2in]
\includegraphics[width=0.4\hsize]{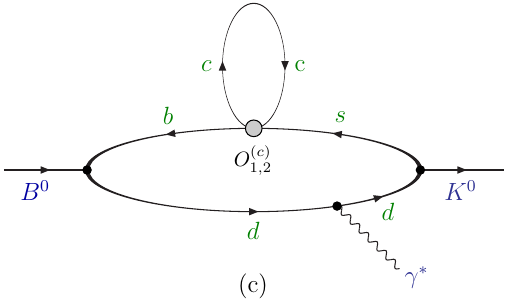}\qquad\quad
\includegraphics[width=0.4\hsize]{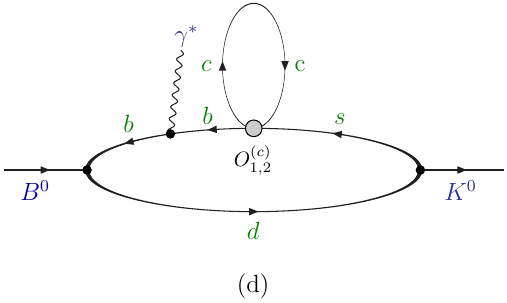}\
\caption{Connected quark-flow diagrams for the process $B\to K\ell^+\ell^-$ which contain a charm-quark loop. The shaded circle, marked $O_{1,2}^{(c)}$, represents either of the two  current-current operators $O_1^{(c)}$ or $O_2^{(c)}$ defined in Eq.\,(\ref{eq:O12def}). The charged leptons $\ell^\pm$, which couple to the virtual photon, $\gamma^\ast$, are not shown. Panels (a)-(d) correspond, respectively, to photon emission from the charm, strange, down, and bottom quarks.
\label{fig:CPBK123}}
\end{center}
\end{figure}

We start by explaining the phenomenological motivation for developing the framework by briefly reviewing the experimental status of $b\to s$ exclusive measurements. The precise measurements of  $B \to K^{(*)} \ell^+ \ell^-$ decays by LHCb \cite{LHCb:2013zuf,LHCb:2013ghj,LHCb:2015svh,LHCb:2023gpo,LHCb:2023gel,LHCb:2024onj}, see also \cite{CDF:2011tds,BaBar:2008fao,Belle:2009zue,CMS:2013mkz,CMS:2015bcy,Belle:2016fev,CMS:2017rzx,ATLAS:2018gqc,CMS:2024atz},  opened a new exciting sector of FCNC phenomenology.  While  the hints of the presence of NP in the ratio of branching fractions
$R_K = B(B^+ \to K^+ \mu^+ \mu^-)/B(B^+ \to K^+ e^+ e^-)$ \cite{LHCb:2014vgu,LHCb:2017avl}, and similarly in 
$R_{K^*}$, $R_{K_S}$, $R_{K^{*+}}$ \cite{LHCb:2019hip,LHCb:2021trn,LHCb:2021lvy},  have disappeared \cite{LHCb:2022qnv,LHCb:2022vje}, there nevertheless remain some tensions between SM estimates  and experimental measurements in absolute $BR$s and angular distributions  \cite{Alguero:2018nvb,Hurth:2020ehu,Altmannshofer:2021qrr,Gubernari:2022hxn,Alguero:2023jeh,Wen:2023pfq,Bordone:2024hui,Guadagnoli:2023ddc} (see however \cite{Ciuchini:2015qxb,Ciuchini:2020gvn,Ciuchini:2021smi,Ciuchini:2022wbq}),  and it is therefore very important to produce reliable theoretical predictions of these quantities in the SM and beyond. In this paper we show how this can be achieved for $B\to K\ell^+\ell^-$ decays, and in particular how the charming penguin contributions, i.e. contributions from diagrams containing charm-quark loops, can be determined (see, for example, Fig.\,\ref{fig:CPBK123}).

The LHCb experiment has set an upper bound for the branching ratio of the $B_s^0\to \gamma\mu^+\mu^- $ decay at large values of $q^2$, where $\sqrt{q^2}$ is the invariant mass of the muon pair, 
\begin{equation}
B(B_s^0\to\mu^+\mu^-\gamma)|_{\sqrt{q^2}>4.9\,\mathrm{GeV}}<2.0\times 10^{-9} \label{eq:LHCb1bound}
\end{equation}
at 95\% confidence level\,\cite{LHCb:2021awg}. This result was obtained from the $\mu^+\mu^-$ invariant mass distribution near the kinematical end-point.  Subsequently the collaboration has searched for fully reconstructed $B_s^0\to\mu^+\mu^-\gamma$ decays and set the following bound at large values of $q^2$
\begin{equation}
B(B_s^0\to\mu^+\mu^-\gamma)|_{\sqrt{q^2}>3.92\,\mathrm{GeV}}<4.2\times 10^{-8} \label{eq:LHCb2bound}
\end{equation}
at 95\% confidence level\,\cite{LHCb:2024uff}. In addition the collaboration has set upper bounds for the branching ratios for bins in $\sqrt{q^2}$ below the charmonium resonances. The LHCb is continuing its search for this decay with the upgraded detector and increased luminosity in LHC's Run 3\,\footnote{See the Cern Courier, May-June 2024, p.6\,.}. In Ref.\,\cite{Frezzotti:2024kqk}, using Lattice QCD, we have evaluated all the contributions to the amplitude and rate from the matrix elements of the reduced effective Hamiltonian in Eq.\,(\ref{eq:Heff7910}) with the exception of those from the charming penguins, i.e. those of the operators $O_1^{(c)}$ and $O_2^{(c)}$ (defined in Eq.\,(\ref{eq:O12def})) and also of the chromomagnetic operator $O_8$ (defined in Eq.\,(\ref{eq:O78def})). 
In that paper we used a phenomenological ansatz, based on Vector Meson Dominance, to estimate the charming penguin contributions. Our final result for the partial branching ratio was
\begin{equation}
B(B_s^0\to\mu^+\mu^-\gamma)|_{\sqrt{q^2}>4.9\,\mathrm{GeV}}=6.9(9)\times 10^{-11}\,, \label{eq:ourresult}
\end{equation}
considerably below the experimental upper bound in Eq.(\ref{eq:LHCb1bound}).
The rate for this decay is dominated by the matrix elements of $O_9$ and $O_{10}$, in particular by the corresponding vector form factor, and the contributions from $O_{1,2}^{(c)}$ were estimated to be relatively small. However, within the range of parameters used in Section V of Ref.\,\cite{Frezzotti:2024kqk}, the inclusion of the operators $O_{1,2}^{(c)}$ had a significant effect on the errors, e.g. from 15\% for $\sqrt{q^2}>4.9$\,GeV, increasing as we approach the charmonium resonances to 30\% for $\sqrt{q^2}>4.2$\,GeV. In the present paper we present a strategy for computing the charming penguin contributions in Lattice QCD, using an extension of the spectral density approach proposed in Ref.\cite{Frezzotti:2023nun}, showing that it is now possible to envisage a reduction of the uncertainties and a coverage of the full kinematic region in $q^2$, where $q^2$ is the invariant mass of the lepton pair.

The decay $B\to K^\ast  \gamma$  is, as yet, an unexplored channel in lattice calculations and presents additional issues related to the instability of the $K^\ast$ meson.  In particular the reconstruction of the Minkowski amplitude from Euclidean lattice correlation functions is more complex and the necessary theoretical framework has not yet been fully developed. For this reason we do not discuss $B\to K^\ast\ell^+\ell^-$ and $B\to K^\ast\gamma$ decays further in this paper, although we anticipate that the method developed in this paper may eventually be an important element in the determination of the corresponding amplitudes. 

Motivated by the challenge to obtain reliable theoretical results with which the experimental results outlined above can be compared, we now begin a discussion of the theory. For real physical quantities, such as leptonic decay constants, semileptonic form factors (between states which are stable under strong interactions) and the $B$-parameters of neutral pseudoscalar-meson mixing, the continuation from Minkowski to Euclidean space poses no difficulty and they can readily be determined from Lattice QCD computations of Euclidean correlation functions\,(see, for example, the comprehensive compilation of results in Ref.\,\cite{FlavourLatticeAveragingGroupFLAG:2024oxs}). 
Since Euclidean correlation functions are real, difficulties in the continuation from Minkowski to Euclidean space arise when the physical quantities of interest are complex. This is particularly severe for the evaluation of the amplitudes for $B$-decays, where a large number of on-shell intermediate states can propagate and contribute to the imaginary part. 
In an attempt to overcome this complication, methods based on the spectral representation have been developed \cite{Barata:1990rn, Hashimoto:2017wqo, Gambino:2020crt, Hansen:2017mnd, Hansen:2019idp, Bulava:2019kbi, Bruno:2020kyl, Frezzotti:2023nun, Patella:2024cto,Bergamaschi:2023xzx} and applied to a variety of physical processes, including Refs.\,\cite{Bulava:2021fre, ExtendedTwistedMassCollaborationETMC:2022sta, Bailas:2020qmv, Gambino:2022dvu, Barone:2023tbl, Kellermann:2025pzt, Evangelista:2023fmt, ExtendedTwistedMass:2024myu, Frezzotti:2024kqk, Bennett:2024cqv, DeSantis:2025qbb, DeSantis:2025yfm, Bonanno:2023ljc, Panero:2023zdr}. In particular, in Ref.\,\cite{Frezzotti:2023nun} a method has been proposed which allows for the determination of both the real and imaginary parts of complex amplitudes in which the hadronic factor is of the generic form: 
\begin{equation}
    H(k)=i\int\dfour x\,e^{ik\cdot x}\bra{F}\,T\big[J_1(x)J_2(0)]\,\ket{I}\,,
    \label{eq:J1J2}
\end{equation}
where $T$ represents ``time ordering", $J_{1,2}$ are two local composite hadronic operators and the initial and final states, $I$ and $F$, are single hadrons or the QCD vacuum. 
The method is based on the spectral representation of the corresponding time-dependent correlation functions and overcomes the difficulties related to the analytic continuation from Minkowski to Euclidean time which arise when there are intermediate states propagating between the two currents with energies smaller than those of the external states contributing to the amplitude. 
In its numerical implementation, the method relies on the Hansen, Lupo, Tantalo (HLT) technique of Ref.\,\cite{Hansen:2019idp} for evaluating the spectral function smeared with a proper kernel. 
In the proposal of Ref.\,\cite{Frezzotti:2023nun} the smearing parameter is provided by the standard $i\epsilon$ prescription used to regularize Feynman integrals. 
For conciseness, in what follows we will refer to the approach developed in Ref.\,\cite{Frezzotti:2023nun} as the Spectral Function Reconstruction (SFR) method.

In this paper we study the decays $B\to K\ell^+\ell^-$ and $\bar{B}_s\to \gamma\ell^+\ell^-$~\footnote{We present the discussion here for  $\bar{B}_s\to\gamma\ell^+\ell^-$ decays with a valence $b$ quark, but of course it can be trivially translated to $B_s$ decays with a valence $\bar{b}$ antiquark.}, focusing particularly (but not exclusively) on contributions from the so-called \emph{charming penguins}, i.e. diagrams with charm-quark loops which contain long-distance contributions in the region of charmonium resonances as well as imaginary parts. 
Charming penguins constitute one of the main sources of power corrections to the decay amplitude computed in the infinite $m_b$ (mass of the $b$-quark) limit using QCD factorization \cite{hep-ph/9905312,Beneke:2000ry,Beneke:2001ev}, since they involve current-current operators with $\mathcal{O}(1)$ Wilson coefficients and leading Cabibbo-Kobayashi-Maskawa (CKM) factors \cite{Ciuchini:1997hb,Ciuchini:1997rj,Ciuchini:2001gv}. 
Indeed, the enhancement due to the Wilson coefficient might overcome the power suppression of the matrix element, resulting in a contribution which may, in principle, be comparable to the one generated by the semileptonic operators at leading power in the $1/m_b$ expansion, and thus be very relevant for phenomenology \cite{Ciuchini:2015qxb,Ciuchini:2020gvn}. 
In particular, charming penguin matrix elements have the same quantum numbers as the semileptonic operator with a vector-lepton coupling, and therefore their contributions might be misidentified as being due to new physics, leading to apparent deviations from the Standard Model in the angular analysis of $B \to K^* \mu^+ \mu^-$ decays \cite{Alguero:2018nvb,Hurth:2020ehu,Altmannshofer:2021qrr,Gubernari:2022hxn,Alguero:2023jeh,Wen:2023pfq,Bordone:2024hui}. 
So far, charming penguin contributions have been estimated in a model-dependent way \cite{Khodjamirian:2010vf,Khodjamirian:2012rm,Gubernari:2022hxn,Isidori:2024lng}, since such contributions cannot be evaluated using traditional lattice QCD techniques. 
In this paper, we demonstrate that the SFR method can be combined with lattice QCD to determine the contributions to the $B\to K\ell^+\ell^-$ decay amplitude from charming penguins and the chromomagnetic operator $O_8$. 
For the process $\bar{B}_s\to\gamma\ell^+\ell^-$, the charming penguin contributions involve matrix elements of three local currents (two electromagnetic ones to which the real and virtual photons couple and the weak $b\to s$ current), and we demonstrate that the SFR method can also be extended to this case, allowing for the evaluation of the amplitude from the computation of Euclidean correlation functions. 
Indeed, as we show in Sec.\,\ref{subsec:generalN}, the method can be generalized to an arbitrary number of operators.
We also note that in the amplitudes for $\bar{B}_s\to\gamma\ell^+\ell^-$ decays, two simultaneous unitarity cuts can be present. 

In evaluating the matrix elements of multilocal operators, such as those in Eq.\,(\ref{eq:J1J2}), in addition to the ultraviolet divergences which arise in the renormalization of each local operator, additional divergences can appear as ``contact terms" when two (or more) operators approach each other. 
This is the case for the decays we are studying in this paper from the region of configuration space where the electromagnetic and weak $b\to s$ currents are close to each other.
Although the renormalization of such contact terms is standard procedure, and in this case results in the appearance of the operators $O_7$ and $O_9$ in the effective Hamiltonian (see Eq.\,(\ref{eq:Hbtos}) below), a subtlety appears
when we apply the spectral density methods which require the contributions from the different time orderings to be considered separately.
In Sec.\,\ref{subsec:contact} we show that the reduction of the superficial degree of divergence from quadratic to logarithmic due to electromagnetic current conservation in diagrams such as those in Fig.\,\ref{fig:CPBK123}(a) does not hold separately for each time ordering, but only when they are combined. 
We present a method for the separation of terms with ultraviolet divergences (and hence, in particular, the cancellation of the power divergences) from the long-distance contributions requiring spectral density techniques. 

The renormalization of each of the bare lattice local operators appearing in the multi-local matrix element is largely a well studied issue and the details depend on the lattice regularization being used. The particular issue for the renormalization of the current-current operators $O_1^{(c)}$ and $O_2^{(c)}$ defined in Eq.\,(\ref{eq:O12def}) below is the non-perturbative subtraction of power divergences, made more difficult by the absence of the GIM (Glashow-Illiopoulos-Maiani) mechanism in the effective theory for $b$-decays. Although this is a more general issue than just for the applications of the HLT and SFR methods, we explain it in detail in Sec.\,\ref{subsec:power} and show how the subtraction of power divergences can be performed non-perturbatively in the particular lattice discretization which we are using in exploratory numerical studies, the details of which are given in Sec.\ref{sec:exploratory}.

In order to begin testing the feasibility of the proposed method, we have carried out an exploratory lattice QCD calculation of the Euclidean correlation functions from which the contribution to the $B \to K\ell^{+}\ell^{-}$ decay amplitude from the charming-penguin diagram in Fig.\,\ref{fig:CPBK123}(a), i.e. the diagram in which the virtual photon is emitted by the $c\bar{c}$ pair, can be extracted using the SFR/HLT method. 
This proof-of-principle study was performed on a single gauge ensemble, corresponding to a lattice spacing $a\simeq 0.08~{\rm fm}$, generated by the Extended Twisted Mass Collaboration (ETMC), with $N_f = 2+1+1$ Wilson--Clover twisted-mass dynamical fermions. 
The calculation was carried out at a single (small) value of the kaon's three-momentum, $|\vec{p}_{K}| \simeq 250~\mathrm{MeV}$, and with a lighter-than-physical $b$-quark mass set to $m_b = 2m_c$, while all other quark masses were set to their physical values. 
Using the SFR/HLT framework, we computed the charming penguin contribution to the smeared amplitude and studied its behaviour across a range of smearing parameters $\varepsilon$, comparing the results to expectations based on a simple model employing the vacuum saturation approximation (VSA). The results obtained and the prospects for a complete, high-precision calculation are discussed in Sec.\,\ref{sec:exploratory}, where we also highlight the numerical challenges posed by the presence of narrow charmonium resonances, such as the $J/\psi$ and $\psi(2S)$. 
We are pleased to note however, that the results for $q^2=(p_{\ell^+}+p_{l^-})^2$ sufficiently above the region of the main charmonium resonances are very encouraging.  
We also outline possible strategies for reliably isolating the contributions from the resonances, including combining model approaches with lattice data, which may enable a controlled extrapolation to the physical $\varepsilon \to 0$ limit, an extrapolation that becomes particularly delicate when the spectral density displays a non-smooth behaviour.

The plan for the remainder of the paper is as follows. 
In Sec.\,\ref{sec:Hbtoseff} we  recall the relevant effective $b\to s$ Hamiltonian mediating  the processes mentioned above. We then explain the reason why spectral density methods are necessary in Sec.\,\ref{sec:motivation} 
and provide an introduction to the SFR method.
The following two sections are devoted to a demonstration that the contributions from charming penguins can indeed be evaluated using this method for $B\to K\ell^+\ell^-$ decays (Sec.\,\ref{sec:latticeK}) and for $\bar{B}_s\to\mu^+\mu^-\gamma$ decays (Sec.\,\ref{sec:Btogammagammastar}). 
In Sec.\ref{sec:neglected} we briefly discuss the remaining contributions to the amplitude and explain how they can be determined using lattice computations. Issues concerning renormalization, and in particular the appearance of ultraviolet divergences due to contact terms as well as the subtraction of power divergences (i.e. terms which diverge as inverse powers of the lattice spacing) are explained in Sec.\,\ref{sec:renormalization}. 
In Sec.\,\ref{sec:exploratory} we present the results of a exploratory numerical calculation of the charming penguin contribution to the $B\to K\ell^{+}\ell^{-}$ decay and discuss the prospects for a full calculation of the hadronic amplitude. 
We present our conclusions in Sec.\,\ref{sec:concs}.
There are two appendices; Appendix\,\ref{sec:appcontact} in which we make some further comments about the renormalization of the contact terms and Appendix\,\ref{sec:appreality} in which we exploit the time reversal and parity symmetries of QCD to demonstrate that the spectral density for $B\to K\ell^+\ell^-$ decays is real.

\section{The effective weak Hamiltonian 
for {\boldmath $B\to K^{(\ast)}\ell^+\ell^-$} and 
{\boldmath $B\to\gamma\ell^+\ell^-$} decays}
\label{sec:Hbtoseff}
 The effective Hamiltonian for $b\to s$ decays can be written as:
 \begin{equation}\label{eq:Hbtos}
 {\cal H}_\mathrm{eff}=-\frac{4G_F}{\sqrt{2}}\Big\{\lambda_u\{C_1(\mu)(O_1^{(c)}(\mu)-O_1^{(u)}(\mu))+C_2(\mu)(O_2^{(c)}(\mu)-O_2^{(u)}(\mu))\}
 +\lambda_t\sum_{n=1}^{10}C_n(\mu)O_n(\mu)\Big\}\,,
 \end{equation}
 where the $C_n$ are Wilson coefficients, $\lambda_q=V_{qb}V^\ast_{qs}$ and $V$ is the CKM matrix. Defining the projectors $P_L=(1-\gamma^5)/2$ and $P_R=(1+\gamma^5)/2$, the operators in Eq.\,(\ref{eq:Hbtos}) are as follows:
  $O_{1,2}^{(q)}$ are the current-current operators
 \begin{equation}\label{eq:O12def}
 O_{1}^{(q)}=\big(\bar{s}^j\gamma^\mu P_L q^i\big)~\big(\bar{q}^i\gamma_\mu P_L b^j\big)\,,\qquad
 O_{2}^{(q)}=\big(\bar{s}^i\gamma^\mu P_L q^i\big)~\big(\bar{q}^j\gamma_\mu P_L b^j\big)\,,
 \end{equation}
 where $q=u$ or $c$ and $i,j$ are color labels. In the sum over $n$ on the right-hand side of Eq.\,(\ref{eq:Hbtos}), $O_1=O_1^{(c)}$ and $O_2=O_2^{(c)}$\,;
 $O_{3-6}$ are the QCD penguin operators,
 \begin{eqnarray}
 O_3=\big(\bar{s}\gamma^\mu P_L b\big)\,\sum_{q}\big(\bar{q}\gamma^\mu P_L q\big)\,,&\qquad&
 O_4=\big(\bar{s}^i\gamma^\mu P_L b^j\big)\,\sum_{q}\big(\bar{q}^j\gamma^\mu P_L q^i\big)\,,\nonumber\\
 O_5=\big(\bar{s}\gamma^\mu P_L b\big)\,\sum_{q}\big(\bar{q}\gamma^\mu P_R q\big)\,,&\qquad&
 O_6=\big(\bar{s}^i\gamma^\mu P_L b^j\big)\,\sum_{q}\big(\bar{q}^j\gamma^\mu P_R q^i\big)\,;
  \end{eqnarray}
 $O_{7,8}$ are the photon and gluon magnetic dipole operators
 \begin{equation}\label{eq:O78def}
 O_{7}=\frac{e}{(4\pi)^2}\,m_b\,\big(\bar{s}\sigma^{\mu\nu} P_R b\big)\,F_{\mu\nu}\,,\qquad
  O_{8}=\frac{g_s}{(4\pi)^2}\,m_b\,\big(\bar{s}\sigma^{\mu\nu} T^a P_Rb \big)\,G^a_{\mu\nu}\,;
 \end{equation}
 and $O_{9,10}$ are the semileptonic operators
 \begin{equation}\label{eq:O910def}
 O_9=\frac{e^2}{(4\pi)^2}\,\big(\bar{s}\gamma^\mu P_L b\big)\,\big(\bar{\ell}\gamma_\mu\ell\big)\,,\qquad
 O_{10}=\frac{e^2}{(4\pi)^2}\,\big(\bar{s}\gamma^\mu P_Lb\big)\,\big(\bar{\ell}\gamma_\mu\gamma^5\ell\big)\,.
 \end{equation}

There is a hierarchy among the different contributions to the decay amplitudes. The dominant contributions are from the operators $O_{7-10}$ (loop-level Wilson coefficients and tree-level matrix elements) and from operators $O_{1,2}$ (Wilson coefficients at $O(1)$\,\footnote{$C_1$ and $C_2$ appear at $O(\alpha_s)$ and tree-level respectively, but the perturbation series for both contains leading logarithms.} and loop-level matrix elements), while the contributions from the QCD penguin operators $O_{3-6}$ are suppressed by the small Wilson coefficients $C_{3-6}$ and by the loop matrix elements. The gluonic dipole operator $O_8$ plays a special role, since it has a loop-level Wilson coefficient but a tree-level matrix element when the gluon interacts with the spectator quark. Finally, the GIM-suppressed combinations $O_{1,2}^{(c)}-O_{1,2}^{(u)}$ are doubly Cabibbo suppressed and therefore negligible with respect to $O_{1,2}^{(c)}$. In this paper, we therefore work with the reduced effective Hamiltonian
\begin{equation}\label{eq:Heff7910}
    \tilde{\cal H}_\mathrm{eff}=-\frac{4G_F}{\sqrt{2}}\,V_{tb}V_{ts}^\ast\Big\{C_1O_1^{(c)}+C_2O_2^{(c)}
    + C_7O_7+C_8O_8+C_9O_9+C_{10}O_{10}\Big\}\,.
\end{equation}
We demonstrate in this paper that the matrix elements of operators $O_1^{(c)}$ and $O_2^{(c)}$, the so-called charming penguins as well as those of the chromomagnetic operator $O_8$, can be calculated using lattice QCD. 
The techniques adopted in this paper for evaluating the amplitude with the reduced effective Hamiltonian in Eq.\,(\ref{eq:Heff7910}) can also be used to evaluate the neglected matrix elements, including those containing the QCD penguin operators $O_3$\,-\,$O_6$ or CKM-suppressed operators. Their evaluation presents no new conceptual issues (although diagrams with different topologies may appear) and therefore for the main body of this paper we focus on the dominant contributions to the amplitude, the matrix element of  $\tilde{\cal H}_\mathrm{eff}$ in Eq.\,(\ref{eq:Heff7910}).

\section{Motivation for the introduction of spectral density methods}\label{sec:motivation}

In this section we explain the necessity of introducing methods based on the spectral density approach and provide an introduction to the particular approach we implement, i.e. the SFR method\,\cite{Frezzotti:2023nun} combined with the HLT procedure\,\cite{Hansen:2019idp}. 
In lattice QCD physical quantities are obtained from the computation of correlation functions at one or more Euclidean times. It is therefore natural, when considering the continuation from Minkowski to Euclidean space of multi-local operators, to divide the time integrations into contributions from different orderings of the time coordinates of the local operators. In Sec.\,\ref{subsec:TOKll} we demonstrate that one of the two time orderings for the process $B\to K\ell^+\ell^-$ contributes an imaginary part to the amplitude and in Sec.\,\ref{subsec:TOgll} we show that three of the six time orderings for the decay $\bar{B}_s\to\gamma\ell^+\ell^-$ do so as well. This motivates the introduction of spectral density techniques. The expressions for the contributions to the hadronic matrix elements are given below in Eqs.(\ref{eq:Hmuplus0}) and (\ref{eq:rhodef0}) for $B\to K\ell^+\ell^-$ decays and Eqs.\,(\ref{eq:H1munu0})\,-(\ref{eq:rho30}) for $\bar{B}_s\to\gamma\ell^+\ell^-$ decays.

The results for $B\to K\ell^+\ell^-$ decays in Eqs.\,(\ref{eq:Hmuplus0})and (\ref{eq:rhodef0}) correspond to matrix elements of two local operators, and hence one time separation between them. Their contribution to the amplitude can be evaluated by implementing the SFR method\,\cite{Frezzotti:2023nun}. On the other hand, the decays $\bar{B}_s\to\gamma\ell^+\ell^-$ correspond to matrix elements of three local operators, with two time intervals between them, and so the equations Eqs.\,(\ref{eq:H1munu0})\,-(\ref{eq:rho30}) represent an extension of the SFR method. In Sec.\,\ref{subsec:generalN} we show that such extensions can be generalized to the matrix elements of an arbitrary number, $n$ say, of operators with $n!$ time orderings and, for each time ordering, $n-1$ time intervals between the operators. 
The generic relations are presented in Eqs.\,(\ref{eq:rhoP}) and (\ref{eq:HP2}) and can be applied to processes with matrix elements of an arbitrary number of operators.

 Since in this section we consider the contributions from matrix elements containing the two current-current operators, $O_1^{(c)}$ and $O_2^{(c)}$ in the effective Hamiltonian of Eq.\,(\ref{eq:Hbtos}) (i.e. the charming penguin contributions), for which the formal discussion is identical, in order to simplify the notation we write $O_{1,2}^{(c)}$ to indicate either $O_{1}^{(c)}$ or $O_{2}^{(c)}$. 

\subsection{Time-ordered diagrams contributing to {\boldmath $B\to K\ell^+\ell^-$} decay amplitudes}
\label{subsec:TOKll}
\begin{figure}[t]
\begin{center}
\includegraphics[width=0.4\hsize]{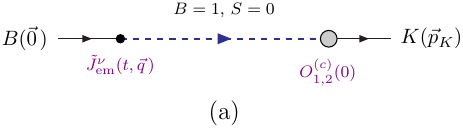}
\qquad\qquad
\includegraphics[width=0.4\hsize]{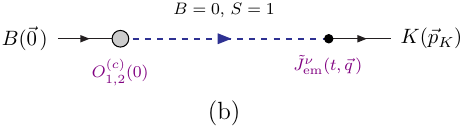}\end{center}
\caption{Schematic illustrations of the two time-orderings in Eq.\,(\ref{eq:H12def0}) contributing to $B\to K\ell^+\ell^-$ decays. The dashed lines between the operators represent all propagating states with the indicated $B$ and $S$ quantum numbers. The emission of the virtual photon is not shown. Panel (a) corresponds to the time ordering $t<0$, while panel (b) corresponds to $t>0$.}
\label{fig:BtoKtimeorderings}
\end{figure}

For $B\to K\ell^+\ell^-$ decays, with the $B$-meson at rest, the hadronic factor in the amplitude is
\begin{eqnarray}
H^\nu_{1,2}(\vecp{q})&=&i\int\dfour x\,e^{iq\cdot x}
\bra{K(\vec{p}_K)}T\big[J_\mathrm{em}^\nu(t,\vecp{x})\,O_{1,2}^{(c)}(0)\big]\ket*{B(\vecp{0})}\nn\\
&=&i\left\{\int_{-\infty}^0\hspace{-8pt} dt~e^{iq_0t}
\bra{K(\vec{p}_K)}O_{1,2}^{(c)}(0)\,\tilde{J}_\mathrm{em}^\nu(t,\vecp{q})\ket*{B(\vecp{0})}
+\int^{\infty}_0\hspace{-6pt} dt~e^{iq_0t}
\bra{K(\vec{p}_K)}\tilde{J}_\mathrm{em}^\nu(t,\vecp{q})\,O_{1,2}^{(c)}(0)\ket*{B(\vecp{0})}
\right\}\,,
\label{eq:H12def0}
\end{eqnarray}
where $T$ denotes time ordering, $\tilde{J}^\nu_\mathrm{em}(t,\vecp{q})=\int \dthree x \,e^{-i\vec{q}\cdot\vec{x}}\,J_\mathrm{em}^\nu(t,\vecp{x})$ and $J^\nu_\mathrm{em}(t,\vecp{x})
=\sum_f Q_f~\bar{f}(t,\vecp{x})\gamma^\nu f(t,\vecp{x})$ ($f$ denotes the flavor quantum number of the quark and $Q_{u,c}=2/3$ and $Q_{d,s,b}=-1/3$)
is the electromagnetic current coupling to the virtual photon with four-momentum $q$.
Note that $q_0=m_B-E_K$, where $E_K$ is the energy of the kaon, $E_K=\sqrt{m_K^2+\vecp{q}^2}$, so that $H^\nu_{1,2}$ is a function of $\vec{q}$ only.
We now consider each of the two terms on the right-hand side of the second line of Eq.\,(\ref{eq:H12def0}) in turn:\\[0.1cm]
i) In the first term on the right-hand side of Eq.\,(\ref{eq:H12def0}), in which $t<0$, the hadronic states propagating between the two operators have Beauty and Strangeness quantum numbers $B=1$ and $S=0$ respectively and three-momentum $-\vec{q}$ (see Fig.\,\ref{fig:BtoKtimeorderings}(a)). They therefore have energies greater than $m_B-q_0$. This contribution to the amplitude is therefore real and can be calculated in the standard way from lattice computations of Euclidean correlation functions (see Eq.\,(\ref{eq:Hmuminus}) below).\\[0.05cm]
ii) In the second term on the other-hand, in which $t>0$ (see Fig.\,\ref{fig:BtoKtimeorderings}(b)), the operators $O_{1,2}^{(c)}(0)$ annihilate the $\bar{b}$-quark and create an $\bar{s}$-quark, so that the states propagating between the two operators have $B=0$ and $S=1$. 
There is a continuous spectrum of such on-shell states with energies less than $m_B$ and so there is an imaginary contribution to the amplitude. 
To be specific, the contribution to $H^\nu_{1,2}(\vecp{q})$ from the region $t>0$ can be written in the spectral density form (see Eq.\,(\ref{eq:Hmuplus}) below)
\begin{equation}
H^{\nu+}_{1,2}(\vecp{q})=\int_{E^\ast}^\infty \frac{dE}{2\pi}\,\frac{\rho^{\nu+}_{1,2}(E,\vecp{q})}{E-m_{B}-i\epsilon}\,.
\label{eq:Hmuplus0}
\end{equation}
where the spectral density is given by
\begin{equation}\label{eq:rhodef0}
\rho_{1,2}^{\nu+}(E,\vecp{q})=\bra{K(-\vecp{q})}J^\nu_\mathrm{em}(0)\,(2\pi)^3\delta(\hat{\bs{P}})\,(2\pi)\delta(\hat{H}-E)\,O_{1,2}(0)\ket*{B(\vecp{0})}\,,
\end{equation}
$\hat{\bs{P}}$ and $\hat{H}$ are the three-momentum and Hamiltonian operators respectively and $E^\ast$ is the lowest energy of the on-shell intermediate states propagating between the two operators.
Since there are on-shell states with $E<m_B$ and (as always) states with $E>m_B$, the pole at $E=m_B$ is present in the region of integration over $E$ and the $i\epsilon$ cannot be dropped. 
The SFR method is designed to determine both the real and imaginary parts of the amplitude. 
It has been proposed and applied for the first time to the the decays $P\to\ell\bar\nu_\ell (\ell^{\prime\,+} \ell^{\prime\,-})$, where $\ell$ and $\ell^\prime$ are charged leptons\,\cite{Frezzotti:2023nun}, and then to the matrix element of $O_7$ in $B\to\gamma\ell^+\ell^-$ decays\,\cite{Frezzotti:2024kqk}. In Sec.\,\ref{sec:latticeK} we apply the method to the matrix elements of $O_{1,2}^{(c)}$ in 
$B\to K\ell^+\ell^-$ decays.

\subsection{Time-ordered diagrams contributing to {\boldmath$\bar{B}_s\to \gamma\ell^+\ell^-$} decay amplitudes}\label{subsec:TOgll}

For $B\to \gamma\ell^+\ell^-$ decays, with the $B$-meson at rest, the hadronic factor in the amplitude is
\begin{equation}\label{eq:HmunuBtogammagamma}
H^{\mu\nu}_{1,2}(\vecp{k})=i\,\int dt\int \dthree x\int dt_W\int\dthree y~\bra*{0}T\big[J^\mu_\gamma(t,\vecp{x})\,J^\nu_{\gamma^\ast}(0,\vecp{y})\,O_{1,2}^{(c)}(t_W,\vecp{0})\big]\ket*{\bar{B}_s(\vecp{0})}\,e^{ik\cdot x}\,e^{i\vec{k}\cdot \vec{y}}\,,
\end{equation}
where $J^\mu_\gamma$ and $J^\nu_{\gamma^\ast}$ are the electromagnetic currents which emit the real and virtual photons with momenta $k$ and $q=p_B-k$ respectively, $k^2=0$ and $p_B=(m_{\bar{B}_s},\vecp{0})$\,\footnote{To simplify the notation we have omitted the label {\footnotesize em} on the currents, leaving it implicit that these are electromagnetic currents defined after Eq.\,(\ref{eq:H12def0}). It is convenient to distinguish the position at which the real and virtual photons are emitted and this we do by the labels $\gamma$ and $\gamma*$ respectively.}. Now the matrix element is of three local operators and there are six possible time-orderings, three of which give contributions to the amplitude containing an imaginary part.

\begin{figure}[t]
\begin{center}
\includegraphics[width=0.45\hsize]{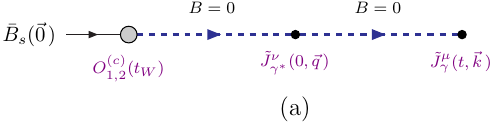}
\qquad\qquad
\includegraphics[width=0.45\hsize]{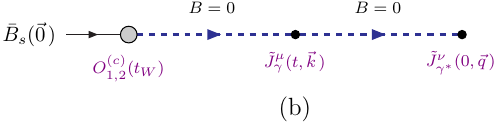}\\[0.1cm]
\includegraphics[width=0.45\hsize]{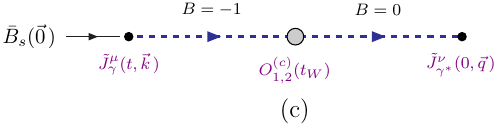}
\end{center}
\caption{Schematic illustrations of the three time-orderings in Eq.\,(\ref{eq:HmunuBtogammagamma}) which contribute an imaginary part to the $B\to \gamma\ell^+\ell^-$ decay amplitude. The dashed lines between the operators represent all propagating states with the indicated $B$ quantum number. The emission of the real and virtual photons is not shown.
Panel (a) corresponds to the time ordering $t_W<0<t$, panel (b) to $t_W<t<0$, while panel (c) to $t<t_W<0$.}
\label{fig:Btogammatimeorderings}
\end{figure}

i) $t_W<0<t$ (Fig.\,\ref{fig:Btogammatimeorderings}(a)): The operators $O^{(c)}_{1,2}(t_W)$ annihilate the $\bar{b}$ antiquark so that the hadronic states propagating between $t_W$ and time $0$ have Beauty quantum number $B=0$ (and Strangeness S=0) and there is a continuum of such on-shell states with energies below $m_{\bar{B}_s}$. 
This contribution therefore has an imaginary part. 
The on-shell states propagating between times $0$ and $t$ also have $B=0$, however they have positive invariant masses and therefore cannot combine into the final-state massless photon. Only one of the two time intervals, $[t_W,0]$, allows for energy-conserving on-shell propagating hadronic states. We will see in Eq.\,(\ref{eq:H1integrations}) in Sec.\ref{sec:Btogammagammastar} that this first contribution can be written in the spectral density form\,\footnote{In order to simplify the notation we have dropped the labels {\scriptsize $1,2$} on $H^\nu$. The subscript {\scriptsize$1$} on $H_1^{\mu\nu}(\vecp{k})$ indicates the first time-ordering and Eq.\,(\ref{eq:H1munu0}) holds for both operators $O_1^{(c)}$ and $O_2^{(c)}$.}
\begin{equation}
H_1^{\mu\nu}(\vecp{k})= \int_{E_1^\ast}^\infty\frac{dE_1}{2\pi}\int_{E_2^\ast}^\infty\frac{dE_2}{2\pi}\frac{\rho_1^{\mu\nu}(E_1,E_2,\vecpp{k})}{(E_1-m_{\bar{B}_s}-i\epsilon)(E_2-k_0)}\,,\label{eq:H1munu0}
\end{equation}
where
\begin{equation}
\rho_1^{\mu\nu}(E_1,E_2,\vecpp{k})
=\bra{0}J^\mu_\gamma(0)\,(2\pi)^4\delta(\hat{\bs{P}}-\vecp{k})\,\delta(\hat{H}-E_2)\,J^\nu_{\gamma^\ast}(0)\,(2\pi)^3\delta^{(3)}(\hat{\bs{P}})\,\delta(\hat{H}-E_1)\,O^{(c)}_{1,2}(0)\ket*{\bar{B}_s(\vecp{0})}\,,
\label{eq:rhoa0}\end{equation}
$\hat{\bs{P}}$ and $\hat{H}$ are again the three-momentum and Hamiltonian operators and $E_{1,2}^\ast$ are the threshold energies for the two channels. 
 The first factor in the denominator, $m_{\bar{B}_s}-E_1+i\epsilon$, arises from the $t_W$ integration whereas the second, $k_0-E_2$ comes from the $t$ integration. Since $k_0<E_2$, we drop the $i\epsilon$ in this factor (although as explained in Sec.\ref{subsec:tw0t} this is not necessary).

ii) $t_W<t<0$ (Fig.\,\ref{fig:Btogammatimeorderings}(b)): As in the previous case, the operators $O^{(c)}_{1,2}(t_W)$ annihilate the $\bar{b}$ antiquark so that the hadronic states propagating between $t_W$ and time $t$ have Beauty quantum number $B=0$ (and Strangeness S=0) and there is a continuum of such on-shell states with energies below $m_{\bar{B}_s}$ leading to an imaginary part in the corresponding propagator. In addition however, the hadronic states propagating between $t$ and $0$ also have $B=0$ and if $q^2$ is greater than the threshold for such states, (e.g. $\pi\pi$ states in a $p$-wave), the corresponding propagator also contains an imaginary part. 

We will see in Eq.\,(\ref{eq:H2integrations}) in Sec.\,\ref{sec:Btogammagammastar} that this second contribution can be written in the form:
\begin{equation}\label{eq:Hmunu20}
H_2^{\mu\nu}(\vecp{k})=-\int_{E_1^\ast}^\infty\frac{dE_1}{2\pi}\,\int_{E_2^\ast}^\infty\frac{dE_2}{2\pi}\,
\frac{\rho_2^{\mu\nu}(E_1,E_2,\vecp{k})}{(E_1
-m_{\bar{B}_s}-i\epsilon)\,(E_2+k_0-m_{\bar{B}_s}-i\epsilon)}\,,
\end{equation}
where
\begin{equation}
\rho_2^{\mu\nu}(E_1,E_2,\vecp{k})=\bra{0}J^\nu_{\gamma^\ast}(0)\,(2\pi)^4\delta^{(3)}(\hat{P}+\vecp{k})\,\delta(\hat{H}-E_2)\,J^\mu_\gamma(0)\,(2\pi)^4\delta^{(3)}(\hat{P})\, \delta(\hat{H}-E_1)\,O_{1,2}^{(c)}(0)\ket*{\bar{B}_s(\vecp{0})}\,.
\end{equation} 
The energy $E_1$ is that of the zero momentum $B=0$ states propagating between $t_W$ and $t$ and therefore the first pole in the denominator in Eq.\,(\ref{eq:Hmunu20}) is always in the range of the $E_1$ integration and, as before, the corresponding $i\epsilon$ cannot be dropped. The energy $E_2$ is that of the $B=0$ states with momentum $\vec{q}=-\vec{k}$, and whether the second pole in the denominator of Eq.\,(\ref{eq:Hmunu20}) is in the range of the $E_2$ integration depends on the value of $\vec{k}$. If $q_0=m_{\bar{B}_s}-k_0$ is greater than $E_2^\ast$ then the second pole will be in the range of integration and again we cannot drop the $i\epsilon$. This is the case, for example, when $q^2$ is above the charmonium resonances and obtaining the amplitude from Euclidean correlation functions requires a generalization of the SFR method in order to use the HLT approach to evaluate the spectral density.

In Eq.\,(\ref{eq:Hmunu20}) we have written the two propagators with the same $i\epsilon$ and the physical result is obtained by taking the limit $\epsilon\to 0$. 
We note that in intermediate stages of the SFT/HLT procedure, different values of $\epsilon$ could be used in each propagator. 
Whether there is any advantage in using this freedom, particularly in reducing lattice artifacts, can be investigated in future numerical computations. 

iii) $t<t_W<0$ (Fig.\,\ref{fig:Btogammatimeorderings}(c)): For this time ordering the emission of the real photon at $t$, implies that the states propagating between $t$ and $t_W$ have $B=-1$ and three momentum $\vec{q}=-\vec{k}$ and hence an energy greater than $m_{\bar{B}_s}$. The corresponding propagator is therefore real. On the other hand the states propagating between $t_W$ and $0$ have $B=0$ and again if $q^2$ is greater than the threshold for such states then there is an imaginary part in the amplitude.

We will see in Eq.\,(\ref{eq:H3}) in Sec.\,\ref{sec:Btogammagammastar} that this third contribution can be written in the form:
\begin{equation}
H^{\mu\nu}_3(\vecp{k})=\int_{E_1^\ast}^\infty \frac{dE_1}{2\pi}\int_{E_2^\ast}^\infty \frac{dE_2}{2\pi}\,\frac{\rho_3(E_1,E_2,\vecp{k})}{(E_1-m_{\bar{B}_s}+k_0)(E_2-q_0-i\epsilon)}\,,\label{eq:H3munu0}
\end{equation}
where $q_0=m_{\bar{B}_s}-k_0$ is the energy of the virtual photon and
\begin{equation}\label{eq:rho30}
\rho_3^{\mu\nu}(E_1,E_2,\vecp{k})=\bra{0}J^\nu_{\gamma^\ast}(0)\,(2\pi)^4\delta^{(3)}(\hat{P}+\vecp{k})\,\delta(\hat{H}-E_2)\,O_{1,2}^{(c)}(0)\,(2\pi)^4\delta^{(3)}(\hat{P}+\vecp{k}) \delta(\hat{H}-E_1)J^\mu_\gamma(0)\ket*{\bar{B}_s(\vecp{0})}\,.
\end{equation}
The form of the denominator in Eq.\,(\ref{eq:H3munu0}) reflects the discussion in the previous paragraph. The energy $E_1$ is that of $B=-1$ intermediate states propagating between $t$ and $t_W$ and is greater than $m_{\bar{B}_s}-k_0$ for all values of $\vec{k}$. The corresponding pole is therefore not in the range of the $E_1$ integration and we have dropped the $i\epsilon$ in the corresponding propagator (although as explained in Sec.\ref{subsec:ttw0} this is not necessary). The energy $E_2$ is that of the $B=0$ intermediate states propagating between $t_W$ and $0$ and for larger values of $q_0$, the pole at $E_2=q_0$ is in the range of the $E_2$ integration and we therefore need to keep the $i\epsilon$.

The remaining 3 time orderings, $0<t_W<t$, $0<t<t_W$ and $t<0<t_W$ do not have energy-conserving intermediate states which can go on shell and hence only give real contributions to the amplitude. Their computation therefore does not require the use of spectral density methods.

\subsection{Generalized spectral density representation}\label{subsec:generalN}
In this paper we are particularly interested in $B\to K\ell^+\ell^-$ and $\bar{B}_s\to\gamma\ell^+\ell^-$ decays, however we now show that the discussion above can be generalized to hadronic factors containing the matrix elements of an arbitrary number $n$ of composite local operators. 
Consider the multilocal hadronic matrix element
\begin{equation}\label{eq:generalH}
H(k_I,k_F;k_1,k_2,\cdots,k_n)=\int\prod_{i=1}^n\big\{\dfour x_i\,e^{ik_i\cdot x_i}\big\}~\bra{F(k_F)}T\big[O_1(x_1)\cdots O_n(x_n)\big]\ket{I(k_I)}\,,
\end{equation}
where the initial and final states, $\ket{I(k_I)}$ and $\ket{F(k_F)}$, are either on-shell single-particle states or possibly the vacuum.  In the discussion above, for $B\to K\mu^+\mu^-$ decays $n=2$ and the initial and final states are the $B$ and $K$ mesons, whereas for $\bar{B}_s\to\gamma\ell^+\ell^-$ decays $n=3$ and the initial and final states are the $\bar{B}_s$-meson and the vacuum respectively. 

For each of the $n!$ time-orderings we can write the contribution to $H(k_I,k_F;k_1,k_2,\cdots,k_n)$ in terms of a spectral density. 
In the above, we have done so explicitly for those time-orderings which contribute an imaginary part to the $B\to K\ell^+\ell^-$ decay amplitude in Eqs.\,(\ref{eq:Hmuplus0}) and (\ref{eq:rhodef0}) or to the $\bar{B}_s\to\gamma\ell^+\ell^-$ amplitude in Eqs.\,(\ref{eq:H1munu0})\,-\,(\ref{eq:rho30}). 
We have not written the contributions to $H$ from the remaining time-orderings because they can be evaluated using standard techniques without requiring spectral density techniques. For the general case in Eq.(\ref{eq:generalH}) we start by writing
\begin{equation}
T\bra{F(k_F)}\, O_1(x_1)\cdots O_n(x_n)\, \ket{I(k_I)}
=
\sum_{P=1}^{n!}\theta(x_{P_1}^0-x_{P_2}^0)\cdots\theta(x_{P_{n-1}}^0-x_{P_n}^0)\,
\bra{F(k_F)}\, O_{P_1}(x_{P_1})\cdots O_{P_n}(x_{P_n})\, \ket{I(k_I)}
\,,
\label{eq:nazsplittimes}
\end{equation}
where the sum runs over the $n!$ permutations of the hadronic operators and $\{P_1,\cdots,P_n\}$ is the ordering of the indices $\{1,\cdots,n\}$ generated by the permutation $P$. We express each matrix element in the sum on the right-hand side of Eq.\,(\ref{eq:nazsplittimes}) as follows:
\begin{eqnarray}
\bra{F(k_F)}\, O_{P_1}(x_{P_1})\cdots O_{P_n}(x_{P_n})\, \ket{I(k_I)}
\nonumber \\[8pt]
&&\hspace{-2in}=
e^{ik_F\cdot x_{P_1}-ik_I\cdot x_{P_n}}\,
\bra{F(k_F)}\, O_{P_1}(0)\, e^{-i\hat{P}\cdot(x_{P_1}-x_{P_2})}\, \cdots 
O_{P_{n-1}}(0)\, e^{-i\hat{P}\cdot(x_{P_{n-1}}-x_{P_n})} O_{P_n}(0)\, \ket{I(k_I)}
\nonumber \\[8pt]
&&\hspace{-2in}=
e^{ik_F\cdot x_{P_1}-ik_I\cdot x_{P_n}}\,
\int
\frac{d^4p_1}{(2\pi)^4}\cdots \frac{d^4p_{n-1}}{(2\pi)^4}~
\Big\{\prod_{j=1}^{n-1}\,e^{-ip_j\cdot(x_{P_j}-x_{P_{j+1}})}\Big\}~
\rho_P(p_1,\cdots,p_{n-1})\,,
\label{eq:nazspectreP}
\end{eqnarray}
where $\hat{P}=(\hat{H},\hat{\mathbf{P}})$ is the four-momentum operator and the generalized spectral density is given by
\begin{equation}\label{eq:rhoP}
\rho_P(p_1,\cdots,p_{n-1})
=
\bra{F(k_F)}\, O_{P_1}(0)\, (2\pi)^{(4)}\delta^{(4)}(\hat{P}-p_1)\,O_{P_2}(0) \cdots 
O_{P_{n-1}}(0)\, (2\pi)^4\delta^{(4)}(\hat{P}-p_{n-1}) \,O_{P_n}(0)\, \ket{I(k_I)}\,.
\end{equation}
We now define
\begin{equation}
\bar{k}_{P_j}=k_F+\sum_{i=1}^j k_{P_j}
\end{equation}
and perform the spatial integrals in Eq.\,(\ref{eq:generalH}) to obtain the contribution to $H(k_I,k_F;k_1, k_2, \cdots, k_n)$ from this time ordering (which we label by $H_P(k_I,k_F;k_1,k_2,\cdots, k_n)$)
\begin{eqnarray}
H_P(k_I,k_F;k_1,k_2,\cdots, k_n)&=&(2\pi)^3\delta^{(3)}\big(\vec{k}_I-\vec{k}_F-\sum_{i=1}^n \vec{k}_i\big) \Big\{\prod_{i=1}^{n-1}\int\!\frac{\dfour p_i}{(2\pi)^4}\,(2\pi)^3
\delta^{(3)}(\vec{p}_i-\vec{\bar{k}}_{P_i})\Big\}\,\rho_P(p_1,\cdots,p_{n-1})
\label{eq:HP1}\\
&&\hspace{-1.7in}\int \!dx^0_{P_1}\!\int dx^0_{P_2}\cdots \int\! dx^0_{P_n}~\theta(x_{P_1}^0-x_{P_2}^0)\cdots\theta(x_{P_{n-1}}^0-x_{P_n}^0)~e^{i(k_F^0x_{P_1}^0-k_I^0x_{P_n}^0)}
\Big\{\prod_{r=1}^{n}\,e^{ik^0_{P_r}x^0_{P_r}}\Big\}
\Big\{\prod_{j=1}^{n-1}\,e^{-ip^0_j(x^0_{P_j}-x^0_{P_{j+1}})}\Big\}\,,\nn
\end{eqnarray}
where we have used $\sum_{j=1}^n\,\vec{k}_{P_j}=\sum_{j=1}^n\,\vec{k}_{j}$\,.

Each of the step functions in eq.(\ref{eq:HP1}) can be written in the representation
\begin{equation}
\theta(x^0_{P_j}-x^0_{P_{j+1}})=\int\frac{dE_j}{2\pi i}~\frac{e^{iE_j(x^0_{P_j}-x^0_{P_{j+1}})}}{E_j-i\epsilon}\,,
\end{equation}
where the $\epsilon\to 0$ limit is implicit. The time integrations can now be performed giving 
\begin{equation}\label{eq:HP2}
H_P(k_I,k_F;k_1,k_2,\cdots, k_n)=(-i)^{n-1}(2\pi)^4\,\delta^{(4)}\big(k_I-k_F-\sum_{i=1}^n k_i\big)\,\Big\{\prod_{i=1}^{n-1}\int\!\frac{\dfour p_i}{(2\pi)^4}\,\frac{(2\pi)^3
\delta^{(3)}(\vec{p}_i-\vec{\bar{k}}_{P_i})}{p_i^0-\bar{k}^0_{P_i}-i\epsilon}\Big\}\,\rho_P(p_1,\cdots,p_{n-1})\,,
\end{equation}
and $H(k_I,k_F;k_1,k_2,\cdots, k_n)=\sum_{P=1}^{n!}H_P(k_1,k_2,\cdots, k_n)$.

\begin{figure}[t]
\begin{center}
\includegraphics[width=0.7\hsize]{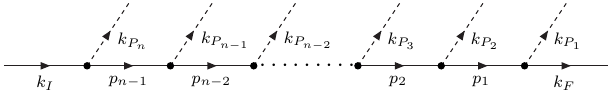}
\end{center}\caption{Schematic diagram illustrating the time ordering described in the text. The dashed lines represent the particles coupling to the operators $O_{P_i},\,(i=1,\cdots,\!n)$ and carrying the momentum $k_{P_i}$.\label{fig:General}}
\end{figure}

The results in Eqs.\,(\ref{eq:rhoP}) and (\ref{eq:HP2}) are illustrated in Fig.\,\ref{fig:General}. Since for each time ordering, the integrals over the spatial coordinates $\vec{x}_{P_i}$ are performed over all space, three-momentum is conserved at each vertex so that $\vec{p}_i=\vec{\bar{k}}_i$ as stated in Eq.\,(\ref{eq:HP2}). Since for each time-ordering energy is not conserved at the vertices, an integral over the energies remains, with a singularity when the energy $p_i^0$ of an internal state is equal to the corresponding external energy $\bar{k}_{P_i}$. As we have seen in the examples in Secs.\,\ref{subsec:TOKll} and \ref{subsec:TOgll}, whether or not the poles are present for a particular time ordering depends on the quantum numbers of the operators and states. In those examples, there is no three-momentum transfer at the weak operator $O_{1,2}^{(c)}$.

For the specific decays and time-orderings considered in Secs.\,\ref{subsec:TOKll} and \ref{subsec:TOgll} and in the following sections, we have used a different notation which however, can readily be seen to be equivalent to the general formula in Eq.\,(\ref{eq:HP2}). Thus for example, in Eq.\,(\ref{eq:H1munu0}) for the first time ordering, $n=3$, $k_I^0=m_{\bar{B}_s}$, $k_F^0=0$, $\bar{k}_{P_1}^0=k^0$, $\bar{k}_{P_2}^0=m_{\bar{B}_s}$ and we have replaced the integration variables $p_1^0$ and $p_2^0$ by $E_2$ and $E_1$ respectively. The expressions for the hadronic factors in Eqs.\,(\ref{eq:Hmuplus0}), (\ref{eq:Hmunu20}) and (\ref{eq:H3munu0}) together with the corresponding spectral densities, can similarly be seen to be specific examples of Eqs.\,(\ref{eq:HP2}) and (\ref{eq:rhoP}).

In the discussion in Secs.\,\ref{subsec:TOKll} and \ref{subsec:TOgll} we have imposed four momentum conservation (i.e. $k_I=k_F+\sum_{i=1}^nk_i$ in the notation of Eq.\,(\ref{eq:generalH})) by hand and used translation invariance to set the position of one of the operators to be at the origin. 
Apart from this, we now see that for the $B\to K\ell^+\ell^-$ decays, Eqs.\,(\ref{eq:rhodef0}) and (\ref{eq:Hmuplus0}) for the $t>0$ time ordering, i.e. the one which requires the spectral density method, is a special case for $n=2$ of the general result in Eqs.\,(\ref{eq:HP2}) and (\ref{eq:rhoP}). In this case, $n=2$, $k_I=(m_B,\vec{0})$ and $k_F=(E_K,-\vecp{q})$, where $\vec{q}$ is the momentum of the virtual photon. Similarly for $\bar{B}_s\to\gamma\ell^+\ell^-$ decays for which $n=3$, Eqs.\,(\ref{eq:H1munu0})\,-\,(\ref{eq:rho30}) correspond to the three time orderings which require the spectral density approach. 

The above discussion has been presented in Minkowski space. The HLT method allows for an evaluation of the spectral densities from Euclidean correlation functions. A fully general analysis of the connection of spectral densities with $S$-matrix elements has been performed in Ref.\,\cite{Patella:2024cto} by generalizing the arguments of Ref.\,\cite{Barata:1990rn}. 

As an illustration of the SFR method proposed for $n=2$ in Ref.\,\cite{Frezzotti:2023nun}, we consider $B\to K\ell^+\ell^-$ decays discussed in Sec.\,\ref{subsec:TOKll}.
We start with Eq.\,(\ref{eq:Hmuplus0}), 
\begin{equation}
H^{\nu+}_{1,2}(\vecp{q})=\int_{E^\ast}^\infty \frac{dE}{2\pi}\,\frac{\rho^{\nu+}_{1,2}(E,\vecp{q})}{E-m_B-i\epsilon}=i\int_0^\infty dt\,e^{iq_0t}\,C_{1,2}^{\nu+}(t,\vecp{q})\,,
\label{eq:HCnuplus0}\end{equation}
and the (Minkowski) correlation function is given in terms of the spectral density 
by
\begin{equation}\label{eq:Cnuplus0}
C^{\nu+}_{1,2}(t,\vecp{q})=\int_{E^\ast}^\infty \frac{dE}{2\pi}\,e^{-i(E-E_K)t}\,\rho^{\nu+}_{1,2}(E)\,,
\end{equation}
where $E_K=\sqrt{m_K^2+\vecpp{q}^2}$ is the energy of the final state kaon. The expression for 
the correlation function in terms of the spectral density can readily be continued to Euclidean space so that it is amenable to lattice computations,
\begin{equation}\label{eq:CnuplusE}
C^{\nu+}_{1,2;E}(t,\vecp{q})=\int_{E^\ast}^\infty \frac{dE}{2\pi}\,e^{-(E-E_K)t}\,\rho^{\nu+}_{1,2}(E)\,,
\end{equation}
where the subscript $E$ on the left-hand side denotes \emph{Euclidean}~\footnote{Note that $H^{\nu+}_{1,2}(\vecp{q})= \int_0^\infty dt\,e^{q_0t}\,C_{1,2;E}^{\nu+}(t,\vecp{q})$ is an incorrect continuation of Eq.\,(\ref{eq:HCnuplus0}) to Euclidean space. This time integral diverges at large $t$ due to the contributions from intermediate states with energies smaller than $m_B$, thus preventing the physical amplitude from being obtained.}. 
The HLT method\,\cite{Hansen:2019idp} is based on the expansion of the smearing kernel in terms of exponentials as follows:
\begin{equation}\label{eq:expansion0}
    \frac{1}{E-m_B-i\epsilon}\simeq\sum_{n}^{n_\mathrm{max}}g_n(m_B,\epsilon)\,e^{-aEn}\,,
\end{equation}
where the $g_n(m_B,\epsilon)$ are complex coefficients and $a$ is the lattice spacing. 
This relies on the key observation of Ref.\,\cite{Barata:1990rn} (generalized in Ref.~\cite{Patella:2024cto}) that by relying on the Stone-Weierstrass theorem such a representation is \emph{exact} in the limit $n_\mathrm{max}\to\infty$.
We now define the smeared hadronic factor
\begin{eqnarray}
\tilde{H}^{\nu+}_{1,2}(\vecp{q};\epsilon)\equiv\int_{E^\ast}^\infty \frac{dE}{2\pi}\,\frac{\rho^{\nu+}_{1,2}(E,\vecp{q})}{E-m_B-i\epsilon}
&\simeq&\sum_{n=1}^{n_\mathrm{max}}g_n(m_B,\epsilon)\int_{E^\ast}^\infty \frac{dE}{2\pi}\,e^{-aEn}\,\rho^{\nu+}_{1,2}(E,\vecp{q})\nn\\&=&
\,\sum_{n=1}^{n_\mathrm{max}}g_n(m_B,\epsilon)\,e^{-aE_Kn}\,C_{1,2;E}^{\nu+}(an,\vecpp{q}).\label{eq:reconstruction0}
\end{eqnarray}
Thus $\tilde{H}^{\nu+}_{1,2}(\vecp{q};\epsilon)$ is given in terms of Euclidean correlation functions which can be computed in Lattice QCD. 
After performing the computations at several values of $\epsilon$, if the $\epsilon\to 0$ extrapolation can be performed reliably, then the physical $H_{1,2}(\vecp{q})$ is obtained.

The issues concerning the limit $\epsilon\to 0$ have been been discussed in detail in Ref.\,\cite{Frezzotti:2023nun} and are sketched in Sec.\,\ref{sec:num_results} below, when we discuss our numerical results. The first issue concerns the fact that the computations are performed on finite
spatial volumes with discrete spectra requiring the condition $\epsilon L\gg 1$ to obtain reliable smeared results.
When this condition is
satisfied, finite volume effects, which are known to be exponentially small in $\epsilon L$~\cite{Bulava:2019kbi}, can be safely kept under control.
The second issue is that
the smallest values of $\epsilon$ that can be reached in practice, depends on both the finite numbers of discrete lattice points at which the
relevant time-dependent correlation function has been computed, i.e. on the temporal extension T of the lattice, and
on the statistical accuracy of the lattice data. In order to ensure good control of the extrapolation $\epsilon\to 0$, at any selected value of $E$, the condition $\epsilon\ll\Delta(E)$ must be satisfied, where $\Delta(E)$ indicates the typical size of
the interval around $E$ in which the hadronic amplitude is significantly varying, such as in a region where narrow resonances contribute to the spectral density. In such cases the condition $\epsilon\ll\Delta(E)$ may be difficult to be fulfilled. In such cases, in order to obtain results with sufficiently small values of $\epsilon$ will require extremely high statistical accuracy of the lattice data. For this reason, in our numerical study in Sec.\,\ref{sec:num_results} we evaluate the amplitude at a value of $q^2$ which is above the squared masses of the narrow charmonium resonances, where the amplitude behaves relatively smoothly. In such an \textit{asymptotic} regime, one can expect that the extrapolation can be performed using a polynomial ansatz.

The decay $\bar{B}_s\to\gamma\ell^+\ell^-$, corresponds to the case $n=3$, and for each time-ordering there are two time intervals with different states propagating in each. In particular, we have seen in Eq.\,(\ref{eq:Hmunu20}), that, depending on the momentum of the real photon, there may be two poles in the ranges of the energy integrations. 
For illustration consider a particular time-ordering $P$ for $n=3$ and the corresponding smeared kernel
\begin{equation}\label{eq:Kepsp1p2}
K_\epsilon(p_1^0,p_2^0)=
K^R_\epsilon(p_1^0,p_2^0)+iK^I_\epsilon(p_1^0,p_2^0)
=
\frac{-i}{(p_1^0-E_F-E_1-i\epsilon)}~\frac{-i}{(p_2^0-E_F-E_1-E_2-i\epsilon)}\;,
\end{equation}
where $E_F=k_F^0$, $E_1=k_{P_1}^0$ and $E_2=k_{P_2}^0$. 
Refs.\,\cite{Barata:1990rn,Patella:2024cto} allow us to write $K_\epsilon(p_1^0,p_2^0)$
in the form
\begin{equation}\label{eq:KrepSW}
K^X_\epsilon(p_1^0,p_2^0)=
\lim_{n_\mathrm{max}\to\infty}\sum_{n_1=1}^{n_\mathrm{max}}\sum_{n_2=1}^{n_\mathrm{max}}\, 
g_\epsilon^X(n_1,n_2)\, e^{-a n_1 p_1^0-a n_2 p_2^0}\;,
\qquad X=\{R,I\}\,,
\end{equation}
where we have introduced the index $X=\{R,I\}$ in order to be able
to deal separately with the real and imaginary components. In order to simplify the notation we have not explicitly shown the dependence of the coefficients $g^X_\epsilon(n_1,n_2)$ upon $E_F$, $E_1$, $E_2$ and $n_\mathrm{max}$. By using the representation of Eq.\,(\ref{eq:KrepSW}), the connection of $H(k_1,k_2)$ with Euclidean lattice correlation functions is readily established:
\begin{equation}
H_P(k_1,k_2)= \lim_{\epsilon\to 0^+}\lim_{n_\mathrm{max}\to \infty}
\sum_{n_1=1}^{n_\mathrm{max}} \sum_{n_2=1}^{n_\mathrm{max}}\, 
\left[g^R_\epsilon(n_1,n_2)+ig^I_\epsilon(n_1,n_2)
\right]\,
C_P(an_1,an_2)\,,
\label{eq:HPexprep}
\end{equation}
where the (amputated) Euclidean correlation function $C_P(an_1,an_2)$ is given by
\begin{flalign}
C_P(an_1,an_2)
&=
e^{-a n_1 p_1^0-a n_2 p_2^0}\,
\rho_P(p_1^0,p_2^0)\,,
\end{flalign}
and $\rho_P(p_1^0,p_2^0)=0$ if $p_1^0< p_1^\ast$ and/or $p_2^0< p_2^\ast$.
We will present explicit expressions for the three time-orderings for $\bar{B}_s\to\gamma\ell^+\ell^-$ decays in Sec.\,\ref{sec:Btogammagammastar} below.

The optimal determination of the coefficients $g^X_\epsilon(n_1,n_2)$ is delicate and is a balance between the systematic uncertainties in the fit of $K^X_\epsilon(p_1^0,p_2^0)$ in terms of exponentials (in which the coefficients generally become large and oscillate in sign) and the statistical uncertainties in the correlation functions. Within the HLT algorithm, explicitly generalized to the multidimensional case in Ref.\,\cite{Patella:2024cto}, this balance is achieved by determining the coefficients from the minimization of the functional
\begin{flalign}
W_X[\vec g] = \frac{A_X[\vecp g]}{A_X[\vecp 0]} + \lambda B[\vecp{g}]\;,
\label{eq:HLT1}
\end{flalign}
where $A_X[\vecp g]$, $B[\vecp g]$ and $\lambda$ are the so-called norm-functional, error-functional and trade-off parameter. The norm-functional is a measure of the distance in functional space of the target kernel $K_\epsilon^X(p_1^0,p_2^0)$ from its approximation in terms of exponentials. A possible definition is
\begin{flalign}
A_X[\vecp g] =
\int_{p_1^\ast}^\infty dp_1^0\, \int_{p_2^\ast}^\infty dp_2^0\,
\left[
K^X_\epsilon(p_1^0,p_2^0) -
\sum_{n_1=1}^{n_\mathrm{max}}\sum_{n_2=1}^{n_\mathrm{max}}\, 
g(n_1,n_2)\, e^{-a n_1 p_1^0-a n_2 p_2^0}
\right]^2\;,
\label{eq:HLT2}
\end{flalign}
where, at this stage, $\vecp g$ is a generic real vector.
The error-functional is the square of the statistical error on  the approximation of $H_P^X(k_1,k_2;\epsilon)$ obtained by using the representation of Eq.\,(\ref{eq:HPexprep}) without taking the limits and by setting $\epsilon>0$ and $n_\mathrm{max}<\infty$. It is given by 
\begin{flalign}
B[\vecp g] =
\sum_{n_1=1}^{n_\mathrm{max}}\sum_{n_2=1}^{n_\mathrm{max}}\,
\sum_{\bar n_1=1}^{n_\mathrm{max}}\sum_{\bar n_2=1}^{n_\mathrm{max}}\,
g(n_1,n_2)\, 
\mathrm{Cov}_P\left(
n_1,n_2 \vert \bar n_1,\bar n_2
\right)
g(\bar n_1,\bar n_2)\;,
\label{eq:HLT3}
\end{flalign}
where $\mathrm{Cov}_P(n_1,n_2 \vert \bar n_1,\bar n_2)$ is the statistical covariance of the lattice correlator $C_P(an_1,an_2)$. By solving the linear system of equations
\begin{flalign}
\left. \vec \nabla_{\vec g}W_X[\vecp g] \right\vert_{\vec g=\vec g_\epsilon^X}=0\;,
\end{flalign}
one gets the coefficients $g^X_\epsilon(n_1,n_2)$ to be used in order to approximate the kernel $K^X_\epsilon(p_1^0,p_2^0)$.

It is important to notice that $B[\vecp g]$ is identically zero in the case of an infinitely precise lattice correlator. Therefore, in this idealized situation the coefficients $g^X_\epsilon(n_1,n_2)$ correspond  to the best fit of the kernel $K^X_\epsilon(p_1^0,p_2^0)$ with respect to the norm used to define $A_X[\vecp g]$. In the realistic situation in which the lattice correlator contains statistical uncertainties, the optimal balance between the systematic and statistical errors on $H_P(k_1,k_2)$ is achieved by starting from large positive values of $\lambda$, i.e.\ from the region in the space of possible solutions where statistical (systematic) errors are small (large), and then progressively reducing the trade-off parameter. Provided that $n_\mathrm{max}$ is sufficiently large, the optimal balance is obtained at the value $\lambda_\star$ such that the results for $H_P(k_1,k_2)$ do not show a statistically significant dependence upon the trade-off parameter for $\lambda\le \lambda_\star$. See Refs.~\cite{Hansen:2019idp,Bulava:2021fre,ExtendedTwistedMassCollaborationETMC:2022sta,Evangelista:2023fmt,ExtendedTwistedMass:2024myu,DeSantis:2025qbb,DeSantis:2025yfm} for more details on this ``stability analysis" and for other applications of the procedure.

\section{Lattice computation of charming penguin contributions to the {\boldmath$B\to K\ell^+\ell^-$} Decay Amplitude}\label{sec:latticeK}

In this section we apply the discussion and results of Sec.\,\ref{subsec:generalN} to the evaluation of the charming penguin contributions to the $B\to K\ell^+\ell^-$ decay amplitude:
\begin{equation}\label{eq:charmingamplitude}
{\cal A}_{1,2}=i\frac{4G_F}{\sqrt{2}}\,e^2\lambda_t\,C_{1,2}^{(c)}\,\big[\bar{u}(p_{\ell^-})\gamma^\mu v(p_{\ell^+})\big]\,D_{\mu\nu}(q)\,\int\dfour x\,e^{iq\cdot x}
\bra{K(\vec{p}_K)}T\big[J_\mathrm{em}^\nu(x)\,O_{1,2}^{(c)}(0)\big]\ket{B(p_B)}\,,
\end{equation}
where $D_{\mu\nu}(q)$ is the propagator of the virtual photon, $p_{\ell^\pm}$ are the momenta of the leptons and the current-current operators $O_{1,2}^{(c)}$ are defined in Eq.\,(\ref{eq:O12def}).  The corresponding four connected diagrams are shown in Fig.\,\ref{fig:CPBK123}. 
In the rest frame of the $B$-meson, from Eq.(\ref{eq:charmingamplitude})  the hadronic matrix elements which need to be evaluated are 
\begin{equation}\label{eq:H12def}
H_{1,2}^\nu(\vecp{q})=i\int\dfour x\,e^{iq\cdot x}
\bra{K(-\vecp{q})}T\big[J_\mathrm{em}^\nu(x)\,O_{1,2}^{(c)}(0)\big]\ket*{B(\vecp{0})}\,,
\end{equation}
where $x=(t,\vecpp{x})$ and $q$ is the four-momentum of the virtual photon.
We note that since electromagnetic current conservation implies that $q_\nu H^\nu(\vecp{q})=0$ so that $H^\nu(\vecp{q})$ is proportional to 
$q^2p_B^\nu-(p_B\cdot q) q^\nu$. Combining $H^\nu(\vecp{q})$ with the photon
propagator and the leptonic current as in Eq.\,(\ref{eq:charmingamplitude}), the contribution to the amplitude is proportional to $p_B^\mu\,\bar{u}(p_{\ell^-})\gamma_\mu v(p_{\ell^+})$ which is often interpreted as a shift in the Wilson coefficient $C_9$, $C_9\to C_9^{\mathrm{eff}}$. The aim of this paper is to develop the necessary framework to quantify this shift.

Separating the contributions from negative and positive $t$, we write
$H^\nu_{1,2}(\vecp{q})=H^\nu_{1,2}(\vecp{q})\theta(-t)+H^\nu_{1,2}(\vecp{q})\theta(t)\equiv H^{\nu-}_{1,2}(\vecp{q})+H^{\nu+}_{1,2}(\vecp{q})$ and consider these two contributions in turn.
In the rest frame of the $B$-meson, at negative $t$ 
\begin{eqnarray}
H^{\nu-}_{1,2}(\vecp{q})&=&i\int_{-\infty}^0 dt\,e^{iq_0 t} C^{\nu-}(t,\vecp{q})\label{Hnuminus0}\\
&=&i\int_{-\infty}^0 dt\,e^{iq_0 t}\int\dthree x\,e^{-i\vec{q}\cdot\vec{x}} \bra{K(-\vecpp{q})}O_{1,2}^{(c)}(0)\,J_\mathrm{em}^\nu(t,\vecp{x})\ket*{B(\vecpp{0})}\nn\\
&=&\int_{E_-^\ast}^\infty\frac{dE}{2\pi}~\frac{\rho^{\nu-}_{1,2}(E,\vecp{q})}{E-m_B+q_0-i\epsilon}
\,,\label{eq:Hmuminus}
\end{eqnarray}
where the label ${\scriptsize -}$ in $E_-^\ast$, $H^{\nu-}_{1,2}(\vecp{q})$ and $\rho^{\nu-}_{1,2}(E,\vecp{q})$ indicates that this expression corresponds to negative times $t<0$. Eq.\,(\ref{eq:Hmuminus}) follows directly from the general equation (\ref{eq:HP2}) for the case $n=2$ which we are considering here\,\footnote{Instead of the overall momentum $\delta$-function in (\ref{eq:HP2}), we impose four momentum conservation directly, $p_K+q=p_B$, where $p_K$ and $p_B=(m_B,0)$ are the four momenta of the $B$ and $K$ mesons respectively, and place the weak operator $O^{(c)}_{1,2}$ at the origin.}. The spectral density is obtained from Eq.\,(\ref{eq:rhoP}),
\begin{equation}\label{eq:rhominus}
 \rho^{\nu-}_{1,2}(E,\vecpp{q})=\bra{K(-\vecpp{q})\,} J^\nu_{\mathrm{em}}(0)
 \,(2\pi)^4\delta^{(3)}(\hat{\mathbf{P}}+\vecp{q})\,\delta(\hat{H}-E)\,O^{(c)}_{1,2}(0)\,\ket*{B(\vecpp{0})}\,,
\end{equation}
where $\hat{\bs{P}}$ and $\hat{H}$ are the three-momentum and Hamiltonian operators respectively.
The energy $E_-^\ast$ in Eq.\,(\ref{eq:Hmuminus}) is that of the lowest on-shell state with Beauty quantum number $B=1$ and three momentum $-\vec{q}$, so that $E^\ast>m_B-q^0$. 
The pole in Eq.\,(\ref{eq:Hmuminus}) is therefore not in the range of the energy integration, so that the $i\epsilon$ in the denominator is unnecessary and can be omitted. The contribution $H^{\nu-}_{1,2}(\vecp{q})$ can therefore be evaluated from the corresponding Euclidean correlation function, without any difficulty from the continuation from Minkowski space.

The situation is different for $t>0$ as we now show. In this case
\begin{eqnarray}
H^{\nu+}_{1,2}(\vecp{q})&=&i\int_0^\infty dt\,e^{iq_0 t}\int\dthree x\,e^{-i\vec{q}\cdot\vec{x}} \bra{K(-\vecpp{q})}J_\mathrm{em}^\nu(t,\vecp{x})\,O_{1,2}^{(c)}(0)\ket*{B(\vecpp{0})}\nn\\
&=&\int_{E_+^\ast}^\infty\frac{dE}{2\pi}~\frac{\rho^{\nu+}_{1,2}(E,\vecp{q})}{E-m_B-i\epsilon}\,,\label{eq:Hmuplus}
\end{eqnarray}
where the label ${\scriptsize +}$ in $E_+^\ast$, $H^{\nu+}_{1,2}(\vecp{q})$ and 
$\rho^{\nu+}_{1,2}$ indicates that this expression corresponds to positive times $t>0$. Here, as illustrated in Fig.\,\ref{fig:BtoKtimeorderings}(b), $E_+^\ast$ is the threshold energy for on-shell states with $B=0$ and $S=1$ so that $E_+^\ast<m_B$. The $i\epsilon$ is therefore necessary to regulate the pole in the integration over $E$. The spectral density is readily obtained from Eq.\,(\ref{eq:rhoP}):
\begin{equation}\label{eq:rhoplus}
\rho_{1,2}^{\nu+}(E,\vecpp{q})=\bra{K(-\vecp{q})}J_\mathrm{em}^\nu(0)\,
(2\pi)^3\delta(\hat{\bs{P}})\,(2\pi)\delta(\hat{H}-E)\,O^{(c)}_{1,2}(0)\ket*{B(\vecp{0})}
\,.
\end{equation}

The SFR method which we propose to use to determine
$H^{\nu+}_{1,2}(\vecp{q})$ from the lattice computation of Euclidean correlation functions has been presented in Sec.\,\ref{subsec:generalN} starting around Eq.\,(\ref{eq:HCnuplus0}), resulting in 
\begin{equation}
H^{\nu+}_{1,2}(q)=\lim_{\epsilon\to 0}\tilde{H}^{\nu+}_{1,2}(q;\epsilon)=
\lim_{\epsilon\to 0}
\,\sum_{n=1}^{n_\mathrm{max}}g_n(m_B,\epsilon)\,e^{-aE_Kn}\,C_{1,2;E}^{\nu+}(an,\vecpp{q}).\label{eq:reconstruction}
\end{equation}

As discussed in detail in Sec.\,III\,A of Ref.\cite{Frezzotti:2023nun} and summarized briefly in Sec.\,\ref{subsec:generalN} above, the optimal determination of the coefficients $g_n$ is delicate and is a balance between the systematic uncertainties in the fit of $1/(E-m_B-i\epsilon)$ in terms of exponentials and the statistical uncertainties in the correlation functions. The details will vary from process to process and the stability of the results will need to be assured case by case, including the one discussed here.

\section{Lattice computation of charming penguin contributions to the {\boldmath$\bar{B}_s\to \gamma\ell^+\ell^-$} Decay Amplitude}\label{sec:Btogammagammastar}

\begin{figure}[t]
\begin{center}
\includegraphics[width=0.3\hsize]{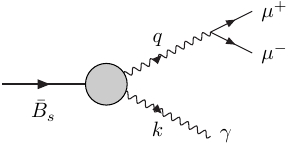}
\end{center}\caption{Schematic diagram of the decay $\bar{B}_s\to\gamma\mu^+\mu^-$, labelled with the notation used in the text.\label{fig:schematic}}
\end{figure}

In this section we discuss the decay $\bar{B}_s\to\gamma\ell^+\ell^-$ illustrated in Fig.\,\ref{fig:schematic} and apply the general procedure described in Sec.\,\ref{subsec:generalN} to this process. We have already anticipated in Sec.\,\ref{subsec:TOgll} that contributions from three of the six time orderings require the introduction of spectral density techniques and we consider each of these in turn.
Although the final results for the spectral densities and the contributions to the hadronic factor can be deduced from Eqs.\,(\ref{eq:rhoP}) and (\ref{eq:HP2}) respectively, it is nevertheless instructive to follow the steps directly for the specific contributions of interest.

We label the four momenta of the real and virtual photons by $k$ and $q$ respectively and introduce the variable $x_\gamma\equiv 2E_\gamma/m_{\bar{B}_s}$, where $E_\gamma=k_0$ is the energy of the real photon in the rest-frame of the $\bar{B}_s$ meson. The physical range for $x_\gamma$ is $0<x_\gamma<1-4m_\mu^2/m_{\bar{B}_s}^2$ and $q^2=m_{\bar{B}_s}^2(1-x_\gamma)$.

The hadronic factor in the  
contribution to the amplitude for the $\bar{B}_s\to\mu^+\mu^-\gamma$ decay from the operators $O_{1,2}^{(c)}$ is given by Eq.\,(\ref{eq:HmunuBtogammagamma}), which we rewrite here for convenience,
\begin{equation}\label{eq:Hmunutwogamma}
H^{\mu\nu}_{1,2}(\vecp{k})=i\,\int dt\int \dthree x\int dt_W\int\dthree y~\bra*{0}T\big[J^\mu_\gamma(t,\vecp{x})\,J^\nu_{\gamma^\ast}(0,\vecp{y})\,O_{1,2}^{(c)}(t_W,\vecp{0})\big]\ket*{\bar{B}_s(\vecp{0})}\,e^{ik\cdot x}\,e^{i\vec{k}\cdot \vec{y}}\,,
\end{equation}
where $T$ is the time-ordering operator and $J^\mu_\gamma$ and $J^\nu_{\gamma^\ast}$ are the electromagnetic currents which emit the real and virtual photons respectively with momenta $k$ and $q=p_{\bar{B}_s}-k$ respectively, $k^2=0$ and $p_{\bar{B}_s}=(m_{\bar{B}_s},\vecp{0})$\,\footnote{To simplify the notation we have omitted the label {\footnotesize em} on the currents, leaving it implicit that these are electromagnetic currents defined after Eq.\,(\ref{eq:H12def0}). It is convenient to distinguish the position at which the real and virtual photons are emitted and this we do by the labels $\gamma$ and $\gamma*$ respectively. We have also omitted the labels {\scriptsize ${1,2}$} labeling the current-current operators on $H^{\mu\nu}$ and on the correlation functions below.} .

As explained in Sec.\,\ref{subsec:TOgll}, of the six possible time orderings, the ones which require the application of the spectral density approach are the two with $t_W<\min[t,0]$ and the one with $t<t_W<0$ and we now consider each of these in turn. The steps are natural applications of the procedure sketched in Sec.\,\ref{subsec:generalN}. 

\subsection{The contribution from the time ordering $t_W<0<t$}\label{subsec:tw0t}
When $t_W<0<t$, as illustrated in Fig.\,\ref{fig:Btogammatimeorderings}(a), the contribution to $H^{\mu\nu}$ is 
\begin{eqnarray}
H^{\mu\nu}_1(\vecp{k})&=&i\,\int_0^\infty\hspace{-6pt} dt\int \dthree x\int_{-\infty}^0 \hspace{-8pt}dt_W\int\dthree y~\bra*{0}J^\mu_\gamma(t,\vecp{x})\,J^\nu_{\gamma^\ast}(0)\,O_{1,2}^{(c)}(t_W,\vecp{y})\ket*{\bar{B}_s(\vecp{0})}\,e^{ik\cdot x}\nn\\
&\equiv&i\int_{-\infty}^0 \hspace{-8pt}dt_W\int_0^\infty\hspace{-6pt} dt~ e^{ik_0 t}~C_1^{\mu\nu}(t,\vec{k};t_W)\,,\label{eq:H1timeintegrals}
\end{eqnarray}
where $k_0=|\vecp{k}|$, the subscript {\footnotesize1} indicates that this is the contribution from the first time-ordering and 
\begin{eqnarray}
C_1^{\mu\nu}(t,\vec{k};t_W)&=&\int\dthree x\,e^{-i\vec{k}\cdot\vec{x}}\int\dthree y ~\bra{0}J^\mu_\gamma(t,\vecp{x})\,J^\nu_{\gamma^\ast}(0)
\,O^{(c)}_{1,2}(t_W,\vecpp{y})\ket*{\bar{B}_s(\vecpp{0})}\nn\\
&=&e^{-im_{\bar{B}_s}t_W}\int\dthree x\,e^{-i\vec{k}\cdot\vec{x}}\int\dthree y ~\bra{0}J^\mu_\gamma(0)\,e^{i\hat{\bs{P}}\cdot\vec{x}}\,e^{-i\hat{H}t}\,J^\nu_{\gamma^\ast}(0)
\,e^{-i\hat{\bs{P}}\cdot y}e^{i\hat{H}t_W}\,O^{(c)}_{1,2}(0)\ket*{\bar{B}_s(\vecpp{0})}\nn\\
&=&\int_{E_1^\ast}^\infty\frac{dE_1}{2\pi}e^{i(E_1-m_{\bar{B}_s})t_W}\int_{E_{2}^\ast}^\infty\frac{dE_2}{2\pi}e^{-iE_2t}\,\rho_1^{\mu\nu}(E_1,E_2,\vecp{k})\,,\label{eq:C1twointegrals}
\end{eqnarray}
where the spectral density is given by
\begin{equation}
\rho_1^{\mu\nu}(E_1,E_2,\vecp{k})=\bra{0}J^\mu_\gamma(0)\,(2\pi)^4\delta^{(3)}(\hat{\bs{P}}-\vecp{k})\,\delta(\hat{H}-E_2)\,J^\nu_{\gamma^\ast}(0)
\,(2\pi)^4\delta^{(3)}(\hat{\bs{P}})\delta(\hat{H}-E_1)\,O^{(c)}_{1,2}(0)\ket*{\bar{B}_s(\vecpp{0})}\,.\label{eq:rho1munu}
\end{equation}

Next we perform the $t$ and $t_W$ integrations as in Eq.\,(\ref{eq:H1timeintegrals}) to obtain,
\begin{equation}
H_1^{\mu\nu}(\vecp{k})=-i\int_{E_1^\ast}^\infty\frac{dE_1}{2\pi}\int_{E_{2}^\ast}^\infty\frac{dE_2}{2\pi}\frac{\rho_1^{\mu\nu}(E_1,E_2,\vecp{k})}{(E_1-m_{\bar{B}_s}-i\epsilon)(E_2-k_0-i\epsilon)}\,.\label{eq:H1integrations}
\end{equation}
Since the square of the invariant mass of the intermediate states propagating between the two currents is positive, $E_2^2-|\vecp{k}^2|=E_2^2-k_0^2>0$, the pole at $E_2=k_0$ is not in the range of the $E_2$ integration and 
the corresponding $i\epsilon$ is not necessary. This is not the case for the $E_1$ integration. The threshold $E_1^\ast$, i.e. the lowest energy of on-shell states in the spectrum of $O^{(c)}_{1,2}\ket*{\bar{B}_s(\vecp{0})}$, is less than $m_{\bar{B}_s}$,  $E_1^\ast<m_{\bar{B}_s}$, and therefore the pole at $E_1=m_{\bar{B}_s}$ is in the range of the $E_1$ integration and the corresponding $i\epsilon$ is necessary.

In Eq.\,(\ref{eq:C1twointegrals}) we have written the Minkowski correlation function in the form
\begin{equation}
C^{\mu\nu}_1(t,\vec{k};t_W)=\int_{E_1^\ast}^\infty\frac{dE_1}{2\pi}\,e^{-i(m_{\bar{B}_s}-E_1)t_W}\int_{E_2^\ast}^\infty\frac{dE_2}{2\pi}\,e^{-iE_2t}\rho_1^{\mu\nu}(E_1,E_2,\vecpp{k})\,
\end{equation}
and therefore the corresponding correlation function in Euclidean space is
\begin{equation}
C^{\mu\nu}_{1E}(t,\vec{k};t_W)=\int_{E_1^\ast}^\infty\frac{dE_1}{2\pi}\,e^{-(m_{\bar{B}_s}-E_1)t_W}\int_{E_2^\ast}^\infty\frac{dE_2}{2\pi}\,e^{-E_2t}\rho_1^{\mu\nu}(E_1,E_2,\vecpp{k})
\,,
\end{equation}
where the subscript $E$ on the left hand side denotes ``Euclidean".
Following HLT\,\cite{Hansen:2019idp} we write 
\begin{eqnarray}
\tilde{H}^{\mu\nu}_1(\vec{k},\epsilon)&=&-a\sum_{t=1}^{t_\mathrm{max}}\int\frac{dE_2}{2\pi}\,e^{-(E_2-k_0)ta} \int\frac{dE_1}{2\pi}\frac{\rho^{\mu\nu}_1(E_1,E_2,\vecpp{k})}{E_1-m_{\bar{B}_s}-i\epsilon}\nn\\
&=&-a\sum_{t=1}^{t_\mathrm{max}}~\sum_{n=1}^{n_\mathrm{max}} g_n(m_{\bar{B}_s},\epsilon) \int\frac{dE_2}{2\pi}
\,e^{-(E_2-k_0)ta}\int\frac{dE_1}{2\pi}\,e^{-E_1na}\rho^{\mu\nu}_1(E_1,E_2,\vecpp{k})\nn\\
&=&-a\sum_{t=1}^{t_\mathrm{max}}~\sum_{n=1}^{n_{\mathrm{max}}} g_n(m_{\bar{B}_s},\epsilon)\,e^{k_0 ta}\,e^{-m_{\bar{B}_s}\!na}\,C^{\mu\nu}_{1E}(t,\vec{k};-na)\,.\label{eq:H1tilde}
\end{eqnarray}
In Eq.\,(\ref{eq:H1tilde}) we have replaced the integral over $t$ by the corresponding discrete sum present in lattice QCD computations. The upper limit $t_\mathrm{max}$ needs to be chosen so that the sum over $t$ has converged with sufficient precision. 
The physical result is given by $H^{\mu\nu}_1(\vecp{k})=\lim_{\epsilon\to 0}\tilde{H}^{\mu\nu}_1(\vec{k},\epsilon)$ in the continuum limit.
The $g_n$ are complex, with the real and imaginary parts to be determined as explained in Sec.\,\ref{subsec:generalN}. 

An alternative to the above computation of the hadronic factor, in which the integral over $t$ is replaced by a discrete sum, would be to apply the HLT procedure to the product of the two propagators in Eq.\,(\ref{eq:H1integrations}) by following the steps outlined in the discussion around Eq.(\ref{eq:Kepsp1p2}). Note that the values of $\epsilon$ in the two propagators can be chosen to be different (the one in the factor $1/(E_2-k_0-i\epsilon)$ can even be set to zero).
The different results for $H^{\mu\nu}_1(\vecp{k})$ will then differ by discretization effects and it is to be investigated whether there is an advantage in practice to using one of the two procedures.

\subsection{The contribution from the time ordering $t_W<t<0$}\label{subsec:twt0}

When the real photon is emitted before the virtual one, as illustrated in Fig.\,\ref{fig:Btogammatimeorderings}(b), the corresponding correlation function is 
\begin{eqnarray}
C^{\mu\nu}_2(t,\vecpp{k};t_W)&=&\int\dthree x\,e^{-i\vec{k}\cdot\vec{x}}\int\dthree y ~\bra{0}J^\nu_{\gamma^\ast}(0)\,J^\mu_\gamma(t,\vecp{x})\,O_{1,2}^{(c)}(t_W,\vecpp{y})\ket*{\bar{B}_s(\vecp{0})}\,,\label{eq:C2Mdef}\\
&=&e^{-im_{\bar{B}_s}t_W}\bra{0}J^\nu_{\gamma^\ast}(0)\,(2\pi)^3\delta(\hat{P}+\vecp{k})\,e^{i\hat{H}t}J^\mu_\gamma(0)\,(2\pi)^3\delta^{(3)}(\hat{P}) e^{-i\hat{H}(t-t_W)}O_{1,2}^{(c)}(0)\ket*{\bar{B}_s(\vecp{0})}\,,\nn\\
&=&e^{-im_{\bar{B}_s}t_W}
\int_{E_1^\ast}^\infty\,\frac{dE_1}{(2\pi)}\,e^{iE_1t}\,
\int_{E_2^\ast}^\infty \frac{dE_2}{2\pi}\,e^{-iE_2^0(t-t_W)}\rho_2^{\mu\nu}(E_1,E_2,\vecp{k})\,,
\end{eqnarray}
where
\begin{equation}
\rho_2^{\mu\nu}(E_1,E_2,\vecp{k})=\bra{0}J^\nu_{\gamma^\ast}(0)\,(2\pi)^4\delta^{(3)}(\hat{P}+\vecp{k})\,\delta(\hat{H}-E_1)J^\mu_\gamma(0)\,(2\pi)^4\delta^{(3)}(\hat{P}) \delta(\hat{H}-E_2)\,O_{1,2}^{(c)}(0)\ket*{\bar{B}_s(\vecp{0})}\label{eq:rho2munu}
\end{equation} 
The subscripts {\footnotesize 2} on $C^{\mu\nu}_2$ and $\rho^{\mu\nu}_2$ indicate that these quantities correspond to the second time ordering.

Performing the $t$ and $t_W$ integrations we obtain
\begin{eqnarray}
H_2^{\mu\nu}(\vecp{k})&=&i\int_{E_1^\ast}^\infty \frac{dE_1}{2\pi}
\int_{E_2^\ast}^\infty\,\frac{dE_2}{(2\pi)}\,\rho_2^{\mu\nu}(E_1,E_2,\vecp{k})\int_{-\infty}^0 dt_W\,e^{i(E_2-m_{\bar{B}_s})t_W}\int_{t_W}^0 dt\, e^{i(E_1+k_0-E_2)t}\nn\\
&=&-i\int_{E_1^\ast}^\infty\frac{dE_1}{2\pi}
\int_{p_2^\ast}^\infty\,\frac{dE_2}{(2\pi)}\,\frac{\rho_2^{\mu\nu}(E_1,E_2,\vecp{k})}{(E_1^0+k_0-m_{\bar{B}_s}-i\epsilon)(E_2^0-m_{\bar{B}_s}-i\epsilon)}\,.\label{eq:H2integrations}
\end{eqnarray}
In this case, as for the first time ordering ($t_W<0<t$) discussed in Subsec.\ref{subsec:tw0t}, the pole at $E_2=m_{\bar{B}_s}$ is always in the range of the $E_2$ integration and we need to keep the corresponding $i\epsilon$. In addition however, depending on value of $k_0$, possible on-shell $B=0$ states may exist with energies smaller than $q_0=m_{\bar{B}_s}-k_0$, where $q_0$ is the energy of the virtual photon (i.e. of the charge-lepton pair). If this is the case,
then the pole at $E_1=m_{\bar{B}_s}-k_0=q_0$ is in the range of the $E_1$ integration and we also need the corresponding $i\epsilon$.

\begin{figure}[t]
\begin{center}
\hspace{-0.3cm}
\includegraphics[width=0.70\hsize]{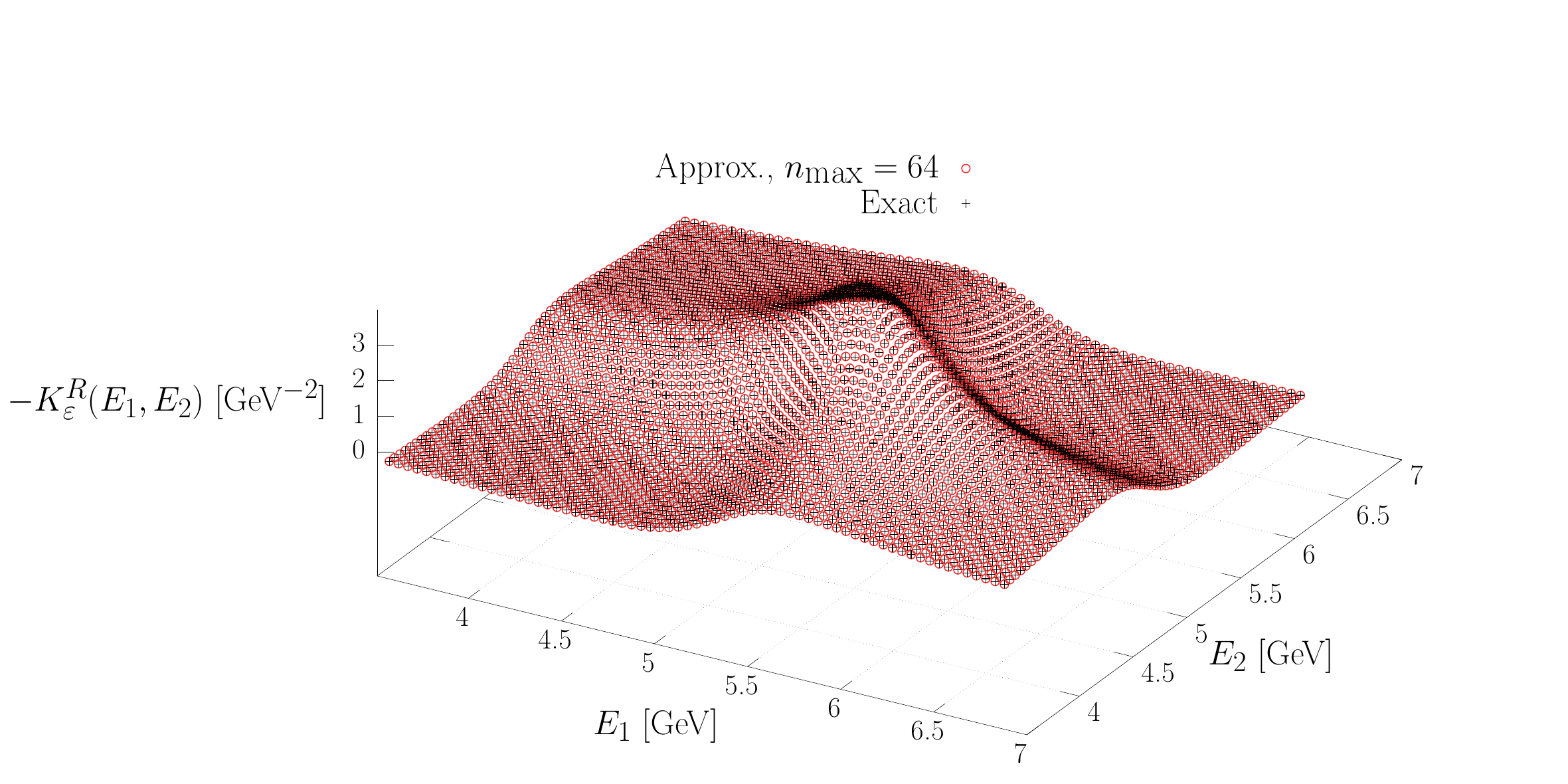}
\hspace{-0.9cm}
\includegraphics[width=0.35\hsize]{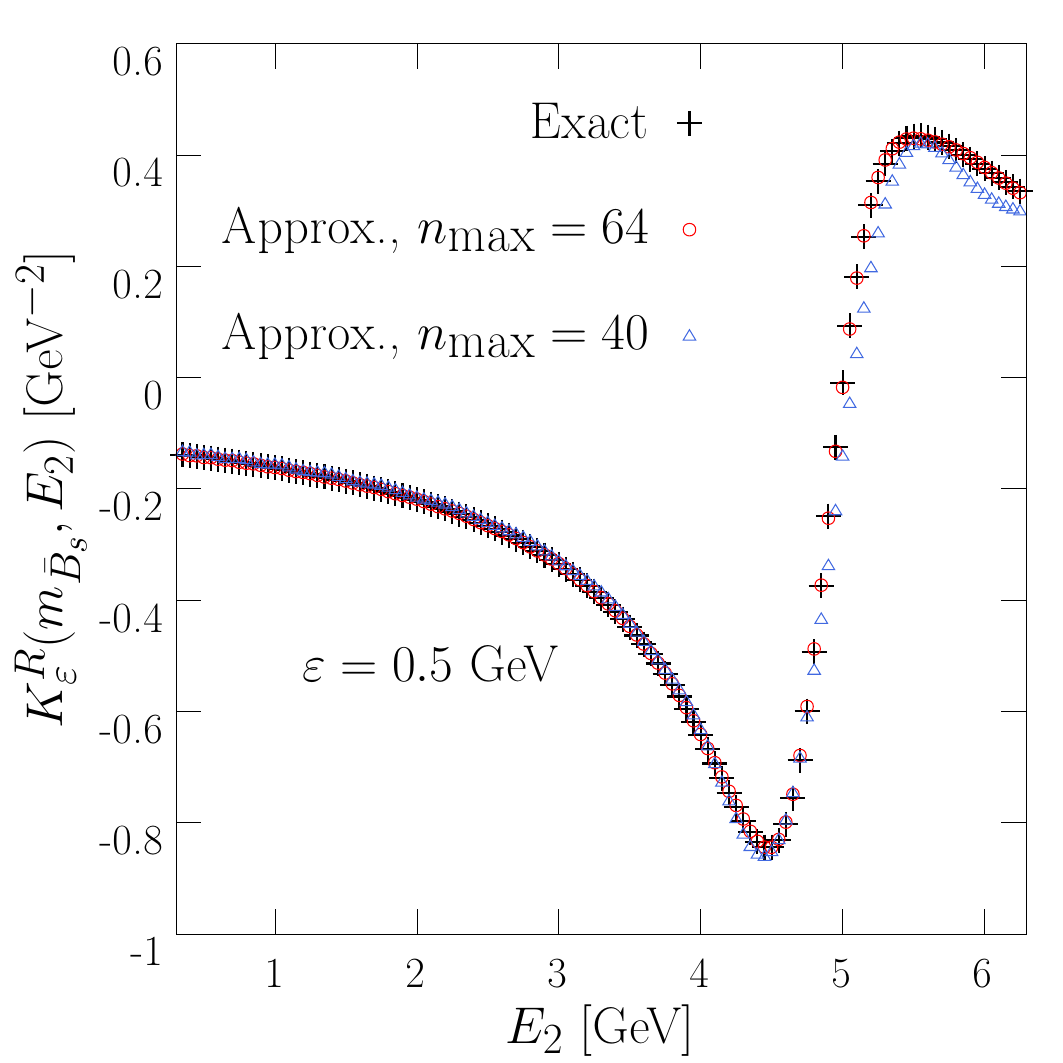}
\\
\hspace{-0.3cm}
\includegraphics[width=0.70\hsize]{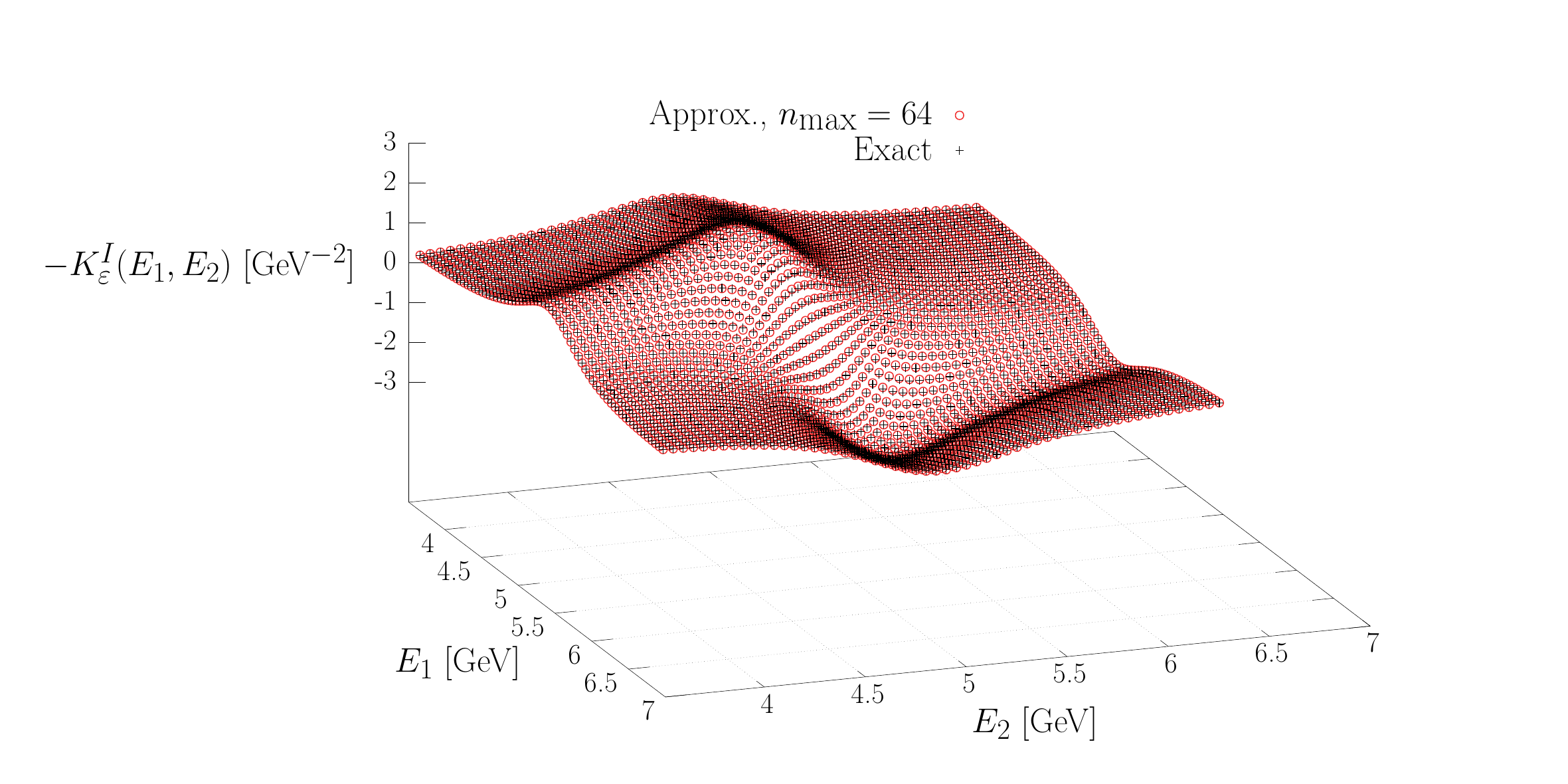}
\hspace{-0.9cm}
\includegraphics[width=0.35\hsize]{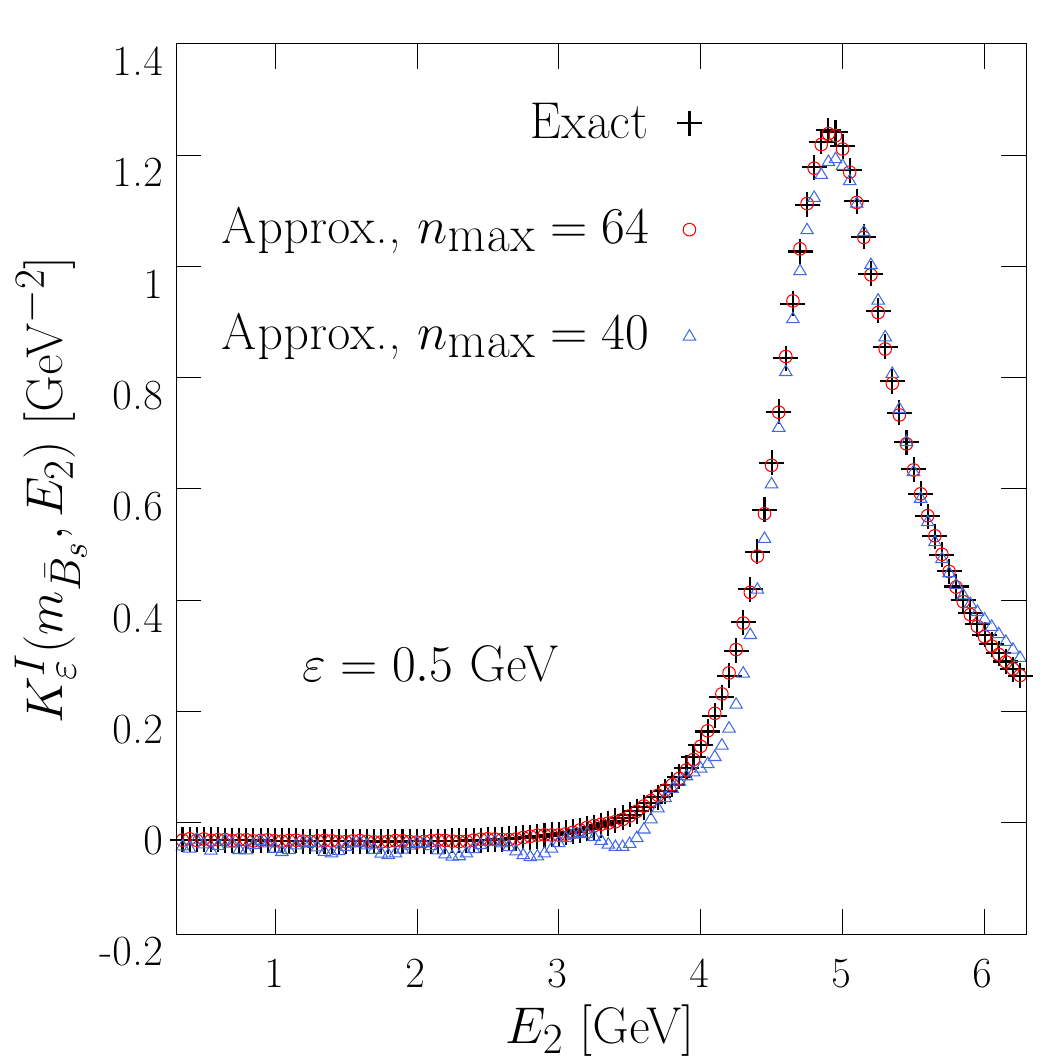}
\end{center}
\caption{In the left column we present examples of the HLT approximations of the kernel $K_\epsilon(E_1,E_2)$ of Eq.\,(\ref{eq:kerneltorec}) with $k^0=0.54$\,GeV and $\epsilon=0.5$\,GeV. In the right-hand column we present $K_\epsilon(m_{\bar{B}_s},E_2)$ i.e. the section at fixed $E_1=m_{\bar{B}_s}$. The plots in the first (second) row correspond to the real (imaginary) part of the kernel. The exact kernel is plotted in black, the HLT approximation with $n_\mathrm{max}=64$ in red and the one with $n_\mathrm{max}=40$ in blue.  
\label{fig:twoekernelrec}}
\end{figure}
The general procedure described at the end of subsection~\ref{subsec:generalN} allows us to express $H_2^{\mu\nu}$ in terms of Euclidean lattice correlators. In this case the smearing kernel is
\begin{flalign}
K_\epsilon(E_1,E_2)
=
\frac{-1}{
(E_1+k_0-m_{\bar{B}_s}-i\epsilon)(E_2-m_{\bar{B}_s}-i\epsilon)}
\label{eq:kerneltorec}
\end{flalign}
and its representation in terms of exponentials is given in Eq.~(\ref{eq:KrepSW}). Therefore, by introducing the amputated Euclidean correlation function\,
\footnote{Note that the trilocal matrix element in the Euclidean correlation function in Eq.\,\ref{eq:C2Edef} is translated in space-time w.r.t. to that in the Minkowski correlator in Eq.\,(\ref{eq:C2Mdef}, so that the real photon is emitted at the origin.}
\begin{flalign}
C_{2E}^{\mu\nu}(an_1,an_2;\vec k)
&=
\int_{E_1^\ast}^\infty \frac{dE_1}{2\pi}
\int_{E_2^\ast}^\infty\,\frac{dE_2}{(2\pi)}\,\rho_2^{\mu\nu}(E_1,E_2,\vecp{k})\,
e^{-an_1E_1-an_2E_2}
\nonumber \\[8pt]
&=
e^{-an_2m_{\bar{B}_s}}
\int d^3x_1\, 
\int d^3x_2\, 
e^{i\vec x_1\cdot \vecp{k}}\,
\bra{0}J^\nu_{\gamma^\ast}(an_1,\vec x_1)\,
J^\mu_\gamma(0)\,
O_{1,2}^{(c)}(-an_2,\vec x_2)\ket*{\bar{B}_s(\vecp{0})}\;,\label{eq:C2Edef}
\end{flalign}
one has
\begin{flalign}
H^{\mu\nu}_2(\vecp{k})=
\lim_{\epsilon\to 0^+}
\lim_{n_\mathrm{max}\to \infty}
\sum_{n_1=1}^{n_\mathrm{max}}
\sum_{n_2=1}^{n_\mathrm{max}}\,
i\left[
g_\epsilon^R(n_1,n_2)
+i
g_\epsilon^I(n_1,n_2)
\right]\,
C_{2E}^{\mu\nu}(an_1,an_2;\vec k)\;.
\end{flalign}
In Fig.\,\ref{fig:twoekernelrec} we show examples of the approximation of the kernel $K_\epsilon(E_1,E_2)$ of Eq.\,(\ref{eq:kerneltorec}) by using the HLT algorithm, i.e. by using Eqs.\,(\ref{eq:HLT1})-(\ref{eq:HLT3}). For this illustration, we have set $k^0=0.54$\,GeV, $\epsilon=0.5$\,GeV, $E_1^\ast=E_2^\ast=m_\pi$, $a=0.05$~fm and considered the ideal situation of an infinitely precise lattice correlation function ($\lambda=0$). The plots in the first row show the real part of the kernel, $K_\epsilon^R(E_{1},E_{2})$, while those in the second row show the imaginary part, $K_\epsilon^I(E_1,E_2)$. The plots on the left show the comparison of the exact kernel (black points) with its HLT approximation obtained with $n_\mathrm{max}=64$ and, as can be seen, the two surfaces are indistinguishable on the scale of the figures. The plots on the right show a comparison of a section of the kernel and have been obtained by setting $E_1=m_{\bar{B}_s}$ and by plotting $K_\epsilon(m_{\bar{B}_s},E_2)$ as a function of $E_2$. As can be seen, for both the real (top) and imaginary (bottom) parts the exact kernel (black line) is approximated very precisely by setting $n_\mathrm{max}=64$ (red points) while some deviations are visible on the scale of the plots in the case of the approximation obtained with $n_\mathrm{max}=40$ (blue points).

We have presented these examples in order to provide evidence that it is also straightforward, even if perhaps numerical subtle, to cope with the mathematics of approximating the kernels in cases where they are functions of two energy variables. 
This does not mean however, that a precise calculation of the physical amplitude will be straightforward. 
Indeed, while in principle there are now no theoretical obstacles to the extraction of the physical amplitude, the precision of its determination, and hence its phenomenological impact, will depend strongly on the statistical accuracy which can be reached in future computations of the required lattice Euclidean correlation functions.

\subsection{The contribution from the time ordering $t<t_W<0$}\label{subsec:ttw0}

The third contribution requiring the implementation of the HLT procedure comes from the time ordering $t<t_W<0$ as illustrated in Fig.\,\ref{fig:Btogammatimeorderings}(c),
\begin{eqnarray}
H^{\mu\nu}_3(\vecp{k})&=&i\,\int^0_{-\infty}\hspace{-6pt} dt\int \dthree x\int_{t}^0 \hspace{-4pt}dt_W\int\dthree y~\bra*{0}J^\nu_{\gamma^\ast}(0)\,O_{1,2}^{(c)}(t_W,\vecp{y})\,J^\mu_\gamma(t,\vecp{x})\ket*{\bar{B}_s(\vecp{0})}\,e^{ik\cdot x}\nn\\
&\equiv&i\int_{-\infty}^0 \hspace{-8pt}dt\int^0_{t}\hspace{-4pt} dt_W~ e^{ik_0 t}~C_3^{\mu\nu}(t,\vec{k};t_W)\,.\label{eq:C3timeintegrations}
\end{eqnarray}
The correlation function $C_3^{\mu\nu}(t,\vec{k};t_W)$ is given by
\begin{eqnarray}
C_3^{\mu\nu}(t,\vec{k};t_W)&=&\int\dthree x\,e^{-i\vec{k}\cdot\vec{x}}\int\dthree y ~\bra{0}J^\nu_{\gamma^\ast}(0)
\,O^{(c)}_{1,2}(t_W,\vecpp{y})\,J^\mu_\gamma(t,\vecp{x})\ket*{\bar{B}_s(\vecpp{0})}\nn\\
&=&e^{-im_{\bar{B}_s}t}\bra{0}J^\nu_{\gamma^\ast}(0)(2\pi)^3\delta^{(3)}(\hat{P}+\vecp{k})\,e^{i\hat{H}t_W}
\,O^{(c)}_{1,2}(0)\,(2\pi)^3\delta^{(3)}(\hat{P}+\vecp{k})\,e^{-i\hat{H}(t_W-t)}J^\mu_\gamma(0)\ket*{\bar{B}_s(\vecpp{0})}\nn\\
&=&e^{-im_{\bar{B}_s}t}\int_{E_1^\ast}^\infty \frac{dE_1}{2\pi}e^{-iE_1(t_W-t)}
\int_{E_2^\ast}^\infty\,\frac{dE_2}{(2\pi)}e^{iE_2t_W}\,\rho_3^{\mu\nu}(E_1,E_2,\vecp{k})\,,
\end{eqnarray}
where
\begin{equation}
\rho_3^{\mu\nu}(E_1,E_2,\vecp{k})=\bra{0}J^\nu_{\gamma^\ast}(0)\,(2\pi)^4\delta^{(3)}(\hat{P}+\vecp{k})\,\delta(\hat{H}-E_2)\,O_{1,2}^{(c)}(0)\,(2\pi)^4\delta^{(3)}(\hat{P}+\vecp{k}) \delta(\hat{H}-E_1)J^\mu_\gamma(0)\ket*{\bar{B}_s(\vecp{0})}\,.\label{eq:rho3munu}
\end{equation} 
The subscripts {\footnotesize 3} on $C^{\mu\nu}_3$ and $\rho^{\mu\nu}_3$ indicate that these quantities correspond to the third time ordering.

Performing the $t$ and $t_W$ integrations we obtain
\begin{eqnarray}
H_3^{\mu\nu}(\vecp{k})&=&i\int_{E_1^\ast}^\infty \frac{dE_1}{2\pi}
\int_{E_2^\ast}^\infty\,\frac{dE_2}{(2\pi)}\,\rho_3^{\mu\nu}(E_1,E_2,\vecp{k})
\int_{-\infty}^0 dt\,e^{i(E_1-m_B+k_0)t}    \int_{t}^0 dt_W\,\,e^{-i(E_1-E_2)t_W} \nn\\
&=&i\int_{E_1^\ast}^\infty \frac{dE_1}{2\pi}
\int_{E_2^\ast}^\infty\,\frac{dE_2}{(2\pi)}\,\frac{\rho_3^{\mu\nu}(E_1,E_2,\vecp{k})}{(E_1-m_{\bar{B}_s}+k_0-i\epsilon)(E_2-q_0-i\epsilon)}\label{eq:H3a}
\end{eqnarray}
where $q_0$ is the energy of the virtual photon, $q_0=m_{\bar{B}_s}-k_0$. The energy $E_1$ is that of states with Beauty and Strangeness quantum numbers $B=-1,\, S=1$ and momentum $-\vec{k}$ and so $E_1>m_{\bar{B}_s}$ and the denominator $m_{\bar{B}_s}-k_0-E_1$ in Eq.\,(\ref{eq:H3a}) does not vanish in the range of the $E_1$ integration for any value of $\vec{k}$. We can therefore drop the $i\epsilon$ in the first factor in the denominator.
However, as was the case for the second time ordering ($t_W<t<0$) (discussed in Subsec.\ref{subsec:twt0}), depending on the value of $\vec{k}$, possible on-shell $B=0$ states may exist with energies smaller than $q_0$. 
This is the case if $q_0>E_2^\ast$ and we then need to keep the corresponding $i\epsilon$.
The hadronic factor is then obtained after taking $\epsilon\to0$ in the continuum limit:
$H^{\mu\nu}_3(\vecp{k})=\lim_{\epsilon\to 0}\,\lim_{a\to0}\tilde{H}^{\mu\nu}_3(\vecp{k},\epsilon,a)$, where
\begin{eqnarray}
\tilde{H}^{\mu\nu}_3(\vecp{k},\epsilon,a)&=&\int_{E_1^\ast}^\infty \frac{dE_1}{2\pi}\int_{E_2^\ast}^\infty \frac{dE_2}{2\pi}\,\frac{\rho_3^{\mu\nu}(E_1,E_2,\vecp{k})}{(E_1-m_{\bar{B}_s}+k_0)(E_2-q_0-i\epsilon)}\nn\\
&=&a\sum_{t=1}^{t_\mathrm{max}}\int_{E_1^\ast}^\infty \frac{dE_1}{2\pi}\int_{E_2^\ast}^\infty \frac{dE_2}{2\pi}\,
e^{-(k_0+E_1-m_{\bar{B}_s})ta}~\frac{\rho_3^{\mu\nu}(E_1,E_2,\vecp{k})}{E_2-q_0-i\epsilon}
\nn\\
&=&a\sum_{t=1}^{t_\mathrm{max}}~e^{-k_0ta}~\sum_{n=1}^{n_\mathrm{max}}e^{-m_{\bar{B}_s}\!na}\,g_n(q_0,\epsilon)\,C_{3E}^{\mu\nu}((t-n)a,\vecp{k};-an)\,.
\label{eq:H3}
\end{eqnarray}
As in Sec.\ref{subsec:tw0t} we have replaced the integral over $t$ by a discrete sum and also in this case, one might consider applying the HLT procedure to the product of both propagators in Eq.\,(\ref{eq:H3a}) by following the steps outlined in the discussion around Eq.(\ref{eq:Kepsp1p2}). 
As in the discussion in Sec.\,\ref{subsec:tw0t} the values of $\epsilon$ in the two propagators can be chosen to be different (the one in the factor $1/(E_1-m_{\bar{B}}+k_0-i\epsilon)$ can even be set to zero).
The different results for $H^{\mu\nu}_3(\vecp{k})$ will differ by discretization effects and it is to be investigated numerically whether there is an advantage in practice to using one of the two procedures.

As mentioned in Sec.\,\ref{subsec:TOgll}, the remaining 3 time orderings, $0<t_W<t$, $0<t<t_W$ and $t<0<t_W$ do not have energy-conserving intermediate states which can go on shell and the computation of their real contributions to the amplitude therefore does not require the use of spectral density methods.

\section{Evaluation of the remaining matrix elements}\label{sec:neglected}

In the previous sections we have discussed how to implement the HLT method to evaluate the matrix elements of bilocal and trilocal operators containing $O_1^{(c)}$ and $O_2^{(c)}$. In this section we briefly consider the matrix elements of the remaining operators in the effective Hamiltonian in Eq.\,(\ref{eq:Heff7910}). 

For $B\to K\ell^+\ell^-$ decays, the contributing matrix elements of the operators $O_7$, $O_9$ and $O_{10}$ can be determined from the computation of the corresponding three-point correlation functions without any complication due the continuation from Minkowski to Euclidean space in analogy with the computation of leptonic and semileptonic form factors.   
Using Lorentz and gauge invariance, the hadronic matrix elements of the local operators can be expressed in terms of real dimensionless invariant form factors:
\begin{eqnarray}
\bra{K(p_K)}\bar{s}\gamma^\mu(1-\gamma^5)b\ket{B(p_B)}&=&\left(p_K+p_B-\frac{m_B^2-m_K^2}{q^2}\,q\right)^{\!\mu} f_+(q^2)+\frac{m_B^2-m_K^2}{q^2}\,q^\mu f_0(q^2)\,,
\label{eq:vectorff}\\ 
\bra{K(p_K)}\bar{s}\sigma^{\mu\nu}b\ket{B(p_B)}&=&i\left\{(p_B+p_K)^\mu q^\nu-(p_B+p_K)^\nu q^\mu\right\}\,\frac{f_T(q^2)}{m_B+m_K}\,.
\label{eq:tensorff}
\end{eqnarray}
The evaluation of these matrix elements can proceed in the standard way by computing Euclidean three-point correlation functions.

For $\bar{B}_s\to\mu^+\mu^-\gamma$ decays, the determination of the matrix elements of $O_7$, $O_9$ and $O_{10}$ has been discussed in detail and implemented numerically in Ref.\,\cite{Frezzotti:2024kqk}. The hadronic matrix elements are now of bilocal operators, in addition to the weak currents there is an electromagnetic current to which the real or virtual photon couples. For $O_9$ and $O_{10}$ this does not introduce any theoretical complication and the hadronic matrix elements can also be written in terms of real form factors
\begin{eqnarray}
    H_9^{\mu\nu}(p,k)&=&H_{10}^{\mu\nu}(p,k) \equiv
    i\int d^4y\,e^{ik\cdot y}\,\mathrm{T}\bra{0}\big(\bar{s}\gamma^\nu P_L b\big)(0)\,J_{\mathrm{em}}^\mu\big(y\big)\ket{\bar{B}_s(p)}\nonumber\\
    &\equiv&-i\big(g^{\mu\nu}k\cdot q-q^\mu k^\nu\big)\frac{F_A}{2m_{\bar{B}_s}}+\epsilon^{\mu\nu\rho\sigma}k_\rho q_\sigma\,\frac{F_V}{2m_{\bar{B}_s}}\,,
\end{eqnarray}
where the projector $P_L=\frac{1-\gamma^5}{2}$ and $\mu$ and $k$ ($\nu$ and $q=p-k$) are the polarization index and momentum of the real (virtual) photon. The important point is that for $t_y\equiv y^0<0$, the on-shell states propagating between the two operators have $B=-1$, $S=1$ and so have energies greater than $m_{\bar \bar{B}_s}$, while for $t_y>0$ they have invariant masses greater than 0 (the invariant mass of the final-state real photon). Therefore there are no on-shell intermediate states propagating between the two operators and hence no imaginary contributions to the amplitude.

\begin{figure}[t]
    \centering
    \includegraphics[width=0.45\linewidth]{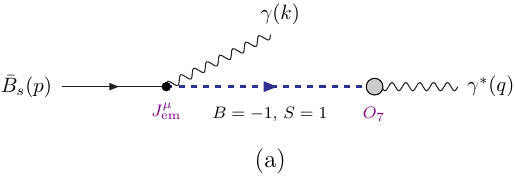}\qquad
    \includegraphics[width=0.45\linewidth]{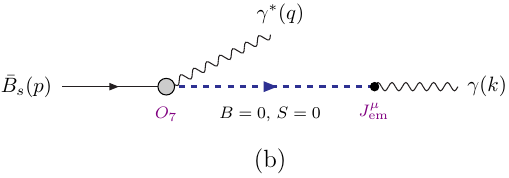}\\[0.2in]
    \includegraphics[width=0.45\linewidth]{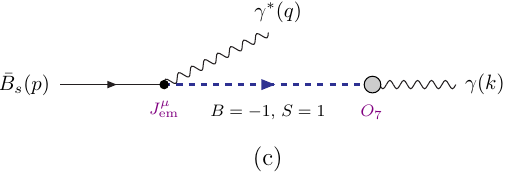}\qquad
    \includegraphics[width=0.45\linewidth]{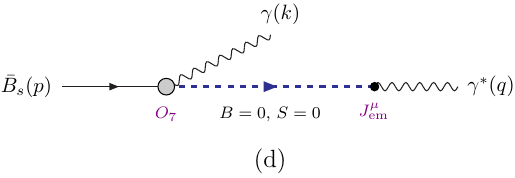}
    \caption{Schematic time-ordered diagrams of matrix elements of the bilocal operator $T[J_{\mathrm{em}}(y)O_7(0)]$ for $\bar{B}_s\to\mu^+\mu^-\gamma$ decays. The four diagrams correspond to: (a) and (b) the virtual photon emitted from $O_7$ with $t_y\equiv y^0<0$ and $t_y>0$ respectively; (c) and (d) the real photon emitted from $O_7$ with $t_y<0$ and $t_y>0$ respectively. 
    }
    \label{fig:TOO7}
\end{figure}
For the corresponding matrix elements of $O_7$ for $\bar{B}_s\to\mu^+\mu^-\gamma$ decays we distinguish two cases depending on whether it is the virtual photon or the real one which is emitted at $O_7$. 
\begin{enumerate}
\item [i)] Following Ref.\,\cite{Frezzotti:2024kqk}, we write the hadronic matrix element for the contribution where the virtual photon is emitted from $O_7$ as
\begin{eqnarray}
    H_{7A}^{\mu\nu}(p,k)&=&i\frac{2m_b}{q^2}\int d^4y~e^{ik\cdot y}~
    \mathrm{T}\bra{0}-i\big[\bar{s}\sigma^{\nu\rho}q_\rho P_R b\big](0)\,J^\mu_{\mathrm{em}}(y)\ket{\bar{B}_s(p)}\nn\\
    &\equiv&-i[g^{\mu\nu}k\cdot q-q^\mu k^\nu]\,\frac{F_{TA}(q^2)m_b}{q^2}+
    \epsilon^{\mu\nu\rho\sigma}k_\rho q_\sigma\,\frac{F_{TV}(q^2)m_b}{q^2}\,,
\end{eqnarray} where $P_R=\frac{1+\gamma^5}{2}$.
Diagrams for the two time-orderings $t_y<0$ and $t_y>0$ are sketched in Fig.\,\ref{fig:TOO7}(a) and \ref{fig:TOO7}(b). We see that in both cases on-shell intermediate states (denoted by the dashed blue line) with the indicated flavor quantum numbers necessarily have invariant masses greater than that of the photon which is emitted later (the virtual photon in Fig.\,\ref{fig:TOO7}(a) and the real photon in Fig.\,\ref{fig:TOO7}(b)). These contributions are therefore real and the invariant form-factors $F_{TV}$ and $F_{TA}$ can be computed using Lattice QCD in the standard way as was done in Ref.\,\cite{Frezzotti:2024kqk}.
\item [ii)] As in Ref.\,\cite{Frezzotti:2024kqk}, we write the hadronic matrix element for the contribution where the real photon is emitted from $O_7$ as
\begin{eqnarray}
    H_{7B}^{\mu\nu}(p,k)&=&i\frac{2m_b}{q^2}\int d^4y~e^{iq\cdot y}~
    \mathrm{T}\bra{0}-i\big[\bar{s}\sigma^{\mu\rho}k_\rho P_R b\big](0)\,J^\nu_{\mathrm{em}}(y)\ket{\bar{B}_s(p)}\nn\\
    &\equiv&-i[g^{\mu\nu}k\cdot q-q^\mu k^\nu]\,\frac{\bar{F}_{TA}(q^2)m_b}{q^2}+
    \epsilon^{\mu\nu\rho\sigma}k_\rho q_\sigma\,\frac{\bar{F}_{TV}(q^2)m_b}{q^2}\,,
\end{eqnarray} where in this case the two form factors are equal $\bar{F}_{TV}(q^2)=\bar{F}_{TA}(q^2)\equiv \bar{F}_T(q^2)$. Diagrams for the two time-orderings $t_y<0$ and $t_y>0$ are sketched in Fig.\,\ref{fig:TOO7}(c) and Fig.\,\ref{fig:TOO7}(d). When $t_y<0$ there are no unitarity cuts present, the contribution is therefore real and can be evaluated without difficulty. When $t_y>0$ and in Fig.\,\ref{fig:TOO7}(d) however, the states propagating between the two operators have quantum numbers $B=0$ and $S=0$ and hence, for sufficiently large values of $q^2$  can be on-shell (e.g. the $\Phi$-meson, $K$-$\bar{K}$ states or even Zweig suppressed lighter states) and therefore contribute an imaginary part. In Ref.\,\cite{Frezzotti:2024kqk}, the HLT method was applied to evaluate this term, and although the uncertainty was found to be large, the computation confirmed the expectation that this contribution is very small.   
\end{enumerate}

The contributions from the penguin operators $O_{3-6}$ and from the
chromomagnetic operator $O_8$ are expected to be numerically small. They
must however be introduced to perform the full renormalization of the
operators in the reduced effective Hamiltonian. Fortunately,
they can be included in the same framework, based on the SFR and HLT
methods,
discussed in detail for charming penguins in the previous sections. For
example, concerning the contribution of $O_8$ in $B\to K \ell^+ \ell^-$
decays, the hadronic factor is given
by the first line of Eq.\,(\ref{eq:H12def0}) with $O^{(c)}_{1,2}$
replaced by $O_8$ and the evaluation of the amplitude follows in the
same way
as for the charming penguins. This is true, in particular, for the
contribution from the time ordering in which the
electromagnetic current is inserted after the weak operator $O_8$, for
which the SFR and HLT methods are required.
Similarly, the contribution of the operator $O_8$ to the hadronic factor
for $\bar{B}_s \to \gamma \ell^+ \ell^-$ decays is given by Eq. (14)
with $O^{(c)}_{1,2}$ replaced by $O_8$ and the evaluation of the
amplitude follows the same steps as for the charming penguins. The
renormalization of the operator $O_8$ is discussed briefly in
Sec.\,\ref{subsec:O8renormalization} below.

In due course, as the computation of the leading contributions becomes
sufficiently precise, the computation of the CKM
suppressed contributions can also proceed following the same steps.

\section{Renormalization}\label{sec:renormalization}
The main aim of this paper is the demonstration, presented in the preceding sections, that, using recent methods based on the spectral density representation of the hadronic amplitudes, the real and imaginary parts of the amplitudes containing charming penguins can be evaluated from computations of Euclidean correlation functions. 
In order to complete the determination of the amplitudes, it is necessary to renormalize the ultraviolet divergences which are present in the diagrams. Although the general techniques required to perform the renormalization can be considered to be standard, some subtleties do appear in the evaluation of the diagrams in Fig.\,\ref{fig:CPBK123} and in this section we discuss how to handle these. 
These include logarithmic divergences in the diagrams of Fig.\,\ref{fig:CPBK123} which appear from the contact term when the electromagnetic current $J_{\mathrm{em}}$ approaches the weak operator $\ocharming$. Although not specific to the decays being studied in this paper, a significant additional complication is the presence of divergences which appear as inverse powers of the lattice spacing due to the mixing of $O_{1,2}^{(c)}$ with operators of lower dimension, including the scalar and pseudoscalar densities, $\bar{s} b$ and $\bar{s}\gamma^5 b$.  We discuss the contact terms in Sec.\,\ref{subsec:contact} and the subtraction of the power divergences in Sec.\,\ref{subsec:power}.

\subsection{The contact terms}\label{subsec:contact}

The non-perturbative renormalization relating local lattice operators containing logarithmic ultraviolet divergences with those defined in continuum schemes, such as the RI-Mom or RI-SMom schemes\,\cite{Martinelli:1994ty,Sturm:2009kb}, is a standard procedure. Since the Wilson coefficients have usually been calculated in the \msbar scheme, which is a purely perturbative one, the matrix elements computed in the intermediate RI-(S)Mom schemes have to be matched to the \msbar scheme and this matching is necessarily performed in perturbation theory. For the purposes of this section we assume that such a renormalization within QCD has been performed for the operators $O_{1,2}^{(c)}$, including the subtraction of the power divergences as described in 
Sec.\,\ref{subsec:power}. 

In addition to the ultraviolet divergences present in the renormalization of the local operators $O_{1,2}^{(c)}$ in QCD, other divergences appear as the electromagnetic current approaches one of the weak operators $O_{1,2}^{(c)}$ and, in particular, lead to the appearance of the operators $O_7$ and $O_9$ in the effective Hamiltonian in Eq.\,(\ref{eq:Hbtos}).
The renormalization of these contact terms is the subject of this section. 
Following the earlier sections, we present the discussion in Minkowski space, but it can readily be translated into Euclidean space by making the replacement $t\to -it$. 

For illustration we start by considering $B\to K\ell^+\ell^-$ decays and the Feynman diagrams of Fig.\,\ref{fig:CPBK123}(a) in which the photon is emitted from a charm quark. 
By dimensional power counting, the charm-quark loop would be quadratically divergent in the ultra-violet cut-off, i.e. would diverge as $1/a^2$ in a Euclidean lattice QCD computation. However, just as for the vacuum polarization in QED, electromagnetic current conservation reduces the degree of divergence by two, so that the divergence is logarithmic. 
In this case, the matrix elements are proportional to $(p_B\cdot q)q^\nu-q^2p_B^\nu$, where $\nu$ is the Lorentz index of the virtual photon, whereas, for example, matrix elements of weak currents in semileptonic decays are proportional to a single power of momentum. 
When applying the SFR method however, we evaluate the diagrams by computing the contributions from the two time-orderings $t<0$ and $t>0$ separately, see Eqs.\,(\ref{eq:Hmuminus}) and (\ref{eq:Hmuplus}).
Each of these terms is separately quadratically divergent, and the divergence cancels when the two contributions are summed, leaving only a logarithmic divergence to be renormalized in the standard way.  
In order to manage the treatment of ultraviolet divergences effectively, we propose to introduce subtractions into the energy integral for $H^{\nu+}_{1,2}(\vecp{q})$ in Eq.\,(\ref{eq:Hmuplus}), so as to separate the divergences at large energy $E$ from the behaviour around the pole at $E\simeq m_B$ which we treat using the SFR method. To this end we introduce the simple identity
\begin{equation}
    \frac{1}{E-m_B-i\epsilon}-\frac3{E-i\epsilon}+\frac3{E+m_B-i\epsilon}
    -\frac1{E+2m_B-i\epsilon}=\frac{6m_B^3}{(E-m_B-i\epsilon)(E-i\epsilon)(E+m_B-i\epsilon)(E+2m_B-i\epsilon)}\,.\label{eq:3subs}
\end{equation}
We see therefore that if we were to replace $1/(E-m_B-i\epsilon)$ in the expression for $H^{\nu+}_{1,2}$ in Eq.(\ref{eq:Hmuplus}) by the four terms on the left-hand side of Eq.\,(\ref{eq:3subs}), and noting that $\rho^{\nu+}(E,\vecp{q})\propto E^2$ at large $E$, the corresponding energy integral would be finite. Moreover the three additional terms on the left-hand side of Eq.\,(\ref{eq:3subs}) have been chosen so as not to have poles in the range of the energy integration\,\footnote{Clearly the choice of such terms is not unique.}. Recalling Eq.\,(\ref{eq:Cnuplus0}), the desired separation can now be readily achieved
\begin{eqnarray} H^{\nu+}_{1,2}(\vecp{q})&=&\lim_{\epsilon\to0}
\int_{E_+^\ast}^\infty\frac{dE}{2\pi}~\frac{\rho^{\nu+}_{1,2}(E,\vecp{q})}{E-m_B-i\epsilon}=3i\int_0^\infty \!\!dt~e^{-iE_Kt}C_{1,2}^{\nu+}(t,\vecp{q})-
3i\int_0^\infty \!\!dt~e^{-i(m_B+E_K)t}C_{1,2}^{\nu+}(t,\vecp{q}) \nn  \\
&&\hspace{2in}
+i\int_0^\infty \!\!dt~e^{-i(2m_B+E_K)t}C_{1,2}^{\nu+}(t,\vecp{q})+
\lim_{\epsilon\to0}H^{\nu+;\,3\mathrm{subs}}_{1,2}(\vec{q},\epsilon)\,,
\label{eq:Hwithsubtractions}
\end{eqnarray}
where 
\begin{eqnarray}
  H^{\nu+;\,3\mathrm{subs}}_{1,2}(\vec{q},\epsilon)&\equiv&  
  \int_{E_+^\ast}^\infty\frac{dE}{2\pi}~
  \bigg\{\frac{1}{E-m_B-i\epsilon}-\frac3{E-i\epsilon}+\frac3{E+m_B-i\epsilon}
    -\frac1{E+2m_B-i\epsilon}\bigg\}\,
  \rho^{\nu+}_{1,2}(E,\vecp{q})\,.\label{eq:H3subs}
\end{eqnarray}
The term $H^{\nu+;\,3\mathrm{subs}}_{1,2}(\vec{q},\epsilon)$, defined in Eq.\,(\ref{eq:H3subs}), has been designed so that it has no ultraviolet divergences, even at finite-values of $\epsilon$. We envisage applying the HLT procedure to each of the four terms in Eq.\,(\ref{eq:H3subs}).
The first three terms on the right-hand side of Eq.\,(\ref{eq:Hwithsubtractions}) compensate for the subtractions introduced in the definition of $H^{\nu+;\,3\mathrm{subs}}_{1,2}(\vec{q},\epsilon)$ given in Eq.\,(\ref{eq:H3subs}). 
They have no poles in the range of the integration over $E$ so that we can set $\epsilon=0$ in these terms. They do however, have quadratic and linear ultraviolet divergences. 
In order to cancel the power divergences, we combine these three terms with the contribution from the region $t<0$ in Eq.\,(\ref{Hnuminus0}) in a correlated way so that only the logarithmic divergence remains.
In a lattice QCD computation this combination is performed through a jackknife analysis, with the expectation that the correlations between the four terms will reduce the resulting uncertainty. 

Note that $H^{\nu+;\,3\mathrm{subs}}_{1,2}(\vec{q},\epsilon)$ defined in Eq.\,(\ref{eq:H3subs}), can be written in the form:
\begin{eqnarray}
 H^{\nu+;\,3\mathrm{subs}}_{1,2}(\vec{q},\epsilon)&=&i\int_0^\infty \!\!dt
 ~\big\{e^{i(q_0+i\epsilon)t}-3e^{-i(E_K-i\epsilon)t}+
 3e^{-i(m_B+E_K-i\epsilon)t}-e^{-i(2m_B+E_K-i\epsilon)t}\big\}
 ~C_{1,2}^{\nu+}(t,\vecp{q})\,,
\end{eqnarray}
so that the integrand vanishes at the point $t=0$. When evaluating the time integration in a Euclidean computation on a discrete lattice the contribution from the point at $t=0$ must therefore be included in the remaining terms.

In Appendix\,\ref{sec:appcontact} we present some additional remarks concerning the contact terms. Specifically, we describe the procedure for renormalizing the logarithmic divergence resulting in the appearance of the operators $O_9$ and $O_7$ in the effective Hamiltonian (\ref{eq:Heff7910}). We also explain how the transverse structure of the amplitude is recovered when the contributions from the two time-orderings are combined.

\begin{figure}[t]
\begin{center}
\includegraphics[width=0.4\hsize]{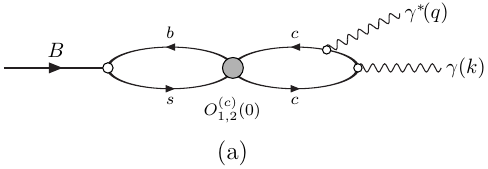}\hspace{0.4in}\includegraphics[width=0.4\hsize]{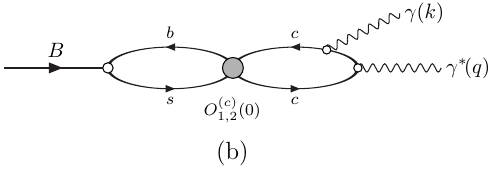}
\end{center}
\caption{Panels (a) and (b) show the two diagrams for the decay $\bar{B}_s\to\mu^+\mu^-\gamma$ in which both the real photon $\gamma$ with momentum $k$ and the virtual photon with momentum $q$ are emitted from the charm-quark loop. \label{fig:2cphotons}}
\end{figure}
 
A corresponding procedure needs to be developed and applied to the contact terms present in the amplitude for $\bar{B}_s\to\gamma\ell^+\ell^-$ decays. We postpone the presentation of the details to a future publication, when we will be in a position to have learned from the experience of computing the charming penguin contributions to $B\to K\ell^+\ell^-$ decay amplitudes.
Here we only note that for the process $\bar{B}_s \to \gamma\ell^+\ell^-$, there are also diagrams in which both the real and virtual photons are emitted from the charm-quark loop as in Fig.\,\ref{fig:2cphotons}. Although the superficial degree of divergence of such diagrams would allow for a
logarithmic ultraviolet divergence, and indeed, even at lowest order, each of the two diagrams in Fig.\,\ref{fig:2cphotons} separately has
such a divergence, electromagnetic gauge invariance ensures that sum of the two contributions from these diagrams is
finite. The requirement that the matrix element in Eq.\,(\ref{eq:Hmunutwogamma}) vanishes when contracted with $k^\mu$ and $q^\nu=(p_{\bar{B}_s}-k)^\nu$, implies that the matrix element is proportional to two powers of the external momenta which is sufficient to make it finite. However, similarly to the above discussion of the cancellation of the power divergences in $B\to K\ell^+\ell^-$ decays, the cancellation of the logarithmic ultraviolet divergences in the diagrams of Fig.8 requires the sum of the contributions from all the time-orderings. This is true since we use the lattice conserved vector current. Had we used  the local vector current a more complicated subtraction procedure would need to be implemented but we do not discuss this here.

\subsection{Subtraction of power divergences }\label{subsec:power}

The renormalization of the relevant lattice operators appearing in Eqs.\,(\ref{eq:O12def})-(\ref{eq:O910def}) and  their matching to the Wilson coefficients of the effective Hamiltonians in Eqs.\,(\ref{eq:Hbtos}) and (\ref{eq:Heff7910}) depends on the lattice regularization of the QCD quark action.  
In  this  subsection we present a general discussion about the renormalization of the composite operators of the Hamiltonian (\ref{eq:Heff7910}) and then, for illustration, we will specialize to the renormalization of the operators using Wilson-Clover twisted-mass fermions at maximal twist\,\cite{Frezzotti:2000nk}, which is the action used in Ref.\,\cite{Frezzotti:2024kqk} and in the ongoing numerical implementation of the methods discussed above.

The two semileptonic operators in Eq.\,(\ref{eq:O910def}) renormalize as combinations of a vector and an axial  vector current, thus in order to match them to the corresponding continuum operators we simply have
\bea O_9=  \frac{e^2}{(4\pi)^2}( Z_V  \hat V^\mu -Z_A \hat A^\mu) (\bar \ell \gamma_\mu \ell)\, ,  \qquad O_{10}=  \frac{e^2}{(4\pi)^2}( Z_V  \hat V^\mu -Z_A \hat A^\mu) (\bar \ell \gamma_\mu\gamma_5 \ell)\, , \eea
where $\hat V^\mu=\bar s \gamma^\mu b$ and $\hat A^\mu=\bar s \gamma^\mu \gamma_5 b$ are the bare lattice vector and axial vector currents.
The constants  $Z_V$ and $Z_A$ depend on the lattice regularization and are usually determined  non perturbatively using the Ward Identities (WI)\,\cite{Bochicchio:1985xa,Martinelli:1993dq} or some non-perturbative renormalization procedure, such as the RI-Mom\,\cite{Martinelli:1994ty}, RI-SMom\,\cite{Sturm:2009kb} or the Schr\"odinger functional\,\cite{Jansen:1995ck} schemes.

At the order in the electromagnetic coupling at which we are working, the operator $O_7$ simply renormalizes as the tensor operator
\bea O_7(\mu)= Z_T(\mu a,\alpha_s)  \hat T^{\mu\nu} F_{\mu\nu} \, , \eea
where $\hat T^{\mu\nu}=\bar s \sigma^{\mu\nu} P_R b$  is the bare lattice operator and $Z_T(\mu a,\alpha_s)$ can be determined using one's preferred non-perturbative method, such as those in Refs.\,\cite{Martinelli:1994ty,Sturm:2009kb,Jansen:1995ck}.
 
The renormalization of the operators $O_{1,2}^{(c)}$ and of the tensor operator $O_8$ is considerably more complicated. The mixing induced by electromagnetic corrections has been described above in  Sec.\,\ref{subsec:contact}, whereas we start here the discussion of the renormalization  due to QCD corrections. 

$O_{1,2}^{(c)}$ can mix with operators of lower dimensions, specifically with the scalar and pseudo-scalar densities and  two tensor operators 
\bea \bar sb \, , \qquad   \bar s  \gamma_5  b  \, , \qquad \bar s   \sigma^{\mu\nu} T^a b \, G^a_{\mu\nu} \, ,  \qquad  \bar s   \gamma_5 \sigma^{\mu\nu}T^a b \, G^a_{\mu\nu} \, , \label{eq:ldops}\eea
with power divergent coefficients. The subtraction coefficients required to remove the power divergences cannot be computed in perturbation theory and a non-perturbative subtraction procedure is required\,\cite{Maiani:1991az}.
For compactness of notation, the last two operators in Eq.\,(\ref{eq:ldops}) will be denoted as $\bar s   \sigma \cdot G b$ and $ \bar s  \gamma_5 \sigma \cdot G b$ in the remainder of this section.    

The chromomagnetic operator $O_8$ can also mix with the scalar and pseudo-scalar densities  with power divergent coefficients whereas the mixing with $\bar s   \sigma \cdot G b$ and $ \bar s  \gamma_5 \sigma \cdot G b$ is at most logarithmically divergent and can be computed either  in perturbation theory or with a non-perturbative method\,\cite{Martinelli:1994ty,Sturm:2009kb,Jansen:1995ck}. The renormalization of $O_8$ will be discussed in the dedicated subsection \ref{subsec:O8renormalization} below. 
 
 For the operators $O_{1,2}^{(c)}$ the degree of divergence of the mixing coefficients is reduced as a consequence of three symmetries of the QCD action:
 \begin{itemize}
 \item[i)] ${\cal C}{\cal S}$ which is the product of charge conjugation,  $b \leftrightarrow s$ flavor exchange and $m_b \leftrightarrow m_s$;  
\item[ii)]  $S_3$ which is the combined transformation of ${\cal P}_5$  defined by
\bea  q_f (x) \to  \gamma_5\gamma_0\,  q_f(x_P) \, , \qquad \bar q_f (x) \to - \bar q_f(x_P)\, \gamma_0\gamma_5 \, ,\label{eq:r5h} \eea
for all flavors $f$, together with $m_f\to -m_f$,  where  $x_P\equiv(t,-\vecp x)$.   ${\cal P}_5$  also requires a parity  transformation of the gauge links\,\cite{Frezzotti:2004wz},  not  given here  because its  form and role is standard.

\item[iii)] $R^{sp}_{5f}$ which, for a given  flavor $f$, is the transformation
\begin{equation}  q_f(x) \to q^\prime_f(x) = \gamma_5  q_f(x)\, , \qquad \bar q_f(x) \to \bar q^{\hspace{2pt}\prime}_f(x) = - \bar   q_f(x)\gamma_5 \, , \qquad  m_f \to - m_f \label{eq:r5g}\,.\end{equation}
\end{itemize}
The corresponding  transformations can  also be defined for the regularized lattice actions (the details of the lattice transformations depend on the lattice regularization).  

We now introduce the parity violating and parity conserving components of the operators $O^\pm\equiv1/2\, (O_{2}^{(c)} \pm O_{1}^{(c)})$  (we omit the label  $(c)$ for the remainder of this section) 
\bea 
O^{\pm }_{VA+AV} &=&  \frac{1}{2} \left[ \left(\bar s \gamma_\mu c \right) \left(\bar c \gamma_\mu \gamma_5b \right) \pm 
\left(\bar s \gamma_\mu b  \right) \left(\bar c \gamma_\mu \gamma_5c \right) \right]  +  \frac{1}{2} \left[ \left(\bar s \gamma_\mu \gamma_5 c \right) \left(\bar c \gamma_\mu   b \right) \pm 
\left(\bar s \gamma_\mu \gamma_5 b  \right) \left(\bar c \gamma_\mu   c \right) \right]\, , \label{eq:VApAVg}\\[0.1cm]
 O^{\pm }_{VV+AA} &=&  \frac{1}{2} \left[ \left(\bar s \gamma_\mu c \right) \left(\bar c \gamma_\mu b \right) \pm 
\left(\bar s \gamma_\mu b  \right) \left(\bar c \gamma_\mu c \right) \right]+  \frac{1}{2} \left[ \left(\bar s \gamma_\mu \gamma_5 c \right) \left(\bar c \gamma_\mu  \gamma_5 b \right) \pm 
\left(\bar s \gamma_\mu \gamma_5 b  \right) \left(\bar c \gamma_\mu \gamma_5  c \right) \right]\,.  \label{eq:VVpAAg}\eea
 
The weak operators $O^{\pm }_{VA+AV}$ are odd under the ${\cal C}{\cal S}$ symmetry whereas  $\bar sb$,  $\bar s  \gamma_5  b$ and  the tensor operators $\bar s   \sigma \cdot G b$ and $ \bar s  \gamma_5 \sigma \cdot G b$ are all even. Thus the  mixing coefficients of these operators to $O^{\pm }_{VA+AV}$ are proportional to $m_b-m_s$. 

Under the\textit{} $S_3$ symmetry, the parity violating operators  $O^{\pm }_{VA+AV}$ are odd, as are $ (m_b-m_s) \bar s  \gamma_5  b$  and $ (m_b-m_s)  \bar s  \gamma_5 \sigma \cdot G b$, whereas  $ (m_b-m_s) \bar sb$ and  $ (m_b-m_s) \bar s   \sigma \cdot G b$  are even. Thus $O^{\pm }_{VA+AV}$ mix with the lower dimensional operators of Eq.\,(\ref{eq:ldops}) multiplied by the appropriate powers of the quark masses: 
\begin{equation}     
(m_b-m_s)\,  \bar s \gamma_5 b\, , \quad  (m_b-m_s) (m_b+m_s)\,  \bar s b\, , \quad
  (m_b-m_s)\,  \bar s \gamma_5  \sigma \cdot G b\, , \quad
 (m_b-m_s) (m_b+m_s)\, \bar s   \sigma \cdot G b \, . \label{eq:mixops} \end{equation}
This demonstrates that  the mixing coefficients with the chromomagnetic operators are finite or at most logarithmically divergent, whereas the coefficients of the operators $ (m_b-m_s)\, \bar s \gamma_5 b$ and $(m_b-m_s) (m_b+m_s)\,  \bar s b$ are quadratically and linearly divergent respectively.

For the parity conserving component we use  the  $R^{sp}_{5b}$ symmetry defined in Eq.\,(\ref{eq:r5h}),  under which 
$O^{\pm }_{VA+AV}\to O^{\pm }_{VV+AA}$, so that $O^{\pm }_{VV+AA}$ mixes with the lower dimensional operators of Eq.\,(\ref{eq:ldops}) multiplied by powers of the quark masses as follows:
\begin{equation}
    (m_b+m_s)  \bar s  b\, , \quad  (m_b-m_s) (m_b+m_s)  \bar s\gamma_5 b\, , \quad
  (m_b+m_s)  \bar s  \sigma \cdot G b\, , \quad
 (m_b-m_s) (m_b+m_s) \bar s   \gamma_5 \sigma \cdot G b \, . \label{eq:mixops1}  \end{equation}
The degree of divergence of the mixing coefficients can be readily derived as in the previous discussion.   

In chirally invariant formulations of lattice  QCD and in regularizations which preserve parity,  half of the mixing terms listed in Eqs.\,(\ref{eq:mixops}) and (\ref{eq:mixops1}),   namely those of $O^\pm_{VA+AV}$ with parity even terms and those of $O^\pm_{VV+AA}$ with parity odd terms,  are absent. They can also be eliminated if the symmetries  $R^{sp}_{5b}$ and  $R^{sp}_{5s}$ are enforced in the correlators computed in the lattice regularised theory (by introducing suitable replicas of the valence fermions $b$ and $s$, see below).

The renormalization of the operators depends on the lattice regularization of the quark action and on the presence or absence of the GIM mechanism. Note that in the  case of $O_{1,2}^{(c)}$, GIM is absent since the top quark is much heavier than the lattice cutoff.

\subsubsection{Renormalization in the presence of the GIM mechanism}
Although for the processes that we are studying we do not have the GIM mechanism,  we can imagine introducing a fictitious top quark with a varying mass $m_t$. The dependence of the value of the matrix 
elements of the operators as a function of $m_t$ is known from the perturbative renormalization group, provided that $m_t \gg \Lambda_{\mathrm{QCD}}$, and can be extrapolated to a renormalization scale $\mu \simeq 1/a$ so as to match  the effective Hamiltonian of Eq.\,(\ref{eq:Heff7910}), as first proposed in a different context in Ref.\,\cite{Dawson:1997ic}\,. This motivates considering also the subtraction of power divergences in the presence of the GIM mechanism.

In  the  presence of GIM the mixing coefficients will acquire an additional factor of $(m_t-m_c)$, thus, for example, the parity violating operators $O^{\pm }_{VA+AV}$, will mix with the lower dimensional operators of Eq.\,\,(\ref{eq:ldops}) multiplied now by powers of the quark masses as follows:
\bea &&    (m_t-m_c)(m_b-m_s)\,  \bar s \gamma_5 b\, , \quad (m_t-m_c) (m_b-m_s)\,  \bar s \gamma_5  \sigma \cdot G b\, ,  \quad\nonumber \\[0.05cm] && 
 (m_t-m_c)(m_b-m_s) \,  \bar s b\, ,  \quad
 (m_t-m_c)(m_b-m_s) \, \bar s   \sigma \cdot G b \, . \label{eq:gimpv}\eea
The symmetries i) and ii) defined above imply that the parity violating operators given in the first line of Eq.\,(\ref{eq:gimpv}) get an additional mass factor, either $(m_t+m_c)$ or $(m_b+m_s)$. This implies that the only residual power divergence of $O^{\pm}_{VA+AV}$ is the linear one, due to the mixing with the scalar density. 
This linear  divergence can be eliminated 
by imposing that the matrix element of the operator $\tilde O^{\pm}_{VA+AV}$ between the B-meson and the vacuum vanishes\,\footnote{We are not assuming that the lattice regularisation preserves parity}
\bea  \langle 0 \vert \tilde O^{\pm}_{VA+AV}\vert B\rangle =  \langle 0 \vert O^{\pm}_{VA+AV} +\frac{C^\pm_{VA+AV}}{a} (m_t-m_c) (m_b-m_s)  \bar s  b \vert B\rangle =0 \, .  \label{eq:pvg}\eea 

For the parity conserving part, $O^{\pm}_{VV+AA}$  the mixing pattern is analogous and the elimination of the residual $1/a$ divergence can be achieved by imposing
\bea  \langle K \vert \tilde O^{\pm}_{VV+AA}\vert B\rangle =  \langle K \vert O^{\pm}_{VV+AA} +\frac{C^\pm_{VV+AA}}{a} (m_t-m_c) (m_b+m_s)  \bar s \gamma_5 b \vert B\rangle =0 \, .  \label{eq:pag}\eea 

If we specialize to Twisted-Mass Fermions one can also use the strategy described in full detail in Ref.\,\cite{Frezzotti:2004wz}, which we will not discuss here, and eliminate the linear divergences using several replicas of  the number of top and charm flavors with different Wilson parameters $r_f$.

\subsubsection{Renormalization without the GIM mechanism}
In this section we specialize to twisted mass fermions although strategies similar to the one explained below can be implemented also with other lattice regularizations. We will give also  the lattice version of the symmetries discussed in i)-iii) above. A more detailed discussion can be found in  Ref.\,\cite{Frezzotti:2004wz}.

Without GIM we can adapt the approach of Ref.\,\cite{Frezzotti:2004wz} in order to completely eliminate the  power divergences non-perturbatively. In order to implement this approach we distinguish the sea quarks, which enter in the action used to generate the gauge field configurations, from the valence Ostervalder-Seiler (OS) quenched quarks,  which  enter  in the weak operators and in the sinks and sources  used to annihilate or create the external particles.

In order to implement the strategy of Ref.\,\cite{Frezzotti:2004wz}, we need the lattice version of several symmetries. We start by recalling some basic facts about twisted mass fermions:
\begin{enumerate}
\item The OS action for the valence quark field $q_f$ is defined as 
\bea 
\label{eq:OS_action}
S_{OS}(m_{f}, r_{f}) = \sum_x \bar q_f (x) \left[ \gamma \cdot \tilde \nabla - i \gamma_5 W_{cr}(r_f) + m_f\right] q_f(x)\, , \eea
with 
\bea \gamma \cdot \tilde \nabla = \frac{1}{2} \sum_\mu \gamma_\mu\left( \nabla_\mu^* + \nabla_\mu \right) \, , \qquad   W_{cr}(r_f) =
-\frac{r_f}{2}  \sum_\mu  \nabla_\mu^* \nabla_\mu +M_{cr} (r_f) \, , \eea
where $\nabla_\mu\psi(x)=\frac1a[U_\mu(x)\psi(x+\hat\mu)-\psi(x)]$, $\nabla^\ast_\mu\psi(x)=\frac1a[\psi(x)-U^\dagger_\mu(x-a\hat\mu)\psi(x-a\hat\mu)]$ and $U_\mu(x)$ is the gauge link. 
The critical mass, $M_{cr}$ satisfies $M_{cr} (-r_f)=-M_{cr} (r_f)$ and the renormalized valence quark mass is defined by
\bea \hat m_f =Z_m(r_f) m_f\, . \eea
We will ignore the transformation properties of the ghost fields, described in Ref.\,\cite{Frezzotti:2004wz},  because they are not relevant to our discussion;
\item Consider the change of  variables 
\bea  R_{5f}: \qquad         q_f(x) \to q^\prime_f(x) = \gamma_5  q_f(x)\, , \qquad \bar q_f(x) \to \bar q^\prime_f(x) = - \bar   q_f(x)\gamma_5\, , \label{eq:r5fr}\eea
then a symmetry of each flavor action is given by  the spurionic $R^{sp}_{5f}$ transformation
\bea  R^{sp}_{5f}=R_{5f} \times \left( r_f \to -r_f\right)   \times \left( m_f \to -m_f\right) \, ;   \label{eq:Rsp} \eea
\item The valence action is invariant under the spurionic symmetry ${\cal P}\times \left(r_f \to -r_f\right)$,  where ${\cal P}$ is the parity operator
\bea
q_f(x) \to \gamma_0  q_f(x_P)\, , \qquad  \bar q_f(x) \to  \bar   q_f(x_P)\gamma_0  \, , \eea
for all the flavors; 
\item We introduce the further spurionic transformation which leaves  the lattice action invariant
\bea S_3  ={\cal P}_5 \times (M \to -M)\, , \eea
where $M \equiv \{all \, f \, , \, m_f\}$ and ${\cal P}_5$ is the transformation
\bea  \qquad q_f (x) \to  \gamma_5\gamma_0  q_f(x_P) \, , \qquad \bar q_f (x) \to - \bar q_f(x_P) \gamma_0\gamma_5 \,; \eea
\item We recall the charge conjugation symmetry, ${\cal C}$, for the valence quarks
\bea {\cal C}: \qquad q_f (x) \to i \gamma_0\gamma_2 \bar q^T_f(x) \, , \qquad \bar q_f (x) \to -q^T_f(x) i \gamma_0\gamma_2 \, ;\eea
\item We will also use the transformations
\bea (f_1 \leftrightarrow  f_2): \qquad  q_{f_1} \leftrightarrow q_{f_2}  \, , \qquad  (f_1 \leftrightarrow f_2)_5: \qquad  q_{f_1}\to \gamma_5 q_{f_2} \, , \qquad \bar q_{f_1}\to - \bar q_{f_2} \gamma_5 \, ; \eea
\item Finally we have the ${\cal CS}$ symmetry ${\cal CS}=(s\leftrightarrow b) \times {\cal C} \times (m_s \leftrightarrow m_b) $ under which both $\bar s b$ and $\bar s \gamma_5 b$ are even. 
 \end{enumerate}

We now discuss the subtraction of the power divergences of the diagrams $(b)$ and $(c)$ of Fig.\,\ref{fig:CPBK123}. 
Following Ref.\,\cite{Frezzotti:2004wz}, in order to reduce the degree of divergence of the operators it is convenient to  consider at a formal level an auxiliary gauge model, $(4s6v)$,  where, in addition to the four lightest sea quark flavors ($u_{\mathrm{sea}},\, d_{\mathrm{sea}}, s_{\mathrm{sea}}$
and $c_{\mathrm{sea}}$)  and the $b$ and $s$ valence quarks,  the charm valence quarks are replicated in four copies $c^{[i]},\,i=0$\,-\,$3$,
together with the corresponding ghosts, required to cancel
the associated valence determinant\,\footnote{In the notation of Ref.\,\cite{Frezzotti:2004wz} $c^{[0]}=c,~c^{[1]}=c^\prime,~ c^{[2]}=c^{\prime \prime},$ and $c^{[3]}=c^{\prime \prime\prime}$.}.

We start by considering the parity violating and parity conserving four-fermion operators 
\bea O^{\pm [k]}_{VA+AV} &=&  \frac{1}{2} \left[ \left(\bar s \gamma_\mu c^{[k]} \right) \left(\bar c^{[k]} \gamma_\mu \gamma_5b \right) \pm 
\left(\bar s \gamma_\mu b  \right) \left(\bar c^{[k]} \gamma_\mu \gamma_5c^{[k]} \right) \right]\nonumber \\  &+&  \frac{1}{2} \left[ \left(\bar s \gamma_\mu \gamma_5 c^{[k]} \right) \left(\bar c^{[k]} \gamma_\mu   b \right) \pm 
\left(\bar s \gamma_\mu \gamma_5 b  \right) \left(\bar c^{[k]} \gamma_\mu   c^{[k]} \right) \right]\, , \label{eq:VApAV}\eea
\bea O^{\pm [k]}_{VV+AA} &=&  \frac{1}{2} \left[ \left(\bar s \gamma_\mu c^{[k]} \right) \left(\bar c^{[k]} \gamma_\mu b \right) \pm 
\left(\bar s \gamma_\mu b  \right) \left(\bar c^{[k]} \gamma_\mu c^{[k]} \right) \right]\nonumber \\  &+&  \frac{1}{2} \left[ \left(\bar s \gamma_\mu \gamma_5 c^{[k]} \right) \left(\bar c^{[k]} \gamma_\mu  \gamma_5 b \right) \pm 
\left(\bar s \gamma_\mu \gamma_5 b  \right) \left(\bar c^{[k]} \gamma_\mu \gamma_5  c^{[k]} \right) \right]\, , \eea
where the label $[k]$ denotes one of the four copies of the valence charm quarks, whose renormalized masses must all be taken to be equal to each other and to the mass of the sea charm quark.
For the parity violating component,  the operators which enter in the correlation function necessary to extract the physical amplitudes are 
\bea O^{\pm}_{VA+AV}= O^{\pm [0]}_{VA+AV}+O^{\pm [1]}_{VA+AV}-\frac{1}{2}O^{\pm [2]}_{VA+AV}-\frac{1}{2}O^{\pm [3]}_{VA+AV}\, , \label{eq:pv}\eea
and in this case we take
\bea r_b =r_s \, , \qquad r_s =r_{c^{[0]}}=-r_{c^{[1]}}=r_{c^{[2]}}=-r_{c^{[3]}}\, . \eea  
Similarly
\bea O^{\pm}_{VV+AA}= O^{\pm [0]}_{VV+AA}+O^{\pm [1]}_{VV+AA}-\frac{1}{2}O^{\pm [2]}_{VV+AA}-\frac{1}{2}O^{\pm [3]}_{VV+AA}\, , \label{eq:pc}\eea
for which we take 
\bea r_b =-r_s \, , \qquad r_s =r_{c^{[0]}}=-r_{c^{[1]}}=r_{c^{[2]}}=-r_{c^{[3]}}\, . \eea
The operators in Eqs.\,(\ref{eq:pv}) and (\ref{eq:pc}) are even for $r_c \to -r_c$ and in Ref.\,\cite{Frezzotti:2004wz} it was  demonstrated that the correlation functions of interest in the model $(4s6v)$ are exactly the same as in standard QCD, up to discretization errors of $O(a^2)$.

 Since $c^{[1]}$ and $c^{[3]}$ have the same Wilson parameter ($r_{c^{[1]}}=r_{c^{[3]}}$) and the same renormalized mass, we do not need four charm flavors, three are sufficient, and we may simplify  the above expressions and write 
\bea O^{\pm}_{VA+AV}&=& O^{\pm [0]}_{VA+AV}+\frac{1}{2} O^{\pm [1]}_{VA+AV}-\frac{1}{2}O^{\pm [2]}_{VA+AV}\, , \nonumber \\ O^{\pm}_{VV+AA}&= &O^{\pm [0]}_{VV+AA}+\frac{1}{2} O^{\pm [1]}_{VV+AA}-\frac{1}{2}O^{\pm [2]}_{VV+AA} \,.\eea
Without GIM we can use ${\cal CS}$ and $ S_3  ={\cal P}_5 \times (M \to -M)$
to show that the parity violating operator mixes with 
\bea  (m_b-m_s)  \bar s \gamma_5 b   \quad {\rm and}\quad  (m_b-m_s)  (m_b+ m_s)  \bar s  b\, , \eea
thus the coefficients of the pseudo-scalar and scalar densities are quadratically and linearly divergent respectively.  The quadratic divergence can be eliminated 
by imposing that the matrix element of the operator  between the B meson and the vacuum vanishes, namely
\bea  \langle 0 \vert \tilde O^{\pm}_{VA+AV}\vert B\rangle =  \langle 0 \vert O^{\pm}_{VA+AV} +\frac{C^\pm_{VA+AV}}{a^2}  (m_b-m_s)  \bar s \gamma_5 b \vert B\rangle =0 \, .  
\label{eq:pvx}\eea 
The linear divergence of the coefficient of the operator  $(m_b-m_s)  \bar s \gamma_5 b $ is odd in $r_f$, being proportional to  $m_c \, r_f$, and, since the action and the operators are even in $r_f$,   it  vanishes. Thus  the next term of the coefficient  of the operator  $(m_b-m_s)  \bar s \gamma_5 b $ corresponds to a  factor proportional to $m^2_c$ and is finite or at most logarithmically divergent. 
The linear divergence due to the mixing of  the operator $O^{\pm}_{VA+AV}$ with the scalar density is absent since  the coefficient of the scalar density is odd in $r_f$  whereas the operator and the action are (or can be made)  even.  The mixing with the operator $ \bar s \gamma_5  \sigma \cdot G b $ is only logarithmically divergent and can either be computed in perturbation theory  or by using the  non-perturbative   RI-(S)Mom or   Schr\"odinger functional schemes.

For the parity conserving operator, the most severe UV divergence is the one with the scalar operator and we replace the condition in Eq.\,(\ref{eq:pvx}) with 
\bea  \langle K \vert \tilde O^{\pm}_{VV+AA}\vert B\rangle =  \langle K \vert O^{\pm}_{VV+AA} +\frac{C^\pm_{VV+AA} }{a^2}  (m_b+m_s)  \bar s  b \vert B\rangle =0 \, .  \label{eq:pvv}\eea  

This completes the explicit demonstration that with Twisted Mass fermions, all the power divergences can be subtracted non-perturbatively. 
Frezzotti and Rossi in  Ref.\,\cite{Frezzotti:2004wz} also considered   the mixing of $O^{\pm}_{VA+AV}$ and $O^{\pm}_{VV+AA}$ with  other operators of dimension 6. They have shown that after the subtraction of the power divergences the renormalization of the subtracted operators is multiplicative. Since the demonstration does not depend on GIM it remains valid also in our case. 
\subsubsection{Renormalisation with an extended symmetry of the valence quarks}
\label{sec:exsy}
Another possibility, which may prove to be more convenient  to implement,   is based on the   symmetry of the fermion action with maximally twisted Wilson quarks, namely  ${\cal P}\times {\cal D}_d \times (M \to -M)$\,\cite{Frezzotti:2004wz}, where ${\cal P}$ is the parity transformation and, for what is relevant for the present discussion, ${\cal D}_d $ is given by 
\bea q_f(x) \to e^{3i\pi/2} q_f(-x) \, , \qquad  \bar q_f(x) \to e^{3i\pi/2} \bar q_f(-x) \, , \quad U_\mu(x) \to U^\dagger_\mu(-x-a\hat\mu)\, , \eea
and on the extension of $R^{sp}_{5b}$  and  $ R^{sp}_{5s}$, see Eq.\,(\ref{eq:Rsp}),  to valence quarks. In practice this possibility amounts to  simply averaging the relevant correlation functions  over $r_b= \pm 1$ and $r_s =\pm 1$ ($(r_b, r_s) = (++, +-, -+, --)$). Following Refs.\,\cite{Frezzotti:2004wz} and \,\cite{Constantinou:2015ela}  one can show that the transformation properties of the different operators under ${\cal CS}$, ${\cal S}_3$,   $R^{sp}_{5b}\times R^{sp}_{5s}$ and ${\cal P}\times {\cal D}_d\times (M \to -M)$  are sufficient to eliminate the mixing of the parity violating operators $O^{\pm}_{VA+AV}$ with the opposite parity operators $\bar s \sigma \cdot G b$ and $\bar s b$ as well as with all other parity violating or conserving operators of dimension 6. The subtraction of the quadratically divergent term proceeds exactly as in Eq.\,(\ref{eq:pvx}).  
Similarly,   one can show that for the parity conserving operators $O^{\pm}_{VV+AA}$,  the extension of $R^{sp}_{5b}\times R^{sp}_{5s}$ to valence quarks is sufficient to eliminate the mixing with the opposite parity operators $\bar s \gamma_5 \sigma \cdot G b$ and $\bar s \gamma_5 b$ as well as with all other parity conserving or violating operators of dimension 6. The subtraction of the quadratically divergent term proceeds exactly as Eq.\,(\ref{eq:pvv}).

In the presence of the GIM mechanism, the average of the correlation functions over $r_b= \pm 1$ and $r_s =\pm 1$ also reduces the degree of divergence of the mixing of the operators $O^{\pm}_{VA+AV}$ and $O^{\pm}_{VV+AA}$ with lower dimensional operators.  In the present  case  the mixing coefficients must have also  the same transformation properties of $O^{\pm}_{VA+AV}$ and $O^{\pm}_{VV+AA}$ with respect to 
$\Pi_{f=all\, f } R^{sp}_{5f}$   including $f=s$ and $f=b$. 
For this reason, for the operator  $O^{\pm}_{VA+AV}$, instead of Eq.\,(\ref{eq:gimpv}) the mixing structure will include
\bea     (m_t^2-m_c^2)(m_b-m_s)\,  \bar s \gamma_5 b\ \, ,\label{eq:gimpv1} \eea
with a logarithmically divergent coefficient, while the coefficient of the operator  
\bea  (m_t^2-m^2_c) (m_b-m_s)\,  \bar s \gamma_5  \sigma \cdot G b\, , \label{eq:gimpv2}\eea
 by dimensional counting,  is  reduced to a lattice artifact of $O(a^2)$.
The coefficients of the mixing with the parity conserving operators $\bar s b$ and $\bar    \sigma \cdot G b$  vanish. 
For the parity conserving operators $O^{\pm}_{VV+AA}$ 
the mixing structure will include
\bea     (m_t^2-m_c^2)(m_b-m_s)\,  \bar s  b\ \, ,\label{eq:gimpv1} \eea
with a logarithmically divergent coefficient while the coefficient of the operator  
\bea  (m_t^2-m^2_c) (m_b-m_s)\,  \bar s  \sigma \cdot G b\, , \label{eq:gimpv2}\eea
is  again reduced to a lattice artifact of $O(a^2)$ by dimensional counting. 
The coefficients of the mixing with the parity violating operators $\bar s \gamma_5 b$ and $\bar    \gamma_5\sigma \cdot G b$  vanish.

For both the parity-violating and parity-conserving cases the subtraction of the logarithmic divergences can be performed either by computing the coefficients of the operators in perturbation theory or non-perturbatively in analogy with Eqs.\,(\ref{eq:pvg}) and (\ref{eq:pag}).

\subsubsection{Renormalization of the operator $O_8$}\label{subsec:O8renormalization}
The elimination of the power divergences of the chromomagnetic operators $\bar s \gamma_5  \sigma \cdot G b$ and $\bar s   \sigma \cdot G b$
is much simpler since they can only mix with the pseudo-scalar and  scalar densities with quadratically divergent coefficients.   Thus, for example, we may eliminate the quadratic divergence  of the operator $\bar s \gamma_5  \sigma \cdot G b$ by imposing  the condition 
\bea   \langle 0 \vert \bar s \gamma_5  \sigma \cdot G b +\frac{C^{\sigma5}}{a^2}   \bar s \gamma_5 b \vert B\rangle =0 \, .  \label{eq:sig5}\eea 
Indeed with twisted mass fermions there is also a possible linear divergence coming from the mixing with the opposite parity operator
$O_{mS}=(m_b r_b+m_s r_s) \bar s b$\,\cite{Constantinou:2015ela},
 which can be eliminated by imposing the condition
\bea   \langle K \vert \bar s \gamma_5  \sigma \cdot G b +\frac{C^{\sigma5}}{a^2}   \bar s \gamma_5 b+\frac{C^{\sigma5}_{mS}}{a}O_{mS} \vert B\rangle =0 \, .  \label{eq:sig512}\eea 

For the parity conserving operator $\bar s   \sigma \cdot G b$ we again use the   $R^{sp}_{5b}$ symmetry and impose the following renormalization conditions 
\bea   \langle K \vert \bar s   \sigma \cdot G b +\frac{C^{\sigma}}{a^2}   \bar s  b \vert B\rangle =0 \, .  \label{eq:sig}\eea 
\bea   \langle 0 \vert \bar s   \sigma \cdot G b +\frac{C^{\sigma}}{a^2}   \bar s  b+\frac{C^{\sigma}_{mP}}{a} O_{mP} \vert B\rangle =0 \, ,  \label{eq:sig12}\eea
where $ O_{mP}=(m_b r_b+m_s r_s) \bar s \gamma_5 b$.

Once that the power divergences have been subtracted, the  chromomagnetic operators must be multiplicatively renormalized either in perturbation theory or non perturbatively.
In the latter case the renormalization procedure is particularly  complicated since the external legs in the Green functions are off-mass-shell and the operators mix with gauge non invariant operators, for more detail see Ref.\,\cite{Constantinou:2015ela}.  

The mixing of chromomagnetic operators of definite parity with  densities of the opposite parity (multiplied by
factors linear in the quark masses) is  removed if one averages the correlators containing the insertion of the chromomagnetic operator over opposite values of $r_b$  and $r_s$. In this case, in contrast to the four-fermion  operators  $O^{\pm}_{AV+VA}$  and  $O^{\pm}_{VV+AA}$ where an average over four cases is necessary,  the form of  $O_{mS}$ and $O_{mP}$ implies that it is sufficient to average over two cases, e.g.~$(r_b,r_s)=(1,1)$ and $(-1,-1)$.

\section{Exploratory lattice QCD calculation of charming penguin contributions to {\boldmath$B\to K\ell^{+}\ell^{-}$} decay amplitudes}\label{sec:exploratory}
To test the effectiveness of the proposed method, we have carried out an exploratory numerical calculation of $C^{\nu+}_{1,2;E}(t,\vecp{q})$ in Eq.\,(\ref{eq:CnuplusE}) on a single gauge ensemble generated by the Extended Twisted Mass Collaboration (ETMC) with $N_{f}=2+1+1$ dynamical Wilson-Clover twisted-mass fermions. This ensemble corresponds to a lattice spacing $a\simeq 0.079\,{\rm fm}$. Further information about the gauge ensemble used in the present calculation is collected in Tab.\,\ref{tab:simudetails}.
 \begin{table}[t]
\begin{ruledtabular}
\begin{tabular}{lcccccc}
\textrm{ID} & $L/a$ & $m_{\pi}$ [\textrm{GeV}] & $a$ [\textrm{fm}] & $am_{\ell}$ & $am_{s}$ & $am_{c}$ \\
\colrule
\textrm{B64} & $64$ & $140.2(3)$ & $0.07948(11)$ & $0.00072$ & $0.0182782$  & $0.231567$ 
\end{tabular}
\end{ruledtabular}
\caption{The $N_{f}=2+1+1$ ETMC gauge ensemble used in the present computation. We present the spatial extent in lattice units $L/a$, the pion mass $m_{\pi}$, the lattice spacing $a$, and the values of the bare light- ($am_{\ell}$), strange- ($am_{s}$), and charm-quark mass ($am_{c}$), in lattice units. The strange and charm quark masses are tuned so as to produce a charged kaon mass $m_{K} = 494.6\,{\rm MeV}$ and a $D_{s}$-meson mass $m_{D_{s}} = 1967\,{\rm MeV}$. As explained in the text, we have taken a lighter-than-physical $b$-quark mass, $m_{b} = 2m_{c}$.
\label{tab:simudetails}}
\end{table}
We employ the mixed-action lattice framework introduced in~Ref.\,\cite{Frezzotti:2004wz}, and described in the appendices of Ref.\,\cite{ExtendedTwistedMass:2024nyi}. The action of the valence quarks is discretized in the OS regularization (see Eq.\,(\ref{eq:OS_action}))
\begin{align}
\label{eq:tm_action}
S = \sum_{f=u,d,s,c,b} S_{\rm{OS}}(m_{f}, r_{f})~,
\end{align}
where $m_{f}$ is the mass of the quark with flavor $f$ (with $m_{u}=m_{d}=m_{l}$), and $r_{f}=\pm 1$ is the sign of the twisted-Wilson parameter  for the flavor $f$ ($r_{u,c}= -r_{d,s,b}=1$ for the present calculation). 

For this numerical test we have limited ourselves to the calculation of the charming penguin diagram in Fig.\,\ref{fig:CPBK123}(a). Since we are only interested here in investigating the potential of the SFR/HLT method, we do not consider issues related to the renormalization of the weak effective Hamiltonian discussed above, and compute the penguin contraction of the bare correlator. 
Note however that, the diagram in Fig.\,\ref{fig:CPBK123}(a) does not have ultraviolet power divergences\,\footnote{This can be seen by interpreting the diagram considered as a correlation function in a partially-quenched theory in which the charm quark field $c(x)$ and the charm antiquark field $\bar{c}(x)$ entering both the four-fermion operators and the electromagnetic current, are considered as distinct fields, $c(x)$ and $\bar{c}'(x)$, sharing the same mass electric-charge and twisted Wilson parameter $r_{c}=r_{c'}$. 
Since the four fermion operators have then $\Delta S=-\Delta B=\Delta C=-\Delta C' = 1$, they cannot mix with lower-dimensional operators.}, in contrast to those in Figs.\,\ref{fig:CPBK123}(b), (c) and (d) which do. This is important, as the presence of unsubtracted power-divergences might hide the physical signal in the bare correlation functions. 
The renormalization of the logarithmic divergence present in the diagram of Fig.\,\ref{fig:CPBK123}(a) is discussed in detail in Sec.\,\ref{subsec:contact} and Appendix\,\ref{sec:appcontact}. In particular, in Eq.\,(\ref{eq:H3subs}) we have defined the quantity $H^{\nu+;\,3\mathrm{subs}}_{1,2}(\vecp{q},\epsilon)$ which has no contact divergences but which contains the terms requiring the application of the spectral density methods. It is $H^{\nu+;\,3\mathrm{subs}}_{1,2}(\vecp{q},\epsilon)$ which is the focus of this numerical study.

We work in the rest frame of the $B$-meson, and consider a single value of the virtual photon's three-momentum $|\vecp{q}|=2\pi/L$ (where $L = 64\,a \simeq 5.1~{\rm fm}$ is the spatial extent of the lattice). In physical units $|\vecp{q}| =|\vecp{p}_K|\simeq 250\,{\rm MeV}$. The gauge ensemble used in the present calculation corresponds to physical light, strange and charm quark masses, but the $b$-quark has a lighter-than-physical mass, $m_{b} = 2m_{c}$. 
The momentum $\vecp{q}$ is chosen to be in the third direction, $\vecp{q} = (0,0, |\vecp{q}|)$. We have evaluated the following four-point correlation functions (retaining only the Wick contraction leading to the quark-connected diagram in Fig.\,\ref{fig:CPBK123}(a))
\begin{align}
\label{eq:four_pt}
C^{\nu}_{1,2; E}(t, \vecp{q} ; t_{K}, t_{B} ) = \sum_{\vecp{x},\vecp{y},\vecp{z}} e^{-i\vecp{q}(\vecp{x}-\vecp{y})}   \big\langle 0 \big| T\left[ \phi_{K}(t_{K},\vecp{y})  \,J^{\nu}_{\rm em}(t,\vecp{x}) \,  O_{1,2}^{(c)}(0) \, \phi^{\dagger}_{B}(-t_{B}, \vecp{z}) \right] \big| 0 \big\rangle~,
\end{align}
where $\phi_{B}^{\dag}$ and $\phi_{K}$ are interpolating operators which create the $B$- and annihilate the $K$-meson, respectively. In order to improve the overlap of the interpolating operators with the initial and final states, we have used Gaussian smeared interpolating fields $\phi^{\dag}_{B}$ and $\phi_{K}$. The time $t_{B}$ in Eq.\,(\ref{eq:four_pt}) has been fixed to $t_{B}= 12a$, and the correlation functions have been evaluated as a function of both $t$ and $t_{K}$. To amputate the external states we have also evaluated the $B$- and $K$-meson two-point correlation functions
\begin{align}
C_{\rm 2pt}^{K}(t) &\equiv \sum_{\vecp{x}} e^{-i\vecp{q}\vecp{x}} \langle 0 | \phi_{K}(t,\vecp{x}) \phi_{K}^\dagger(0, \vecp{0}) | 0 \rangle = \frac{|Z_{K}|^2}{2 E_{K}} \left( e^{-E_{K} t} + e^{-E_{K} (T-t)}   \right) + \ldots \\
    C_{\rm {2pt}}^{B}(t) &\equiv \sum_{\vecp{x}} \langle 0 | \phi_{B}(t, \vecp{x}) \phi_{B}^\dagger(0,\vecp{x}) | 0 \rangle = \frac{|Z_{B}|^2}{2 m_{B}} \left( e^{-m_{B} t} + e^{-m_{B} (T-t)}   \right) + \ldots~,
\end{align}
where $T=2L$ is the temporal extent of the lattice and the ellipses represent subleading terms in the limit of large times. For $t_{B},t_{K} \to +\infty$ one has
\begin{align}
C^{\nu}_{1,2; E}(t, \vecp{q} ; t_{K}, t_{B}) &= \frac{Z_{B}}{2 m_{B}} \frac{Z_{K}}{2 E_{K}} e^{-m_{B} t_{B}} e^{-E_{K}t_{K}}  \times \langle K(-\vecp{q}) \big| T  \left[ \tilde{J}_{\rm em}^{\nu}(t, \vecp{q}) \, O_{1,2}^{(c)}(0) \right]  \big| B(\vecp{0}) \rangle       ~ \nonumber \\[8pt]
&= \frac{Z_{B}}{2 m_{B}} \frac{Z_{K}}{2 E_{K}} e^{-m_{B} t_{B}} e^{-E_{K}t_{K}}  \times C_{1,2; E}^{\nu}(t,\vecp{q})\nonumber\\[8pt]
&= \frac{Z_{B}}{2 m_{B}} \frac{Z_{K}}{2 E_{K}} e^{-m_{B} t_{B}} e^{-E_{K}t_{K}}  \times \left[ \theta(t)C^{\nu+}_{1,2; E}(t,\vecp{q}) + \theta(-t)C^{\nu-}_{1,2; E}(t,\vecp{q})   \right]
\,.
\end{align}
In evaluating the Euclidean correlation functions we have taken $J^\nu_{\mathrm{em}}(t,\vecp{x})$ to be the exactly-conserved point-split electromagnetic current given in Eq.\,(B10) of Ref.\,\cite{Desiderio:2020oej}. 
The computations have been performed using approximately four hundred gauge configurations. 
On each configuration, in order to improve the signal, we exploited the invariance under space-time translations and repeated the calculation by placing the four-fermion operators in $O(100)$ randomly-selected space-time positions, using point sources for the quark propagators which begin or end at the weak operators. Moreover, we have found it numerically advantageous to average the correlation functions computed with oppositely-oriented three-momenta $\vecp{q}$ and $-\vecp{q}$. The correlated average of correlation functions with opposite three-momenta leads to a reduction of the statistical uncertainties by a factor larger than two. In order to simplify the notation, in the following we drop the suffix $E$ in the Euclidean correlation functions.

We find it convenient to perform a Fierz rearrangement of the operators $O_{1}^{(q)}$ and $O_{2}^{(q)}$ in Eq.~(\ref{eq:O12def}), and write them in the form
\begin{align}
O_{1}^{(q)}=\big(\bar{s}^j\gamma^\mu P_L b^j\big)~\big(\bar{q}^i\gamma_\mu P_L q^i\big)~,\qquad
O_{2}^{(q)}=\big(\bar{s}^j\gamma^\mu P_L b^i\big)~\big(\bar{q}^i\gamma_\mu P_L q^j\big)~. 
\end{align}
We separate the operators $O_{1,2}^{(q)}$ into their vector-vector ($VV$), axial-axial ($AA$) and axial-vector ($VA+AV$) components
\begin{align}
O_{1,2}^{(q)} = O_{1,2;VV}^{(q)} + O_{1,2;AA}^{(q)} + O_{1,2;AV+VA}^{(q)}\,,
\end{align}
where 
\begin{align}
O_{1;VV}^{(q)} &= \big(\bar{s}^j\gamma^\mu  b^j\big)  ~\big(\bar{q}^i\gamma_\mu q^i\big)~, \qquad O_{1;AA}^{(q)} = \big(\bar{s}^j\gamma^\mu \gamma^{5} b^j\big)  ~\big(\bar{q}^i\gamma_\mu \gamma^{5} q^i\big)~,  \\[10pt]
O_{2;VV}^{(q)} &= \big(\bar{s}^j\gamma^\mu  b^i\big)  ~\big(\bar{q}^i\gamma_\mu q^j\big)~,\qquad O_{2;AA}^{(q)} = \big(\bar{s}^j\gamma^\mu \gamma^{5} b^i\big)  ~\big(\bar{q}^i\gamma_\mu \gamma^{5} q^j\big)~.
\end{align}
The operators $O_{1,2; VA+AV}^{(q)}$ are defined similarly, however, being parity-odd, they do not contribute to the $B\to K \ell^{+}\ell^{-}$ amplitude (see also Appendix\,\ref{sec:appreality}), and will therefore not be considered in the following.

The Euclidean correlators corresponding to the vector-vector and axial-axial components of $O_{1,2}^{(c)}$ are given by
\begin{align}
\label{eq:C}
C^{\nu}_{1,2; VV,AA}(t,\vecp{q}) \equiv \langle K(-\vecp{q}) \big| T  \left[ \tilde{J}_{\rm em}^{\nu}(t, \vecp{q}) \, O^{(c)}_{1,2; VV,AA}(0) \right]  \big| B(\vecp{0}) \rangle\,.
\end{align}
The motivation for separating the correlators into their vector-vector (VV) and axial-axial (AA) components is that in the vacuum-saturation-approximation (VSA) only the $C^{\nu}_{1,2;VV}(t,\vecp{q})$ are non-zero. Moreover, in that approximation $C^{\nu}_{1;VV}(t,\vecp{q}) = N_{c} \, C^{\nu}_{2;VV}(t,\vecp{q})$, where $N_c=3$ is the number of colors. Therefore, if the vacuum-saturation-approximation is a reasonable description of the correlation functions, one expects $C^{\nu}_{1;VV}(t,\vecp{q})$ to be the largest of the four contributions. 
\begin{figure}
    \centering
    \includegraphics[width=0.8\linewidth]{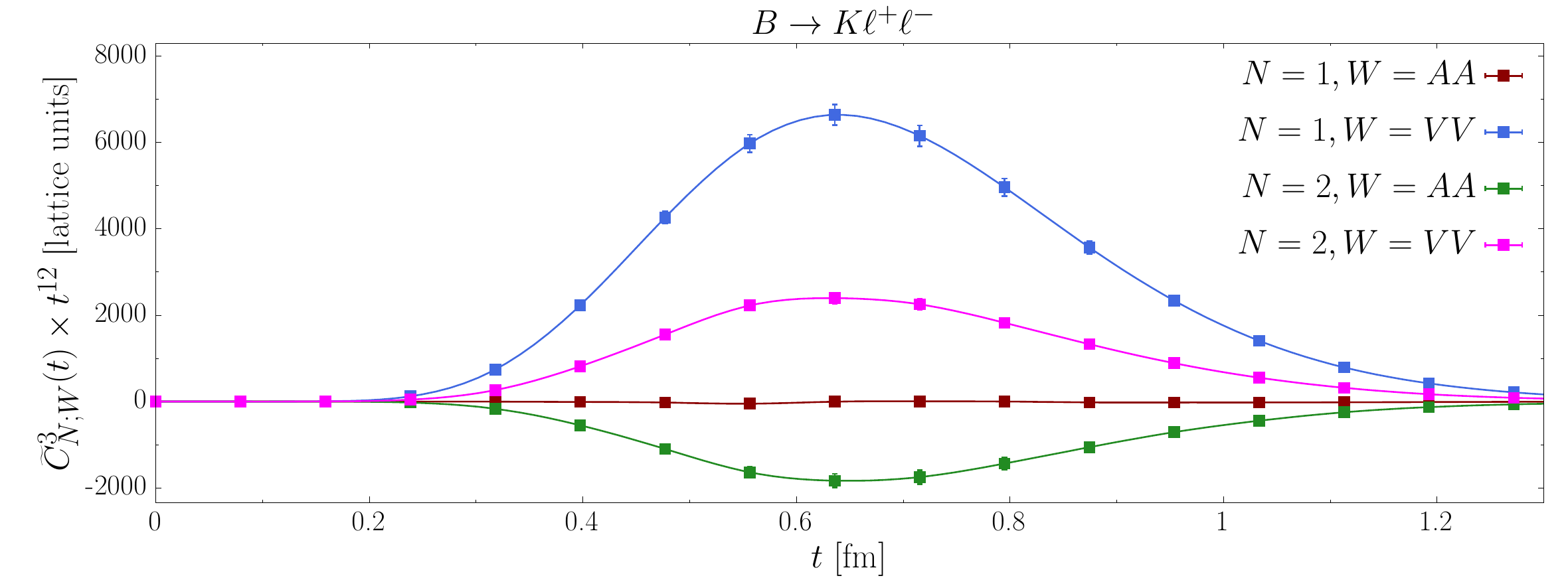}\\
    \includegraphics[width=0.8\linewidth]{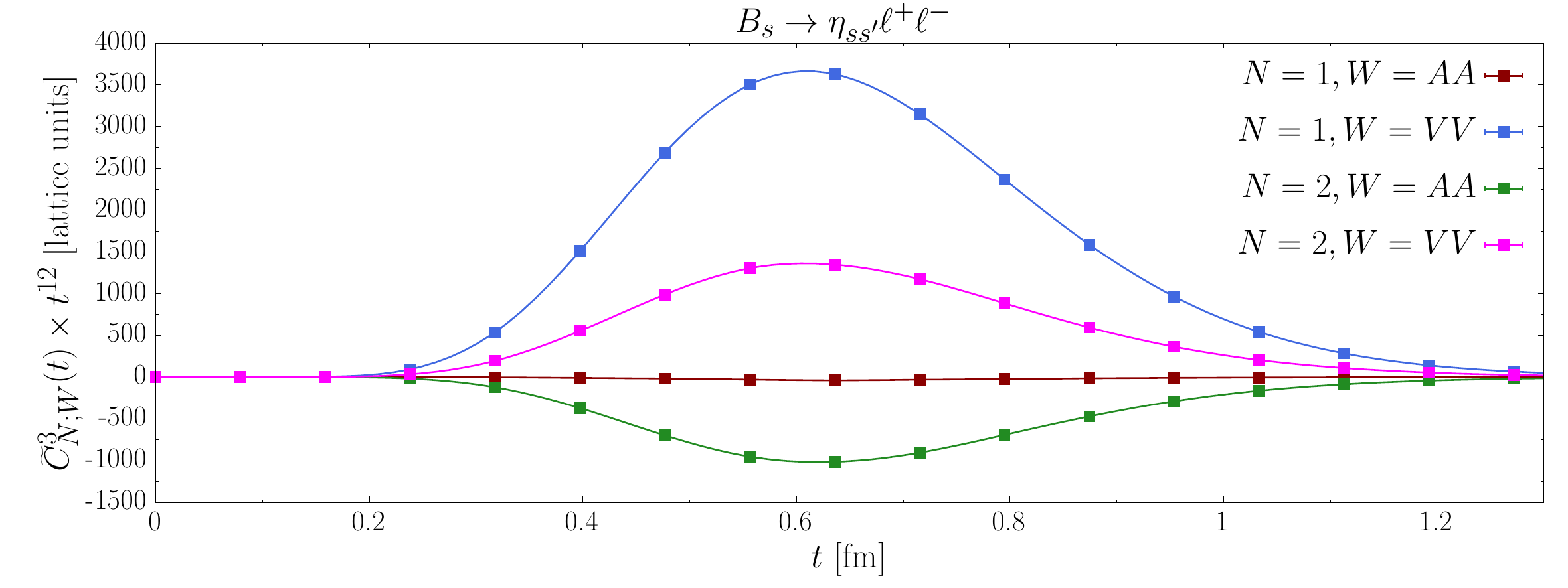}
    \caption{{\it Top Panel:} Bare correlators $\widetilde{C}^{3}_{1,2;VV,AA}(t,\vecp{q})$, defined in Eqs.~(\ref{eq:C})-(\ref{eq:tilde_C}), corresponding to the $B\to K\ell^{+}\ell^{-}$ decay. {\it Bottom:} The correlators $\widetilde{C}^{3}_{1,2;VV,AA}(t,\vecp{q})$  
    corresponding to the $B\to \eta_{ss'} \ell^{+}\ell^{-}$ decay, obtained by replacing the (spectator) light-quark in the $B\to K \ell^{+}\ell^{-}$ correlation function by a strange quark as described in the text. 
    $N=1$ and $N=2$ correspond to the operators $O_1^{(c)}$ and $O_2^{(c)}$ respectively. In each case the results have been multiplied by $t^{12}$, i.e. $t$ to the power 12, for visualization purposes.
    \label{fig:corr}}
\end{figure}
In Fig.\,\ref{fig:corr}, we show the longitudinal components, $C^{3}_{1,2; VV,AA}(t,\vecp{q})$, of the bare correlators in the second time ordering $t>0$, in which the issue of the analytic continuation of the physical Minkowski amplitude to Euclidean spacetime is present. The figure shows the correlators 
\begin{align}
\label{eq:tilde_C}
\widetilde{C}_{1,2; VV,AA}^{3}(t,\vecp{q}) = e^{-E_{K}t}\, C_{1,2;VV,AA}^{3}(t,\vecp{q})~, \qquad t > 0
\end{align}
which are related to the vector and axial components of the spectral densities $\rho^{\nu+}_{1,2}$ introduced in Sec.\,\ref{subsec:TOKll} by a Laplace-transform,
\begin{align}
\widetilde{C}_{1,2; VV,AA}^{\nu}(t,\vecp{q}) = \int_{E^{*}}^{\infty} \frac{dE}{2\pi} e^{-Et} \rho^{\nu+}_{1,2;VV,AA}(E,\vecp{q})~,\qquad \rho^{\nu+}_{1,2}(t,\vecp{q}) = \rho^{\nu+}_{1,2;VV}(t,\vecp{q}) + \rho^{\nu+}_{1,2;AA}(t,\vecp{q})~. 
\end{align}
For improved visualization at Euclidean times of \( \mathcal{O}(0.5-1\,\text{fm}) \), the correlators \( \widetilde{C}_{1,2;VV,AA}^{3}(t,\vec{q}) \) have been multiplied by a factor of $t^{12}$ in the figure.
The Euclidean correlators relevant for the $B\to K \ell^{+}\ell^{-}$ decay are shown in the top panel of Fig.\,\ref{fig:corr}. As the figure shows the largest contribution comes from the $C_{1;VV}^{3}(t,\vecp{q})$ term, in line with the expectation based on the VSA. We also note that $C_{2;VV}^{3}(t,\vecp{q})$ is approximately three times smaller than $C_{1;VV}^{3}(t,\vecp{q})$, again in line with the expectation based on the VSA. However, in the axial channel, while $C_{1;AA}^{3}(t,\vecp{q})$ is zero within the uncertainties (which is the prediction of the VSA), this is not the case for $C_{2;AA}^{3}(t,\vecp{q})$, which is similar in magnitude but opposite in sign to $C_{2;VV}^{3}(t,\vecp{q})$.   

In the bottom panel of Fig.~\ref{fig:corr}, we also show the Euclidean correlation functions obtained by replacing the spectator light-quark with a strange quark. These correlators allow us to determine the charming penguin contributions to the $B_{s}\to \eta_{ss'}\ell^{+}\ell^{-}$ decay amplitude, where $\eta_{ss'}$ is a fictitious pseudoscalar meson of mass $m_{\eta_{ss'}} = 689.9(5)\,{\rm MeV}$~\cite{Borsanyi:2020mff}, composed of two mass degenerate quarks $s$ and $s'$, each having mass equal to that of the strange quark. The main reason to consider these correlators is their markedly smaller statistical uncertainty, approximately a factor of five smaller than in the  $B\!\to\!K\,\ell^{+}\ell^{-}$ case, which allows for a more stringent test of the potential of the SFR/HLT method. 
Because the present study is exploratory and intended as a proof of principle, we have not yet carried out a high-statistics computation of the $B\!\to\!K\,\ell^{+}\ell^{-}$ correlators. Their precision can be substantially improved in future work, and in this respect the analysis of the charming penguin contributions to $B_s\!\to\!\eta_{ss'}\,\ell^{+}\ell^{-}$ offers a benchmark for the accuracy that might be achieved once substantial computational resources are devoted to the calculation of the Euclidean correlators relevant for the  charming--penguin contribution to $B\!\to\!K\,\ell^{+}\ell^{-}$.

\begin{figure}
    \centering
    \includegraphics[width=0.7\linewidth]{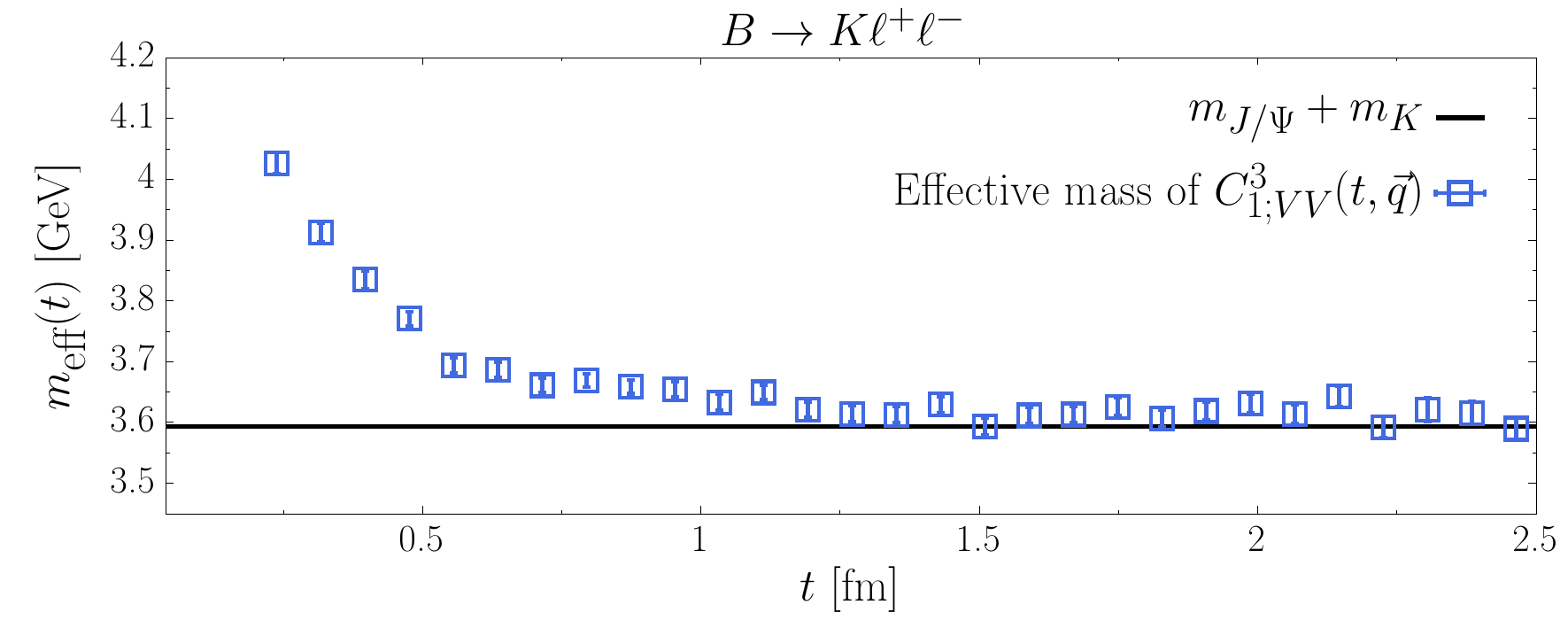}
    \includegraphics[width=0.7\linewidth]{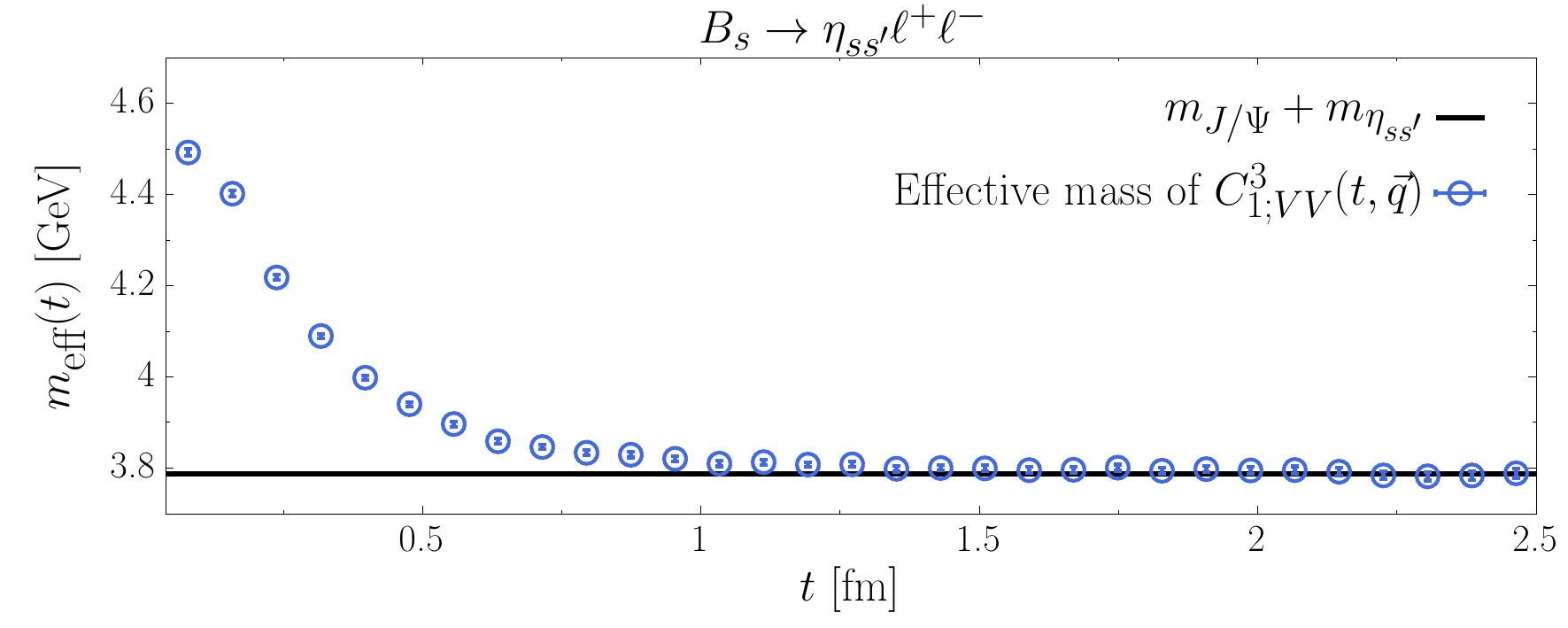}
    \caption{{\it Top}: Effective mass of the correlator $C_{1;VV}^{3}(t,\vecp{q})$ for the $B\to K\ell^{+}\ell^{-}$ decay. {\it Bottom}: effective mass of the correlator $C_{1;VV}^{3}(t,\vecp{q})$ for the $B_{s}\to \eta_{ss'}\ell^{+}\ell^{-}$ decay. In both panels the horizontal black lines correspond to the energy of the expected lowest-lying intermediate state.}
    \label{fig:eff_masses}
\end{figure}

We have also computed the effective masses of the correlators $\widetilde{C}_{1,2;VV,AA}^{3}(t,\vecp{q})$. In the large-time limit, the effective mass has a plateau at a value corresponding to the energy of the lowest-lying intermediate state contributing to the correlation function. For $B\to K\ell^{+}\ell^{-}$ correlators this is a $J/\psi + K$ state, while for $B_{s}\to \eta_{ss'}\ell^{+}\ell^{-}$ decays it is a $J/\psi + \eta_{ss'}$ state. In Fig.\,\ref{fig:eff_masses} we show the effective mass of the correlator $\widetilde{C}^{3}_{1,VV}(t,\vecp{q})$. The top panel in the figure corresponds to the $B\to K\ell^{+}\ell^{-}$ decay, while the bottom panel to $B_{s}\to \eta_{ss'}\ell^{+}\ell^{-}$. The horizontal lines in the top and bottom panel correspond respectively to the value of $m_{J/\psi} + m_{K}$ and $m_{J/\psi} + m_{\eta_{ss'}}$. As is clear from the figures, the effective masses converge towards the expected values.
\subsection{From the Euclidean correlators to the amplitude using the SFR/HLT method}

In this section we discuss the determination of the Minkowski amplitude from the Euclidean correlation functions discussed above.
Since for this proof-of-principles calculation we are only interested in showing that the infrared problem related to the analytic continuation to Euclidean spacetime can be handled using the SFR/HLT technique, we focus on the calculation of the three-times-subtracted amplitude, $H^{\nu+;3{\rm subs}}(\vecp{q},\varepsilon)$, defined in Eq.\,(\ref{eq:H3subs}), which, as already discussed, is free of contact divergences. As explained in detail in Sec.\,\ref{subsec:contact}, in order
to obtain the full Minkowski amplitude, including the contribution from the first time ordering, one must add to $H^{\nu+;3\text{subs}}$ a set of terms that do not require spectral density methods and which we do not evaluate here.  

As previously mentioned, we have performed the calculation of the Euclidean correlators at a single value of the heavy-quark mass, $m_{h}=2m_{c}$, which corresponds to a mass of the heavy-light meson of $m_{H}\simeq 2.9\,{\rm GeV}$. A complete calculation of the charming penguin contributions requires (in addition to performing the non-perturbative renormalization) the determination of the Euclidean correlators for multiple heavy-quark masses followed by an extrapolation of the results to the physical point, $m_{H}=m_{B}\simeq 5.280\,{\rm GeV}$. Making explicit the dependence of the Euclidean correlators (and of the underlying spectral densities) on the heavy-light meson mass $m_{H}$, we define
\begin{align}
\label{eq:corr_mh_dep}
\tilde{C}_{1,2}^{\nu+}(t,\vecp{q}; m_{H} ) &= \int_{E^{*}}^{\infty} \frac{dE}{2\pi} e^{-Et} \rho^{\nu+}_{1,2}(E,\vecp{q}; m_{H})~, \\
\label{eq:amp_mh_dep}
H^{\nu+;3{\rm subs}}_{1,2}(\vecp{q},\varepsilon; m_{H}, m) &= \int_{E^{*}}^{\infty} \frac{dE}{2\pi} \rho^{\nu+}_{1,2}(E,\vecp{q}; m_{H}) \, K_{\varepsilon}(E, m)~,
\end{align}
where the kernel function $K_{\varepsilon}(E,m)$ is given by
\begin{align}
\label{eq:three_times_sub_kernel}
K_{\varepsilon}(E,m) &= \frac{1}{E-m -i\varepsilon} + \frac{3}{E+m-i\varepsilon} - \frac{3}{E-i\varepsilon} -\frac{1}{E+2m-i\varepsilon} \nonumber \\
&= \frac{6m^3}{(E-m-i\epsilon)(E-i\epsilon)(E+m-i\epsilon)(E+2m-i\epsilon)}
~. 
\end{align}
The amplitude $H^{\nu+;3{\rm subs}}(\vecp{q},\varepsilon)$ defined in Eq.~(\ref{eq:H3subs}) is then recovered in the double limit
\begin{align}
H_{1,2}^{\nu+;3{\rm subs}}(\vecp{q},\varepsilon) = \lim_{m_{H}\to m_{B}} \lim_{m \to m_{B}} H_{1,2}^{\nu+;3{\rm subs}}(\vecp{q},\varepsilon; m_{H},m)~.
\end{align}
The introduction of the additional mass parameter $m$ in the kernel function $K_{\varepsilon}(E,m)$ might seem unnecessary; naturally one would set $m=m_{H}$. 
However, setting $m=m_{H}$ results in a non-smooth dependence of the amplitude on the meson mass $m_{H}$, as will be explained in Sec.\,\ref{sec:num_results}. 
Before doing so, we discuss the determination of the smeared amplitudes $H_{1,2}^{\nu+;3{\rm subs}}(\vecp{q}, \varepsilon; m_{H}, m)$ for general $\varepsilon$ and $m$, and for fixed values of $m_{H}\simeq 2.9\,{\rm GeV}$ and $|\vecp{q}|=|\vecp{p}_K|\simeq 250\,{\rm MeV}$. 

To determine the smeared amplitude in Eq.\,(\ref{eq:amp_mh_dep}), we use the HLT method, modifying the discussion in Sec.\,\ref{subsec:generalN} to the kernel function $K_{\varepsilon}(E, m)$ in Eq.\,(\ref{eq:three_times_sub_kernel}). The goal is to find, for each non-zero value of the smearing parameter $\varepsilon$, the best approximation of $K_{\varepsilon}(E, m)$ in terms of the basis function $\{ e^{-aE n}
\}_{n=1,\ldots, n_{\rm max} }$, namely
\begin{flalign}
\label{eq:basis_expansion}
{\rm Re} [K_{\varepsilon}(E, m) ] \simeq \sum_{n=1}^{n_{\rm max}} g^{R}_{n}(m,\varepsilon) e^{-aE n}~,\qquad {\rm Im}[K_{\varepsilon}(E, m)] \simeq \sum_{n=1}^{n_{\rm max}} g^{I}_{n}(m,\varepsilon) e^{-aE n}~,
\end{flalign} 
where $a$ is the lattice spacing and $n_{\rm max}$ is the dimension of the exponential basis. In our calculation we take $n_{\rm max} a \simeq 2.7\,{\rm fm}$ ($n_{\rm{max}}=34$) and the resulting uncertainty from the truncation of the summation in $n$ is much smaller than the statistical error. In this way,  once the coefficients $g_{n}^{R}$ and $g_{n}^{I}$ are known, the smeared amplitude can be obtained from the knowledge of the Euclidean correlators similarly to Eq.\,(\ref{eq:reconstruction0})
\begin{align}
\label{eq:HLT_RE_IM}
H_{1,2}^{\nu+;3{\rm subs}}(\vecp{q},\varepsilon; m_{H}, m) &= \int_{E^{*}}^{\infty} \frac{dE}{2\pi} \left( K^{R}_{\varepsilon}(E,m) + i K^{I}_{\varepsilon}(E,m)\right)\rho^{\nu+}_{1,2}(E,\vecp{q}; m_{H}) \nonumber \\[10pt]
&\simeq \sum_{n=1}^{n_{\rm max}} \left( g_{n}^{R}(m,\varepsilon) +i g_{n}^{I}(m,\varepsilon)\right) \int_{E^{*}}^{\infty} \frac{dE}{2\pi} e^{-aE n} \rho^{\nu+}_{1,2}(E,\vecp{q}; m_{H}) \nonumber \\
&= \sum_{n=1}^{n_{\rm max}} \left( g_{n}^{R}(m,\varepsilon) + ig_{n}^{I}(m,\varepsilon)\right) ~ \widetilde{C}_{1,2}^{\nu+}(an, \vecp{q}; m_{H})~,
\end{align}
where $K^{R}_{\varepsilon}$ and $K^I_{\varepsilon}$ are the real and imaginary parts respectively of the kernel function.
As explained in detail in Ref.~\cite{Hansen:2019idp}, and sketched in Sec.\,\ref{subsec:generalN}, the determination of the coefficients $g^{R}_{n}$ and $g^{I}_{n}$ presents a certain number of technical difficulties. 
Any determination of the smeared hadronic vector based on Eqs.\,(\ref{eq:basis_expansion}) and (\ref{eq:HLT_RE_IM}) will be inevitably affected by both systematic errors (due to the inexact reconstruction of the kernels) and statistical uncertainties (due to the fluctuations of the correlator $\widetilde{C}_{1,2}^{\nu+}(t, \vecp{q}; m_{H})$), which need to be controlled simultaneously. Requiring an overly-precise kernel reconstruction leads to strongly oscillating coefficients $g_{n}^R$ and $g_{n}^I$, which magnify the statistical errors of the correlators when evaluating the sum in the last line of Eq.\,(\ref{eq:HLT_RE_IM}). 
The HLT method enables an optimal balance between statistical and systematic errors to be found by minimizing each of the two linear combinations
\begin{align}
\label{eq:func_W}
W_{R}^{\beta}[\boldsymbol{g}] \equiv \frac{A_{R}^{\beta}[\boldsymbol{g}]}{A_{R}^{\beta}[\boldsymbol{0}]} + \lambda B[\boldsymbol{g}]\,,\qquad
W_{I}^{\beta}[\boldsymbol{g}] \equiv \frac{A_{I}^{\beta}[\boldsymbol{g}]}{A_{I}^{\beta}[\boldsymbol{0}]} + \lambda B[\boldsymbol{g}]\,,\qquad
\end{align}
where the functionals $A^{\beta}_{R,I}$ are given by
\begin{align}
\label{eq:func_A}
A^{\beta}_{R,I}[\bs{g}]=
\int_{E_{\mathrm{th}}}^\infty dE ~ e^{\beta a E} ~  \bigg| \sum_{n=1}^{n_{\rm max}}~g_{n} e^{-aE n} - K^{R,I}_{\varepsilon}(E, m)\,\bigg|^2~, \quad \bs{g}= (g_{1},\ldots,g_{n_{\rm max}}), \qquad E_{\rm th} \leq E^{*}~. 
\end{align}
and are a measure of the quality of the reconstruction of the real and imaginary parts of the kernel. The error-functional $B[{\bs g}]$ is defined by
\begin{align}
B[\boldsymbol{g}] = \frac{1}{(\widetilde{C}_{1,2}^{\nu+}(0,\vecp{q}; m_{H}))^{2}} \sum_{n_1 , n_2 = 1}^{n_{\rm max}} g_{n_{1}}\, g_{n_{2}}~ {\rm{Cov}}(an_1 , an_2 )~, 
\end{align}
where ${\rm Cov}$ is the covariance matrix of the Euclidean lattice correlator $\widetilde{C}_{1,2}^{\nu+}(t,\vecp{q}; m_{H})$, and $\lambda$ is the so-called \textit{trade-off} parameter. In Eq.\,(\ref{eq:func_A}), $\beta$ and $E_{\mathrm{th}}$ are algorithmic parameters which we set to $\beta= 1.99$ and $E_{\rm th} = 3.5\,{\rm GeV}$ ($3.7~{\rm GeV}$) in the case of $B\to K\ell^{+}\ell^{-}$ ($B_{s}\to\eta_{ss'}\ell^{+}\ell^{-}$).\footnote{For the convergence of the integral $\beta$ must be smaller than 2.} In the presence of statistical errors, the functional $B$ disfavours coefficients $\bs{g}$ leading to  large statistical uncertainties in the reconstructed smeared amplitude. The balance between having small systematic errors (small $A_R^{\beta}[\bs{g}]$ and $A_I^{\beta}[\bs{g}]$) and small statistical errors (small $B[\bs{g}]$) depends on the tunable parameter $\lambda$. 
Its optimal value $\lambda^{\rm opt}$ is determined by monitoring the evolution of the reconstructed smeared amplitude for different values of $\lambda$. 
The optimal value is chosen to be in the statistically-dominated regime, where $\lambda$ is sufficiently small that the systematic error due to the kernel reconstruction is smaller than the statistical error (therefore, in this region, the results are stable under variations of $\lambda$ within statistical uncertainties), but still large enough to have reasonably small statistical uncertainties. The kernel functions that we use at non-zero lattice spacing are obtained from Eq.~(\ref{eq:three_times_sub_kernel}) by the replacement
\begin{align}
\frac{1}{x-i\varepsilon} \to \frac{a}{\sinh{a(x-i\varepsilon)}}~,
\end{align}
and therefore differ from the continuum counterparts by $O(a^{2})$ discretization effects.
\begin{figure}
    \centering
\includegraphics[scale=0.35]{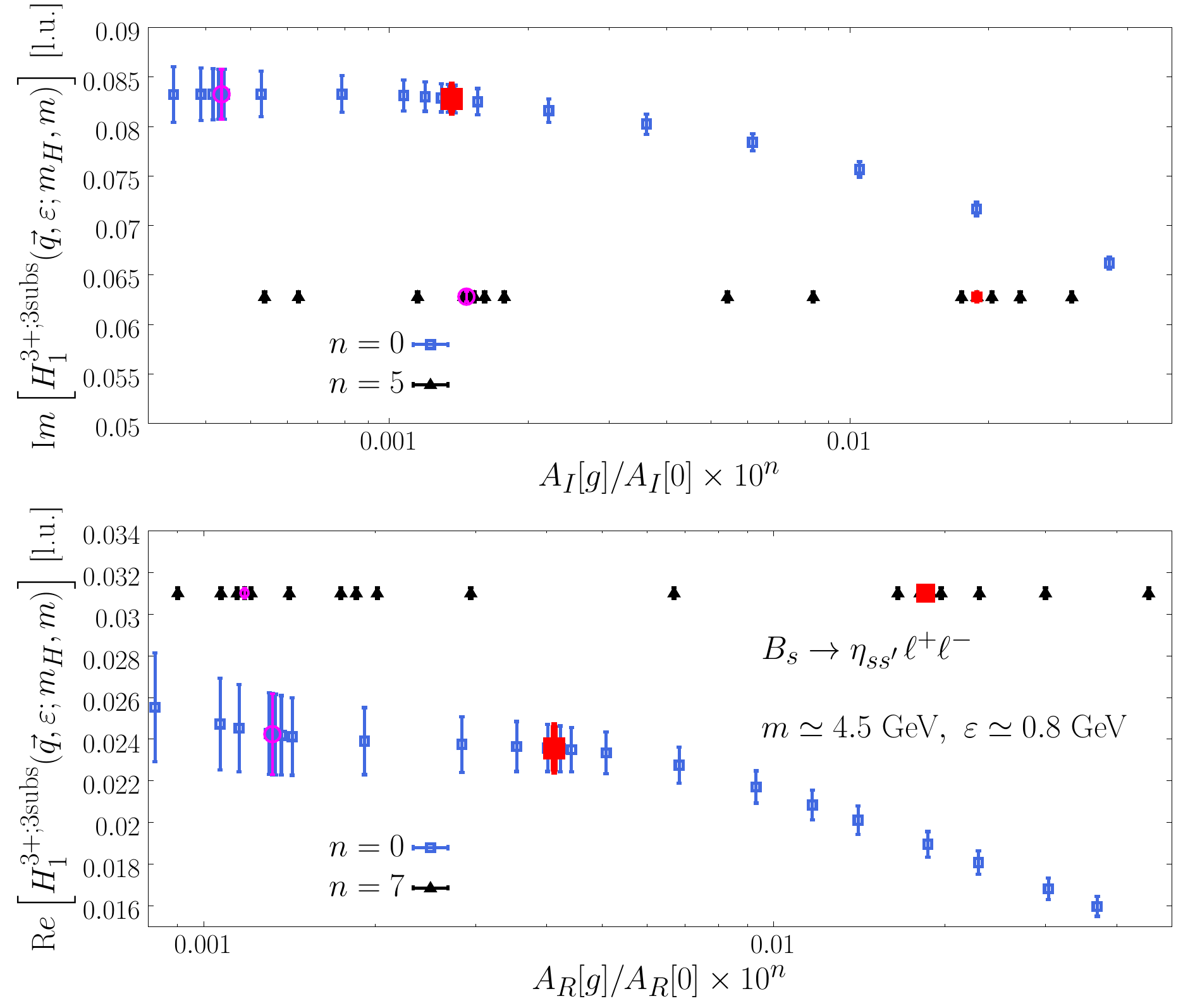}
    \caption{Real (bottom panel) and imaginary (top panel) parts of $H^{3+;3{\rm subs}}_{1}(\vec{q}, \varepsilon; m_{H}, m)$ for $m = 4.5~\mathrm{GeV}$ and $\varepsilon = 0.8~\mathrm{GeV}$, shown in lattice units as a function of the ratio $A_{R,I}[\bs{g}]/A_{R,I}[\bs{0}]$, which is a measure of the goodness of the
kernel reconstruction. The plot illustrates a representative example of the stability analysis. Blue and black points correspond respectively to the contributions from the first term and the remaining three terms of the kernel function in the first line of Eq.\,(\ref{eq:three_times_sub_kernel}). As described in the text, the sign of the latter has been inverted and its imaginary part scaled by a factor of six for better visualization. Moreover, the  values of $A_{W}[\bs{g}]/A_{W}[\bs{0}]$ corresponding to the black points in the figure have been rescaled by factors of $10^7$ and $10^5$ for $W = R$ and $W = I$, respectively. Filled squares and empty circles indicate reconstructions performed with $\lambda = \lambda^{\rm opt}$ and $\lambda = \lambda^{\rm syst}$, respectively; see text for details.}
    \label{fig:stab_analysis}
\end{figure}
\begin{figure}
    \centering
    \includegraphics[scale=0.40]{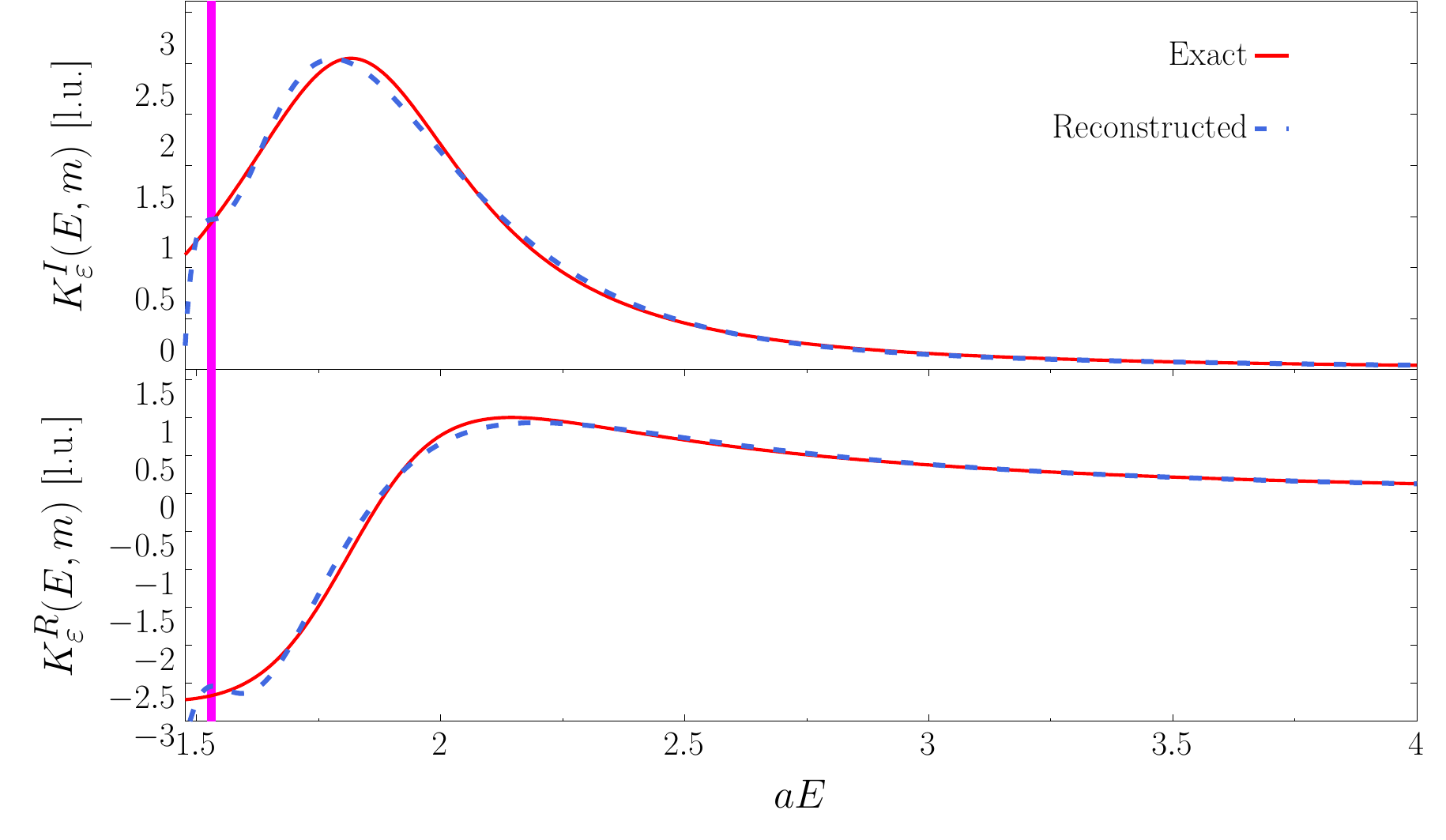}
    \caption{The reconstructed kernels $K^{R/I}_{\varepsilon}(E, m)$ obtained in the HLT reconstruction  with $\lambda=\lambda^{\rm opt}$ shown in Fig.~\ref{fig:stab_analysis} (dashed lines), are compared to the exact ones (solid lines). The vertical line corresponds to the value of the algorithmic parameter $E_{\rm th} \leq E^{*}$ in Eq.~(\ref{eq:func_A}). The results are given in lattice units.}
    \label{fig:kernel_reco}
\end{figure}

Exploiting the linearity of the procedure, we divide the HLT reconstruction of the kernel into two parts.
The first, and the more challenging, part is the HLT reconstruction of the first term on the right hand side of the first line of Eq.\,(\ref{eq:three_times_sub_kernel}).
The second part is the reconstruction of the remaining three terms. This latter step poses no difficulty, as these terms do not develop a pole within the integration range in the limit $\varepsilon \to 0$, allowing for highly accurate kernel reconstruction.
Once both components are reconstructed, we sum their contributions.
In Fig.\,\ref{fig:stab_analysis}, we show an illustrative example of the \textit{stability analysis} that is performed in order to obtain the smeared amplitude using the HLT method. The results shown correspond to the contribution from the operator $O_{1}^{(c)}$ to $B_{s}\to \eta_{ss'}\ell^{+}\ell^{-}$ for $m=4.5\,{\rm GeV}$ and $\varepsilon= 0.8\,{\rm GeV}$. The figure presents the stability analyses of both contributions discussed above: the blue points correspond to the reconstruction of the first term of the kernel function, while the black points correspond to the combined contribution of the remaining three terms. In the figure we have inverted the sign of the latter and multiplied its (small) imaginary part by a factor of six for visualization purposes. The HLT reconstruction of the final three terms is exceptionally precise; in the figure, the corresponding values of $A_{W}[g]/A_{W}[0]$, where $A_{W}[g]\equiv A_{W}^{\beta=0}[g]$, have been rescaled by factors of $10^{7}$ and $10^{5}$ for $W = R$ and $W = I$, respectively.
The red data points (filled squares) appearing in each of the two plots of Fig.\,\ref{fig:stab_analysis}, correspond to the chosen optimal value $\lambda^{\rm opt}$, while the magenta ones (empty circles) correspond to the values $\lambda^{\rm syst}$ determined imposing the condition
\begin{align}
\frac{B[\bs{g}_{\lambda^{\rm syst}}]}{A_{R/I}[ \bs{g}_{\lambda^{\rm syst}}]} = 10\frac{B[\bs{g}_{\lambda^{\rm opt}}]}{A_{R/I}[ \bs{g}_{\lambda^{\rm opt}}]}\,.
\end{align}
The difference between the reconstructions obtained using $\lambda=\lambda^{\rm opt.}$ and $\lambda=\lambda^{\rm syst.}$ is added as a systematic uncertainty in the final error. Finally, in Fig.~\ref{fig:kernel_reco} we show a comparison between the exact and reconstructed kernels $K_{\varepsilon}(E, m)$ obtained using the HLT procedure in the illustrative case shown in Fig.~\ref{fig:stab_analysis}. We will now discuss our results for the smeared amplitude as a function of the kernel parameters $m$ and $\varepsilon$.

\FloatBarrier

\subsection{Numerical results}
\label{sec:num_results}
We have performed the analysis outlined in the previous section for several values of $m$ in the range $(3.5,5.5)\,{\rm GeV}$ and for several values of $\varepsilon$. In all cases, we have reconstructed both the real and imaginary part of the full kernel function in Eq.\,(\ref{eq:three_times_sub_kernel}). For each choice of the parameter $m$ in the kernel, we have evaluated the smeared amplitude for several values of $\varepsilon$, chosen as follows:
\begin{align}
\label{eq:scaling_epsilon_m}
\varepsilon(m)=\alpha\,(|m-E_{\text{th}}|+\delta),\qquad
\alpha = 0.5,\,0.6,\,0.8,\,1,\,1.25,\,1.5,\,2, \qquad \delta = 150~{\rm MeV}.
\end{align}
This $m$-dependent choice for the simulated values of $\varepsilon$ is made for two reasons. Firstly, the numerical difficulty of the HLT reconstruction depends mainly on the ratio
$|m-E_{\text{th}}|/\varepsilon$ for $m>E_{\text{th}}$; the scaling in
Eq.\,(\ref{eq:scaling_epsilon_m}) therefore keeps the difficulty of the HLT reconstruction approximately $m$-independent for fixed~$\alpha$. The presence of the offset $\delta$ is to avoid $\varepsilon$ getting too small when $m\simeq E_{\rm th}$. 
Very small values of $\varepsilon $ typically correspond to very large statistical and systematic errors. 
Secondly, as has been shown in Ref.\,\cite{Frezzotti:2023nun}, the smeared amplitude can be described by a low--order polynomial in $\varepsilon$ of the form 
\begin{align}
\label{eq:scaling_regime}
H_{1,2}^{\nu+;3\text{subs}}(\vec q,\varepsilon; m_{H},m)
= H_{1,2}^{\nu+;3\text{subs}}(\vec q,0; m_{H},m)
  + A\,\varepsilon + B\,\varepsilon^{2} + \ldots~,
\end{align}
where the ellipses are $O(\varepsilon^{3})$ terms that can be neglected, only when
\begin{align}
\varepsilon \ll \Delta(m),
\end{align}
where $\Delta(m)$ denotes the characteristic energy gap around $m$ over which the spectral density
$\rho^{\nu+}_{1,2}(E,\vec q;m_{H})$ varies appreciably. Only for $\varepsilon\lesssim\Delta(m)$ can one safely extrapolate to the $\varepsilon\to0$ limit using the Ansatz in Eq.\,(\ref{eq:scaling_regime}), and the authors of Ref.\,\cite{Frezzotti:2023nun} refer to this region as to the \emph{asymptotic regime}. The spectral densities are expected to become progressively smoother as one increases $m$ above the region of the main charmonium resonances, so increasingly larger $\varepsilon$ values are sufficient to enter the \emph{asymptotic regime}, which justifies the use of larger $\varepsilon$ for larger values of $m$. In contrast, near the resonance peaks, $\Delta(m)$ can be very small, so very small values of $\varepsilon$ may be required before Eq.\,(\ref{eq:scaling_regime}) applies.
\begin{figure}
    \centering
    \includegraphics[width=0.7
    \linewidth]{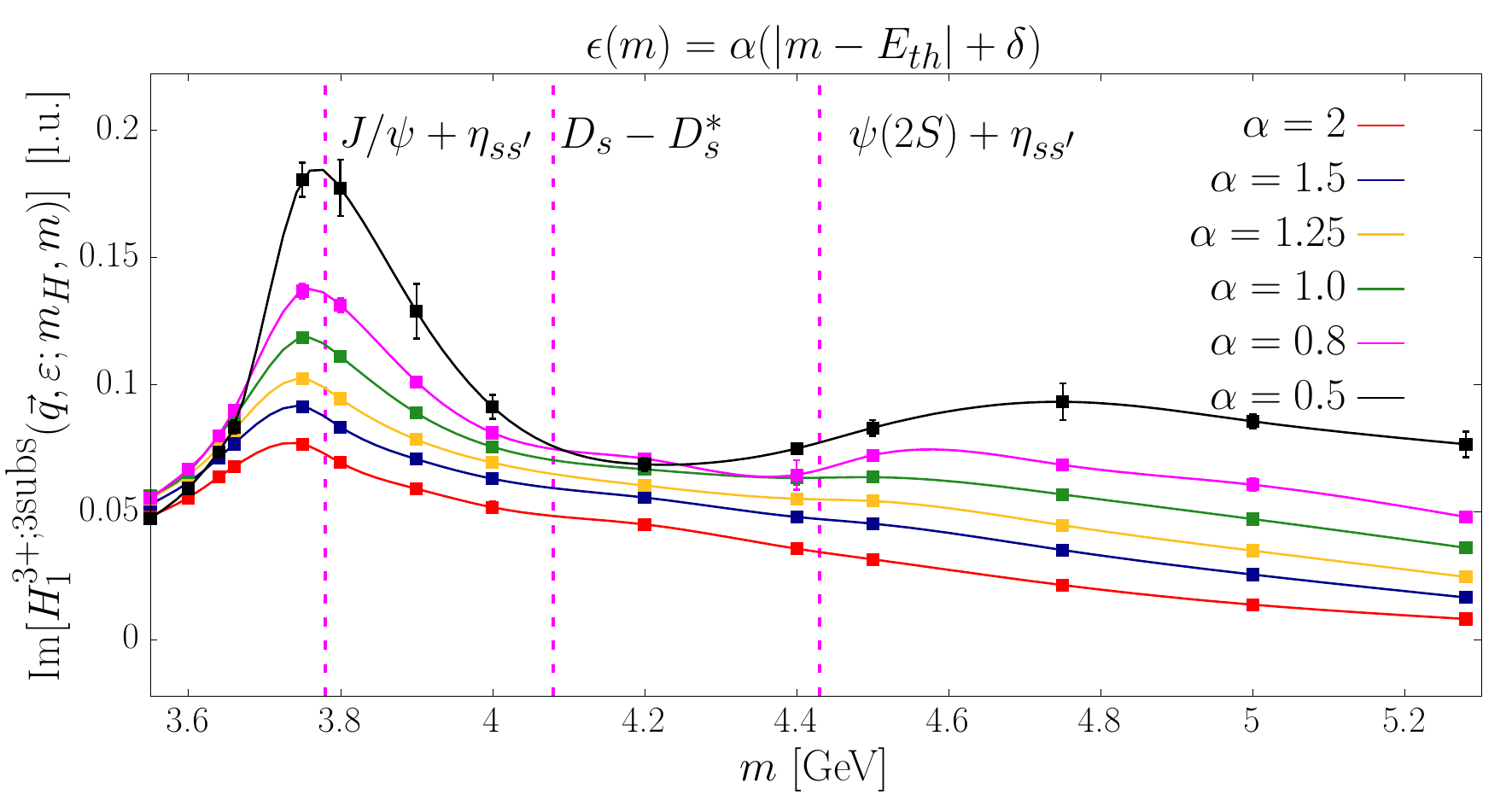}\\
    \includegraphics[width=0.7\linewidth]{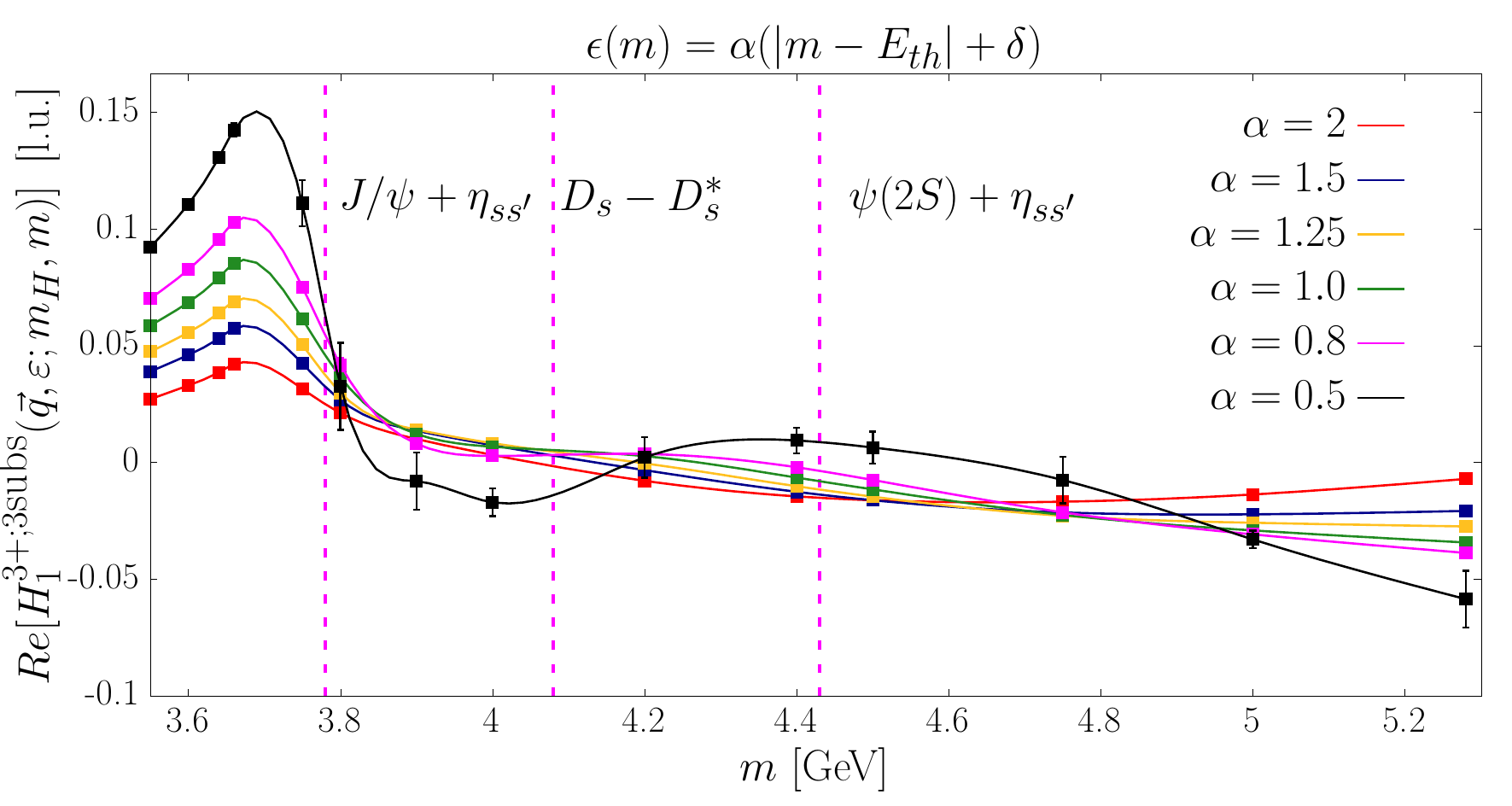}
    \caption{The real (bottom) and imaginary (top) components of the smeared amplitude $H_{1}^{3+;3{\rm subs}}(\vecp{q}, \varepsilon; m_{H}, m)$, as a function of $m$, for some simulated values of $\alpha$ in Eq.\,(\ref{eq:scaling_epsilon_m}). The continuous lines correspond to spline interpolations of the lattice data.}
    \label{fig:O1_Bs}
\end{figure}
\begin{figure}
    \centering
    \includegraphics[width=0.7\linewidth]{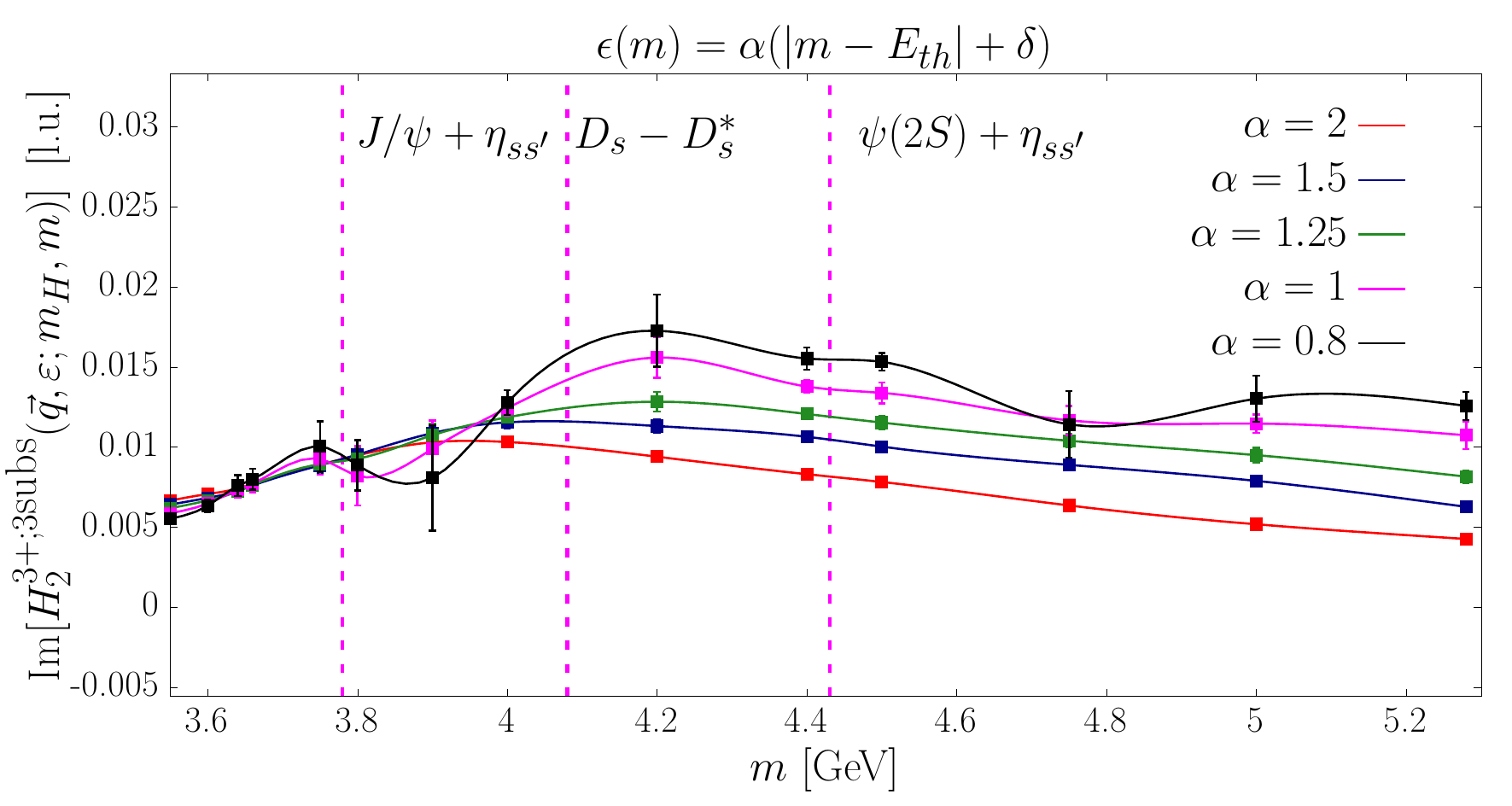}\\
    \includegraphics[width=0.7\linewidth]{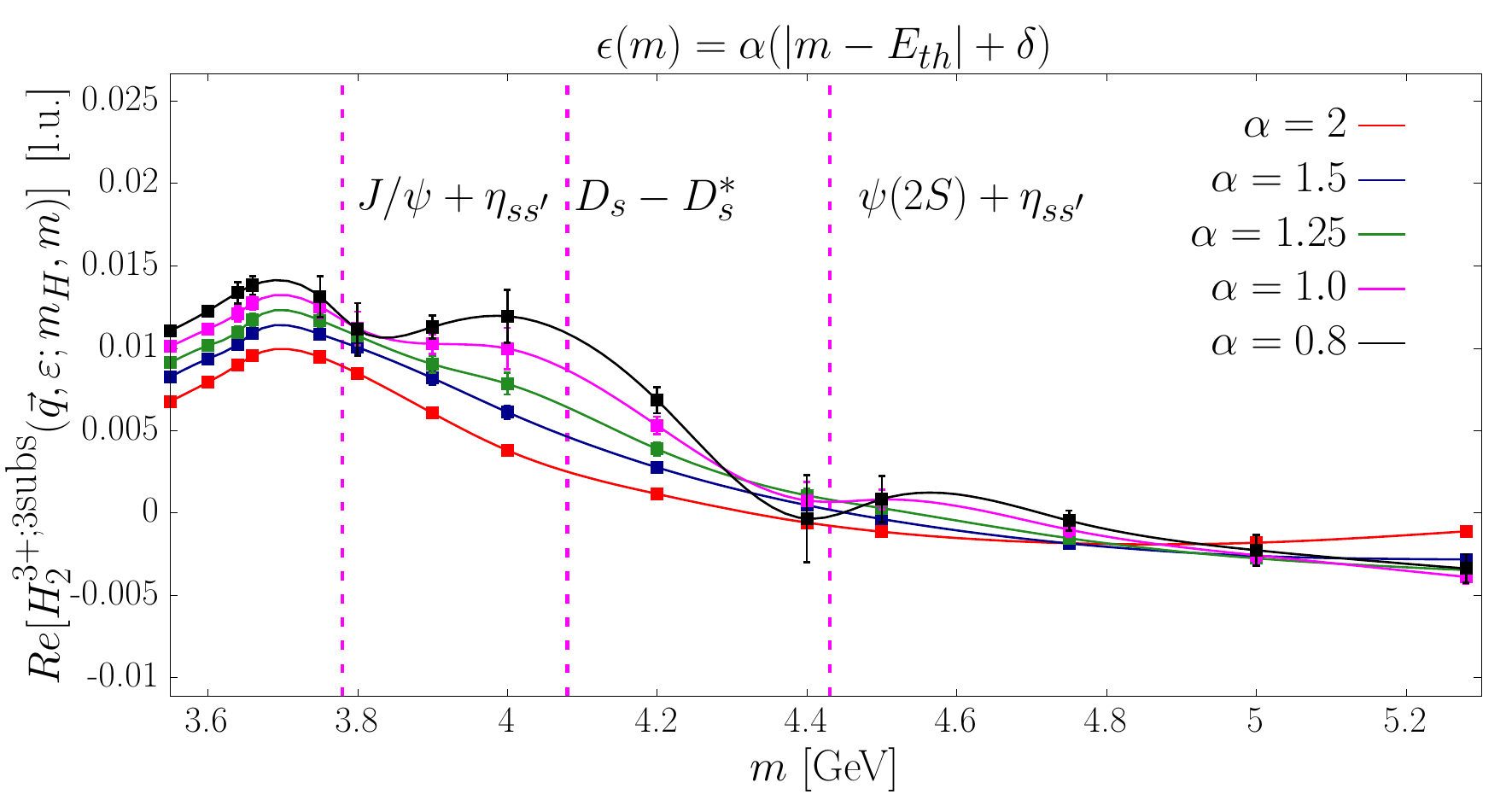}
    \caption{The real (bottom) and imaginary (top) components of the smeared amplitude $H_{2}^{3+;3{\rm subs}}(\vecp{q}, \varepsilon; m_{H}, m)$, as a function of $m$, for some of the simulated values of $\alpha$ in Eq.~(\ref{eq:scaling_epsilon_m}). The continuous lines correspond to spline interpolations of the lattice data.}
    \label{fig:O2_Bs}
\end{figure}

\begin{figure}
    \centering
    \includegraphics[width=0.7\linewidth]{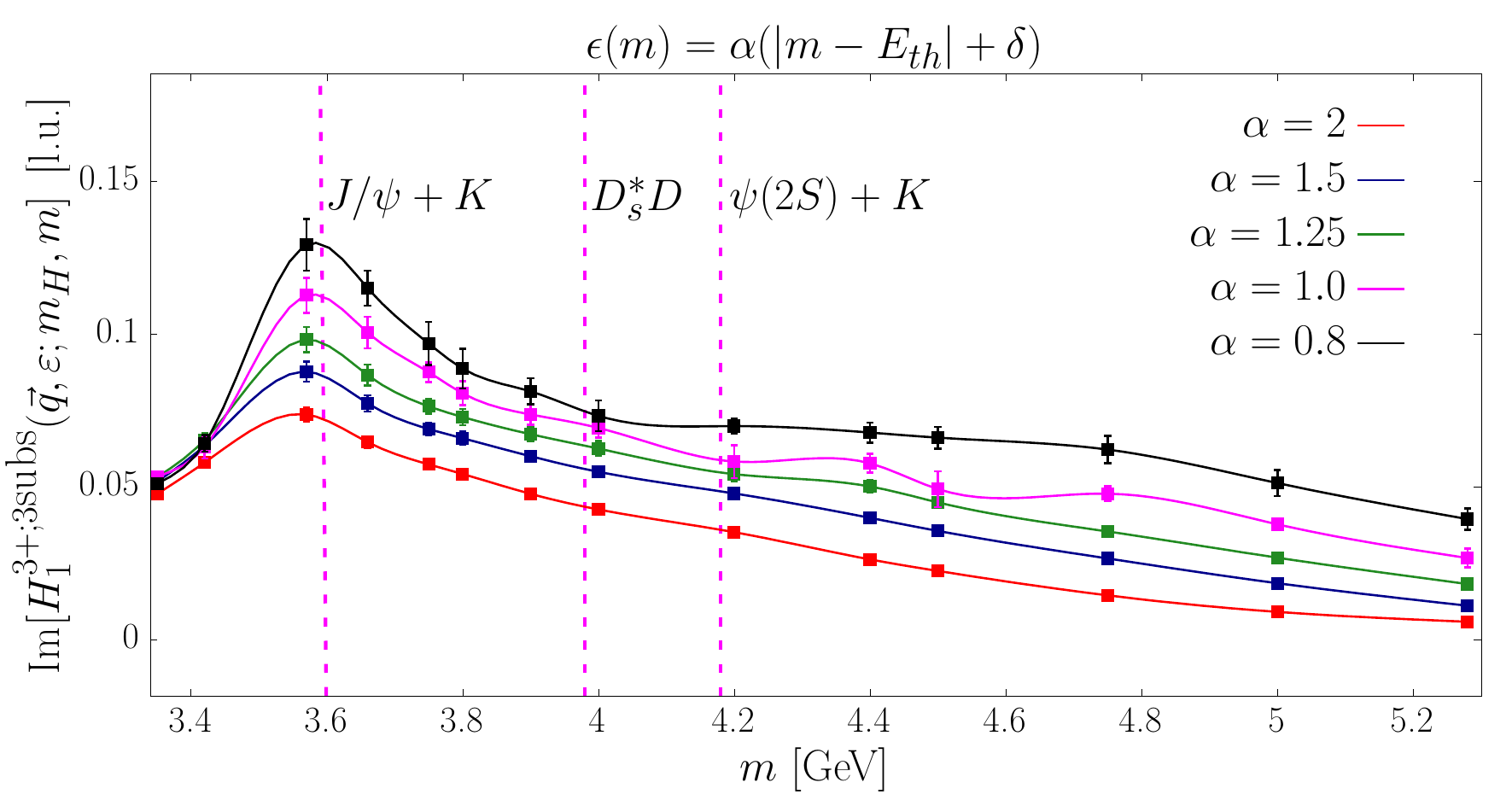}\\
    \includegraphics[width=0.7\linewidth]{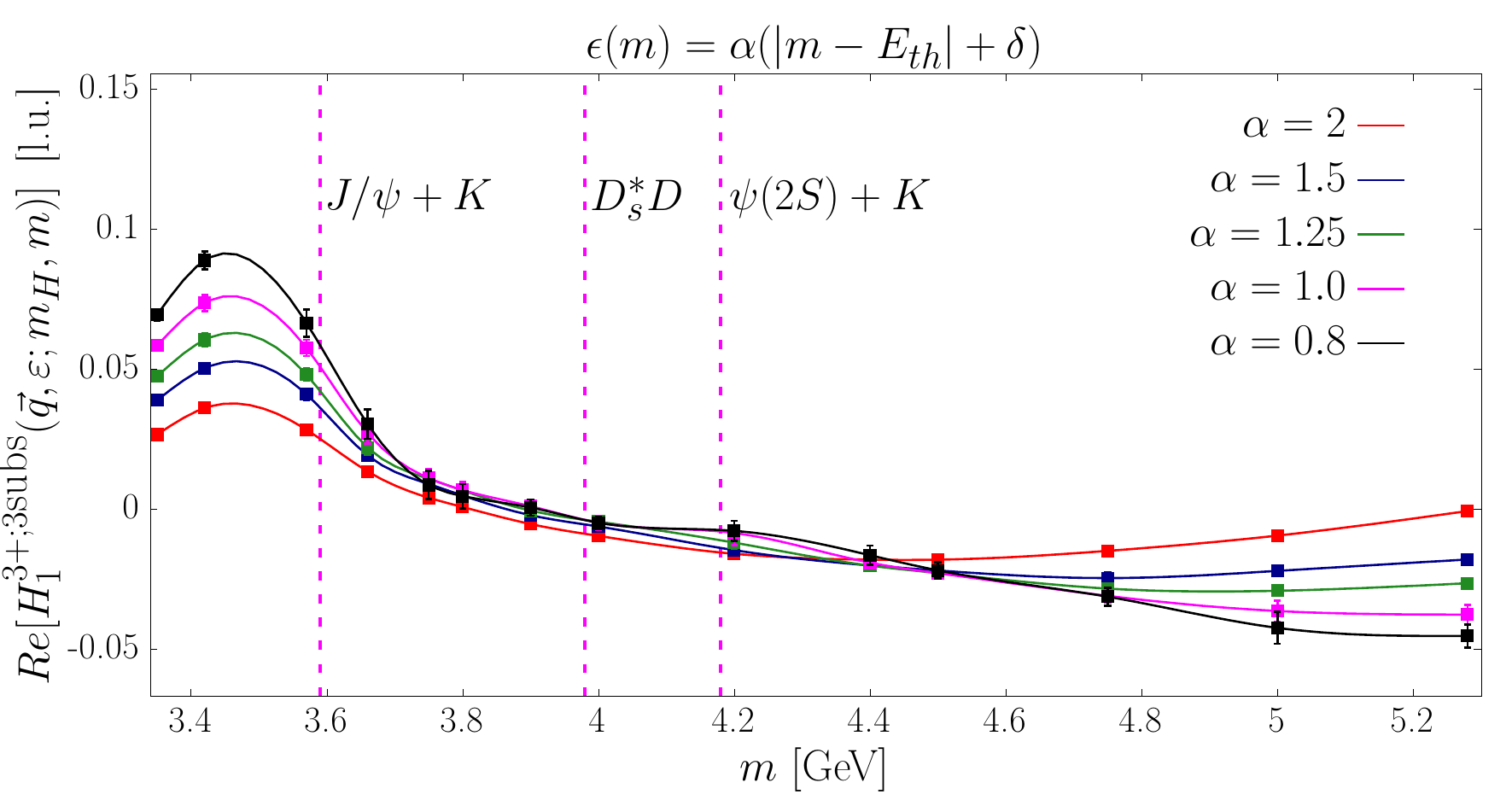}
    \caption{The real (bottom) and imaginary (top) components of the smeared amplitude $H_{1}^{3+;3{\rm subs}}(\vecp{q}, \varepsilon; m_{H}, m)$, relevant for the $B\to K\ell^{+}\ell^{-}$ decay, as a function of $m$, for some of the simulated values of $\alpha$ in Eq.~(\ref{eq:scaling_epsilon_m}). The continuous lines correspond to spline interpolations of the lattice data.}
    \label{fig:O1_B}
\end{figure}

In Fig.\,\ref{fig:O1_Bs} we show the real and imaginary part of the smeared amplitude $H_{1}^{
3+;3{\rm subs}}(\vecp{q}, \varepsilon; m_{H}, m)$ for $|\vecp{q}| \simeq 250\,{\rm MeV}$ and $m_{H}\simeq 2.9\,{\rm GeV}$, as a function of $m$ and for the different choices of the parameter $\alpha$ in Eq.\,(\ref{eq:scaling_epsilon_m}). The figure corresponds to the charming penguin contribution to the $B_{s}\to \eta_{ss'}\ell^{+}\ell^{-}$ decay. The analogous plots for the $O_{2}^{(c)}$ contribution to the $B_{s}\to\eta_{ss'}\ell^{+}\ell^{-}$ decay are shown in Fig.\,\ref{fig:O2_Bs}, and for the $O_{1}^{(c)}$ contribution to the $B\to K \ell^{+}\ell^{-}$ decay in Fig.\,\ref{fig:O1_B}.\footnote{The $O_{2}^{(c)}$ contribution to the $B\to K\ell^{+}\ell^{-}$ decay is qualitatively similar to the $O_{2}^{(c)}$ contribution to the $B_{s}\to \eta_{ss'}\ell^{+}\ell^{-}$ decay, however, it has substantially larger statistical uncertainties. For this reason we do not find it useful to present it here.} 

For the $O_{1}^{(c)}$ contribution to the $B_{s}\to \eta_{ss'}\ell^{+}\ell^{-}$ decay, as seen in Fig.\,\ref{fig:O1_Bs}, the imaginary part of the reconstructed subtracted amplitude shows a clear peak around the energy of the $J/\psi + \eta_{ss'}$ system, while the real part changes sign after passing the resonance peak, which is the expected behaviour in presence of a sharp resonance. As $\alpha$ decreases (and hence $\varepsilon$ decreases), the local structure of the underlying spectral density emerges, with the peaks becoming more pronounced. After passing the $\psi(2S)+\eta_{ss'}$ peak, the dependence on $\alpha$ becomes smoother, in line with expectations. Note that a hint of the presence of the $\psi(2S)$ resonance can be observed in the smeared amplitude, as both its real and imaginary parts begin to rise again when approaching the expected location of the resonance. However, since the smearing resolution is not as good as in the case of the $J/\psi$, the corresponding feature appears significantly broader and less pronounced, making a precise identification of the $\psi(2S)$ more challenging.

The $O_{2}^{(c)}$ contribution to $B_{s}\to \eta_{s s'}\ell^{+}\ell^{-}$ in Fig.\,\ref{fig:O2_Bs} does not show instead the pronounced peak that is visible in the $O_{1}^{(c)}$ contribution at the same values of $\alpha$, while the $O_{1}^{(c)}$ contribution to $B\to K\ell^{+}\ell^{-}$ in Fig.\,\ref{fig:O1_B} is qualitatively similar to that of $B_{s}\to \eta_{ss'}\ell^{+}\ell^{-}$. Note that the $O_{2}^{(c)}$ contribution to $B_{s}\to \eta_{ss'}\ell^{+}\ell^{-}$ is approximately one order of magnitude smaller than the corresponding $O_{1}^{(c)}$ contribution. This suppression is larger than that present in the vacuum‐saturation approximation (VSA), which predicts only a $1/N_{c}$ reduction. This is because the axial–axial (AA) component of the $O_{2}^{(c)}$ correlator, which is predicted to vanish in the VSA, partially cancels the vector–vector (VV) contribution, making larger the net suppression relative to $O_{1}^{(c)}$. Finally, the $O_{1}^{(c)}$ contribution to $B\to K\ell^{+}\ell^{-}$, shown in Fig.\,\ref{fig:O1_B}, is similar to the corresponding contribution to $B_{s}\to \eta_{ss'} \ell^{+}\ell^{-}$ in Fig.\,\ref{fig:O1_Bs}. As expected, the uncertainties are however larger, due to the noise induced by the light-quark propagator.

We are now in a position to explain the motivation for the introduction of the mass parameter $m$. The spectral density $\rho^{\nu+}(E,\vecp{q};m_{H})$ is not expected to change qualitatively by increasing the mass of the heavy meson. Indeed, since we work at physical light, strange and charm quark masses, the positions of the resonance peaks are clearly independent of $m_{H}$. What depends on the heavy-meson mass are the heights of the peaks, namely the matrix elements
\begin{equation}
A_{n} = \langle n \,|\, O_{1,2}^{(c)} \,|\, B_{(s)}\rangle\, ,
\end{equation}
where $n$ labels each allowed intermediate state propagating between the $O_{1,2}^{(c)}$ operators and the electromagnetic current. Therefore, one does not expect a sharp, non-smooth dependence of the spectral density on $m_{H}$. However, by setting $m=m_{H}$, an artificial non-smooth dependence on $m_{H}$ would be introduced. 
To understand this point, let us assume that, in order to extrapolate to the physical $b$-quark mass, as typically done, a series of simulations for various $m_{H}$ values, starting from $m_{D_{(s)}}$, are performed. 
By computing the smeared amplitude with $m=m_{H}$ at each point, one finds that, at the lowest simulated heavy-meson mass, the smeared amplitude is evaluated \emph{below} the main charmonium peaks; then as $m_{H}$ increases, one \emph{surfs} over the different charmonium resonances before reaching $m=m_{B_{(s)}}$. Because the behaviour of the smeared amplitude below (or close to) the main $c\bar{c}$ resonances is expected to differ markedly from that at much larger energies of $\mathcal{O}(m_{B_{(s)}})$, the resulting mass scaling obtained with $m=m_{H}$ would be highly complicated and difficult to handle. An alternative possibility is to always set $m=m_{B_{(s)}}$, but this would induce large discretization effects\footnote{In the figure, we present results up to $m \simeq m_B$, where discretization effects can become significant. This proof-of-principle study is restricted to a single lattice spacing, which limits our ability to assess these effects. In future work, simulations at multiple lattice spacings will allow us to systematically monitor the size of the cut-off effects as a function of $m$, and to exclude the region of large $m$ where these effects are too large to permit a reliable continuum-limit extrapolation.}. In our previous paper on $B_{s}\to \mu^{+}\mu^{-}\gamma$ decays\,\cite{Frezzotti:2024kqk}, a similar problem arose when computing the tensor form factor $\bar{F}_{T}$. In that case, exploiting the freedom to adopt any functional dependence $m(m_{H})$ satisfying $m(m_{B}) = m_{B}$, we considered
\begin{equation}
m(m_{H}) = (1-r)\,m_{H} + r\,m_{B},
\end{equation}
with $r \in [0,1]$. For $r=1$ one has $m=m_{B}$, and for $r=0$ one has $m=m_{H}$. Intermediate values of $r$ interpolate between these extremes. The guiding principle is to choose $r$ small enough to avoid large discretization effects, yet large enough that, at the lowest simulated heavy-meson mass $m_{H}^{\rm min}$, the quantity $m^{\rm min} \equiv r\,m_{H}^{\rm min} + (1-r)\,m_{B}$ lies \emph{above} the main charmonium resonances. The effectiveness of this strategy will be explored in future studies, where the heavy-quark-mass extrapolation to the physical $b$-quark mass will be performed.

\FloatBarrier

\subsection{Comparison with a simple model for the spectral density and the extrapolation to {\boldmath$\varepsilon=0$}}
\label{sec:comp_model}
Having presented our results for both the real and imaginary parts of the smeared amplitude, we now compare them to those obtained from a simple model for the spectral density. The main motivation for this comparison is to gain a first qualitative understanding of the lattice results. The spectral-density model we construct is based on the VSA. Within the VSA approximation, the spectral density $\rho^{\nu}_{1}(E,\vecp{q},m_{H})$ for $B\to K\ell^+\ell^-$ decays is  
\begin{equation}
\label{eq:VSA_model}
\rho^{\nu+;\mathrm{VSA}}_{1}(E,\vecp{q},m_{H}) = q_{c}
\langle K(-\vecp{q}) \,|\, \bar{s}\gamma^{\mu} b \,|\, H(\vecp{0}) \rangle\,
\langle 0 \,|\, V_{\mu}(0)\,\delta\bigl(E-E_{K}-\hat{H}\bigr)\,V^{\nu}(0,\vecp{q}) \,|\, 0 \rangle ,
\end{equation}
where $V^{\mu}\equiv\bar{c}\gamma^{\mu}c$. For $B_s\to\eta_{ss'}\ell^+\ell^-$ decays, one makes the obvious replacements in Eq.\,(\ref{eq:VSA_model}): $K\to\eta_{ss'}$, $E_K\to E_{\eta_{ss'}}$ and $H\to H_s$.
The weak $b\!\to\! s$ matrix element for $B\to K\ell^+\ell^-$ decays can be written in terms of the standard semileptonic form factors $f_{0}$ and $f_{+}$ as  
\begin{equation}\label{eq:semileptonic}
\langle K(-\vecp{q}) \,|\, \bar{s}\gamma^{\mu} b \,|\, H(\vecp{0}) \rangle =
f_{+}(q^{2})\!\left(p_{H}^{\mu}+p_{K}^{\mu}-\frac{m_{H}^{2}-m_{K}^{2}}{q^{2}}\,q^{\mu}\right)+
f_{0}(q^{2})\frac{m_{H}^{2}-m_{K}^{2}}{q^{2}}\,q^{\mu},
\end{equation}
where $p_{H}$ and $p_{K}$ are the four–momenta of the heavy meson and kaon respectively and $q=p_{H}-p_{K}$. For $B_s\to\eta_{ss'}\ell^+\ell^-$, Eq.\,(\ref{eq:semileptonic}) holds with the same obvious replacements as above.

The charm–current spectral density is modelled as a sum of Breit–Wigner distributions corresponding to the leading charmonium resonances supplemented by a tree–level perturbative continuum:  
\begin{equation}
\begin{aligned}
\label{eq:cc_model}
\langle 0 \,|\, V_{\mu}(0)\,\delta\bigl(E-\hat{H}\bigr)\,V^{\nu}(0,\vecp{q}) \,|\, 0 \rangle &=
\sum_{V}\frac{f_{V}^{2}m_{V}^{2}}{2E_{V}}
\left(\delta_{\mu}^{\;\nu}-\frac{k_{V\mu}k_{V}^{\nu}}{m_{V}^{2}}\right)
\frac{\Gamma_{V}}{2\pi}\,
\frac{1}{\bigl(E-E_{V}\bigr)^{2}+(\Gamma_{V}/2)^{2}}  \\
&\quad\hspace{1in}+\,
\bigl(\delta^{\nu}_{\;\mu}\,\tilde{q}^{2}-\tilde{q}^{\nu}\tilde{q}_{\mu}\bigr)\,
\rho_{\rm pert}^{cc}(\tilde{q}^{2}) ,
\end{aligned}
\end{equation}
where the sum extends over the $J/\psi$, $\psi(2S)$, $\psi(3770)$, $\psi(4040)$ and $\psi(4170)$ resonances.  
In Eq.\,(\ref{eq:cc_model}), $m_{V}$, $\Gamma_{V}$ and $f_{V}$ are the mass, total width and decay constant of the resonance, $E_{V}=\sqrt{m_{V}^{2}+|\vecp{q}|^{2}}$, $\tilde{q}=(E,\vecp{q})$, and $k_{V}=(E_{V},\vecp{q})$.  
The quantities $m_{V}$, $\Gamma_{V}$ and $f_{V}$ are taken from the PDG~\cite{ParticleDataGroup:2022pth} (with the exception of the $J/\psi$ mass which is taken from our lattice data), whereas the form factors $f_{+}(q^{2})$ and $f_{0}(q^{2})$ are obtained from our lattice data at the same lattice spacing, heavy–quark mass and three–momentum $\vecp{q}$ used in the charming–penguin calculation. In Tab.\,\ref{tab:model_pars} we collect the values of $m_{V}, f_{V}$ and $\Gamma_{V}$ we used for each charmonium resonance $V$.

 \begin{table}[t]
\begin{ruledtabular}
\begin{tabular}{lccccc}
 & $J/\psi$ & $\psi(2S)$  & $\psi(3770)$ & $\psi(4040)$ & $\psi(4170)$ \\
\colrule
$m_{V} [\rm{GeV}]$ &  $3.05$ & $3.69$ & $3.77$ & $4.04$ & $4.19$   \\
$f_{V} [\rm{GeV}]$ &  $0.412$ & $0.296$ & $0.100$ & $0.187$ &  $0.142$  \\
$\Gamma_{V} [\rm{MeV}]$  &  $0.093$ & $0.294$ & $27.2$ & $80$ & $70$  
\end{tabular}
\end{ruledtabular}
\caption{ The input parameters $m_{V}$, $f_{V}$ and $\Gamma_{V}$ used in the spectral density model in Eqs.\,(\ref{eq:VSA_model})-(\ref{eq:cc_model}) All the parameters are taken from the PDG\,\cite{ParticleDataGroup:2022pth} except for the value the mass of $J/\psi$ which has been taken from our lattice data at the lattice spacing $a\simeq 0.0795\,{\rm fm}$.
\label{tab:model_pars}}
\end{table}

The perturbative spectral function has the leading–order ($O(\alpha_{s}^{0})$) form  
\begin{equation}
\rho_{\rm pert}^{cc}(\tilde{q}^{2}) =
\frac{3}{12\pi^{2}}
\Bigl(1-\frac{s_{\mathrm{th}}}{\tilde{q}^{2}}\Bigr)^{\!1/2}
\Bigl(1+\frac{s_{\mathrm{th}}}{2\tilde{q}^{2}}\Bigr)
\theta\!\bigl(\tilde{q}^{2}-s_{\mathrm{th}}\bigr)~,
\end{equation}
where $s_{\mathrm{th}}$ marks the onset of the perturbative continuum, which must lie above the main charmonium resonances.  
In perturbative QCD one has $s_{\mathrm{th}}=4m_{c}^{2}$; in the present study we consider $s_{\mathrm{th}}\simeq(3.7\,\text{GeV})^{2}$ and $s_{\mathrm{th}}\simeq(4.1\,\text{GeV})^{2}$ and quote half of the difference between the corresponding results as an estimate of the model uncertainty.  
Within the VSA one has $\rho_{1}^{\nu+;\mathrm{VSA}}=3\,\rho_{2}^{\nu+;\mathrm{VSA}}$. Note that our model does not include  $D_{s}D^{*}$ and $D_{s}^{*}D$ (or $D_{s}D_{s}^{*}$ in the $B_{s}\to\eta_{ss'}$ case) rescattering contributions,
which are purely non factorizable. 

We now use Eq.\,(\ref{eq:amp_mh_dep}) to evaluate the three–times–subtracted smeared amplitude $H_{1,2}^{\nu+;3\text{subs}}(\vec q,\varepsilon; m_{H},m)$ as a function of $m$ and $\varepsilon$ within this model of the spectral density.
Since our first–principles calculation is carried out at a single lattice spacing, the energy integral in Eq.\,(\ref{eq:amp_mh_dep}) is effectively restricted to $E\lesssim\mathcal{O}(1/a)$.  
To mimic this in the model we introduce an ultraviolet cutoff $E^{\mathrm{UV}}=n/a$ and analyse the cases $n=2.5$ and $n=4$, taking the half-difference between the two results as an approximate estimate of cutoff effects.  
As already mentioned, in our lattice calculation,
the smeared amplitude is computed with the bare operators $O_{1}^{(c)}$ and $O_{2}^{(c)}$, which are subject to multiplicative renormalization and operator mixing. As an attempt to take this into account,  
we will fix the overall normalization of the model spectral density in such a way that it reproduces our lattice data below the $J/\psi$ threshold, as we will explain below.

\begin{figure}
    \centering
    \includegraphics[width=0.7
    \linewidth]{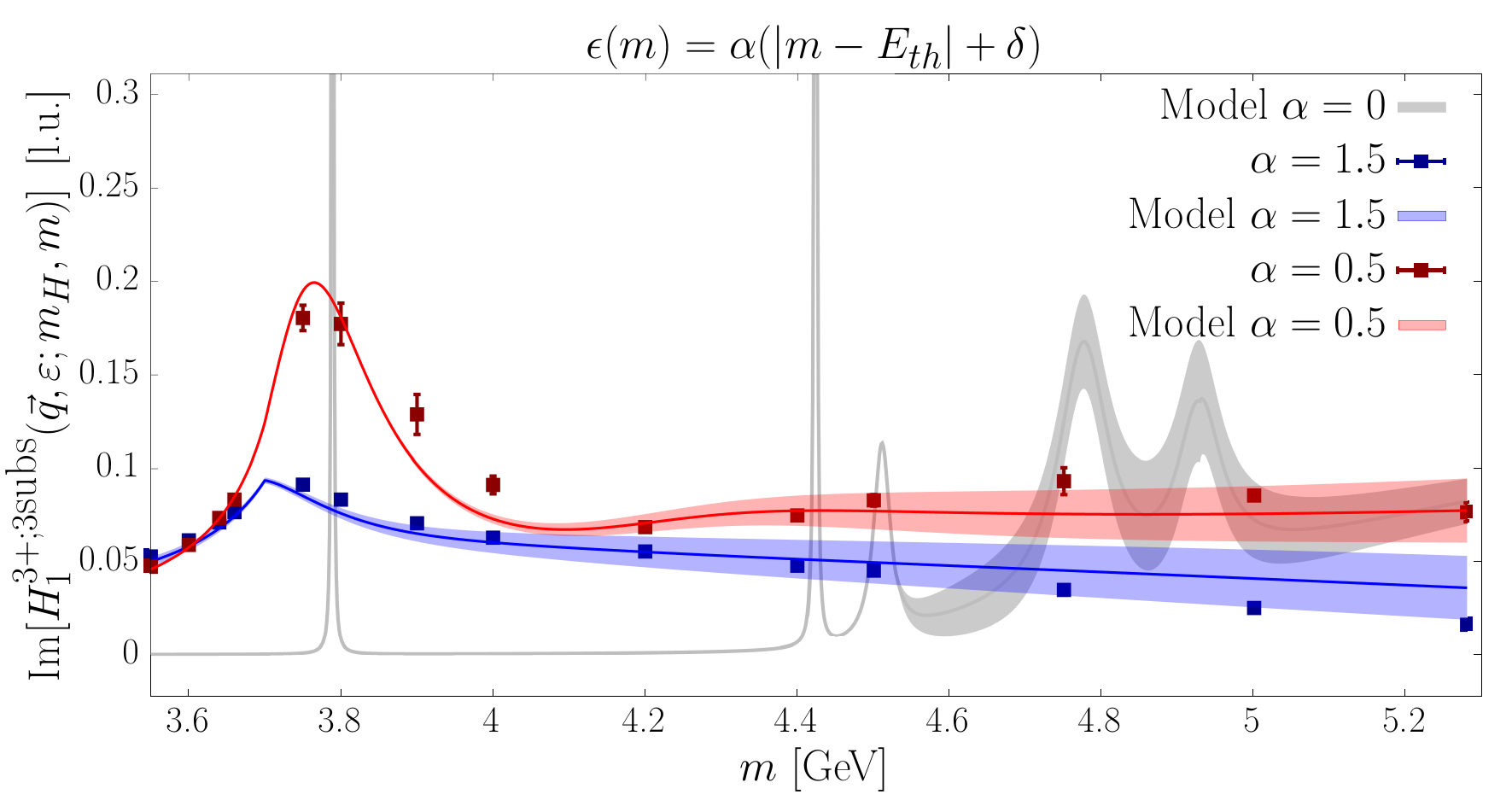}\\
    \includegraphics[width=0.7\linewidth]{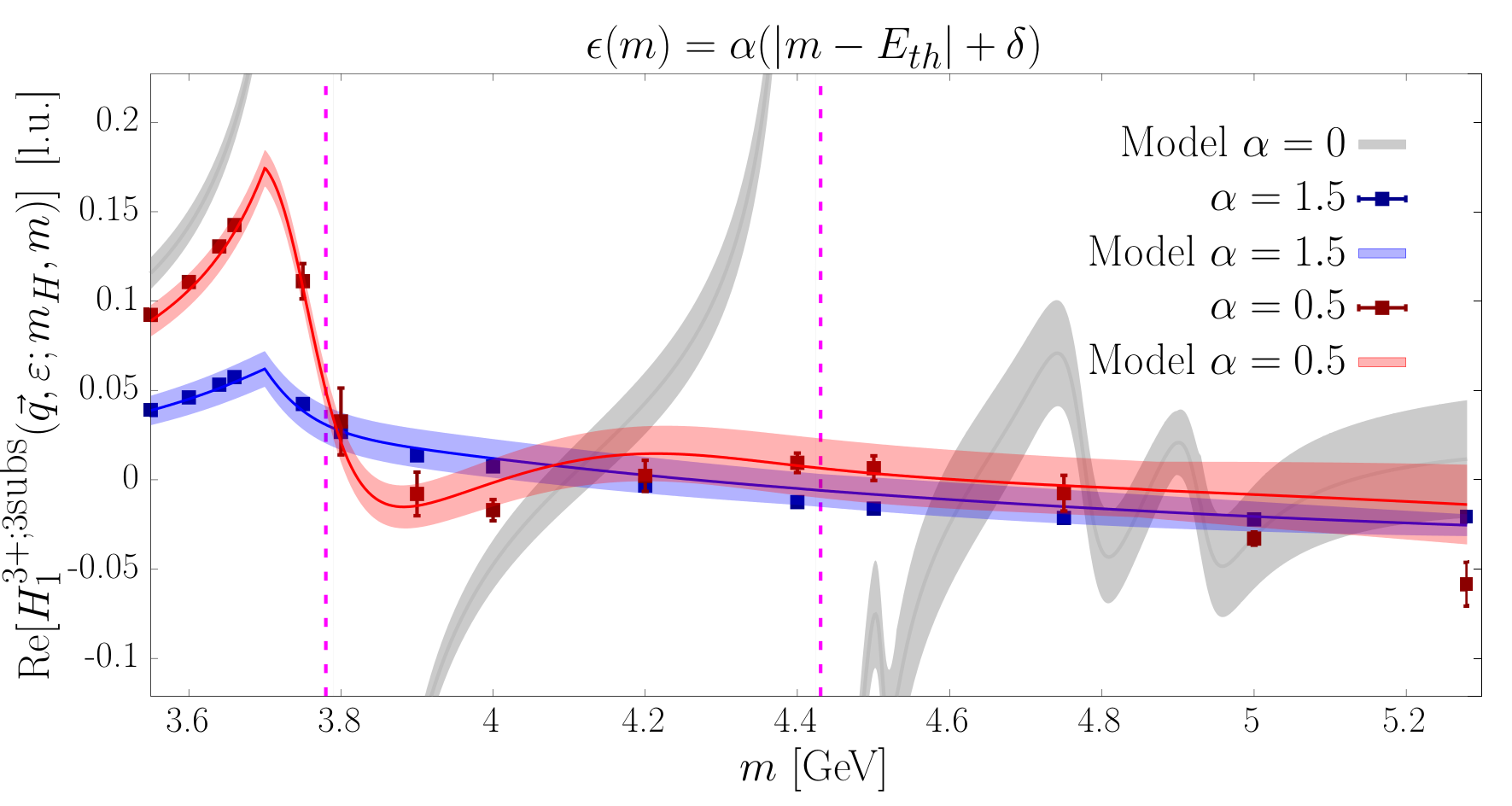}
    \caption{The real (bottom) and imaginary (top) part of the smeared amplitude $H_{1}^{3+;3{\rm subs}}(\vecp{q}, \varepsilon; m_{H}, m)$ in the case of $B_{s}\to \eta_{ss'}\ell^{+}\ell^{-}$ decays, as a function of $m$, for $\alpha=1.5$ (blue) and $\alpha=0.5$ (red). The red and blue colored bands correspond to the model predictions, while the data points are from our lattice computation. The gray semi-transparent bands correspond to the results of
    the model calculation in the $\varepsilon\to0$ limit. The dashed vertical bands in the bottom plot indicate the position of the $J/\psi+\eta_{ss'}$ and $\psi(2S)+\eta_{ss'}$ states. The model uncertainty has been estimated as discussed in the text.}
    \label{fig:model_O1}
\end{figure}

We now compare our first-principles lattice results with the model predictions. 
We do this for the $B_{s}\to \eta_{ss'}\ell^{+}\ell^{-}$ amplitude for which the lattice results have significantly smaller errors than those for the $B\to K\ell^{+}\ell^{-}$ amplitude. 
In Fig.\,\ref{fig:model_O1}  we show the comparison between our lattice results and the model predictions for the contribution from the operator $O_{1}^{(c)}$ to $H_{1}^{3+;3{\rm subs}}(\vecp{q}, \varepsilon; m_{H}, m)$ for the $B_{s}\to \eta_{ss'}\ell^{+}\ell^{-}$ decay. The lattice results for $f_{+}(q^{2})$ and $f_{0}(q^{2})$ which we input into the model for the spectral density are\footnote{Within the model, we use only the central values of $f_{+}(q^{2})$ and $f_{0}(q^{2})$, neglecting their uncertainties.}
\begin{align}
f_{+}(q^{2}) \simeq  1.44(15)  ~,\qquad f_{0}(q^{2}) =   0.87(1)~.
\end{align}
The comparison is shown for two different values of $\alpha$, $\alpha=1.5,0.5$ and for both the real (bottom panel of Fig.\,\ref{fig:model_O1}) and imaginary part (top panel of Fig.\,\ref{fig:model_O1}) of the (three-times-subtracted) smeared amplitude. 

\begin{figure}
    \centering
    \includegraphics[width=0.7
    \linewidth]{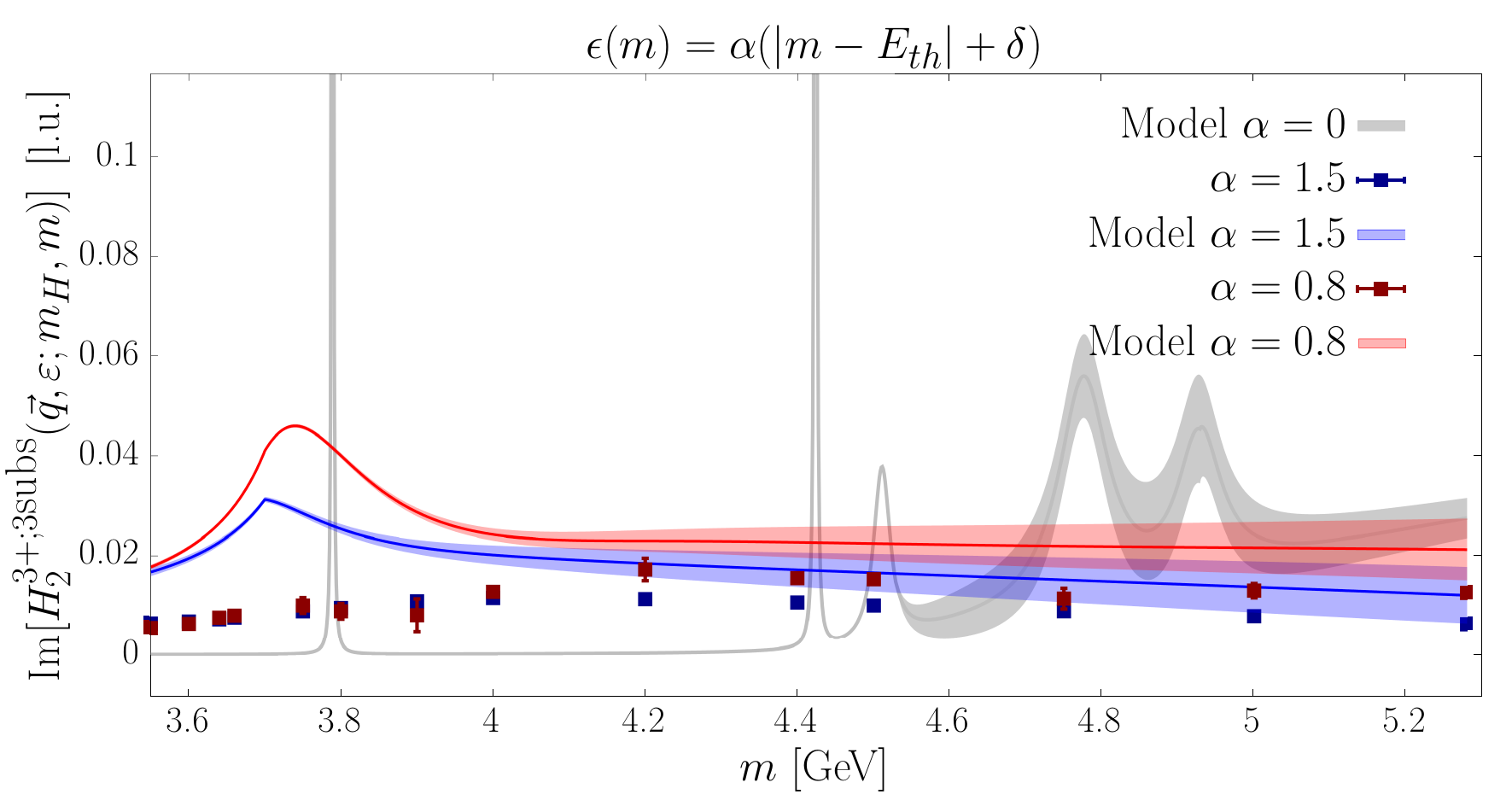}\\
    \includegraphics[width=0.7\linewidth]{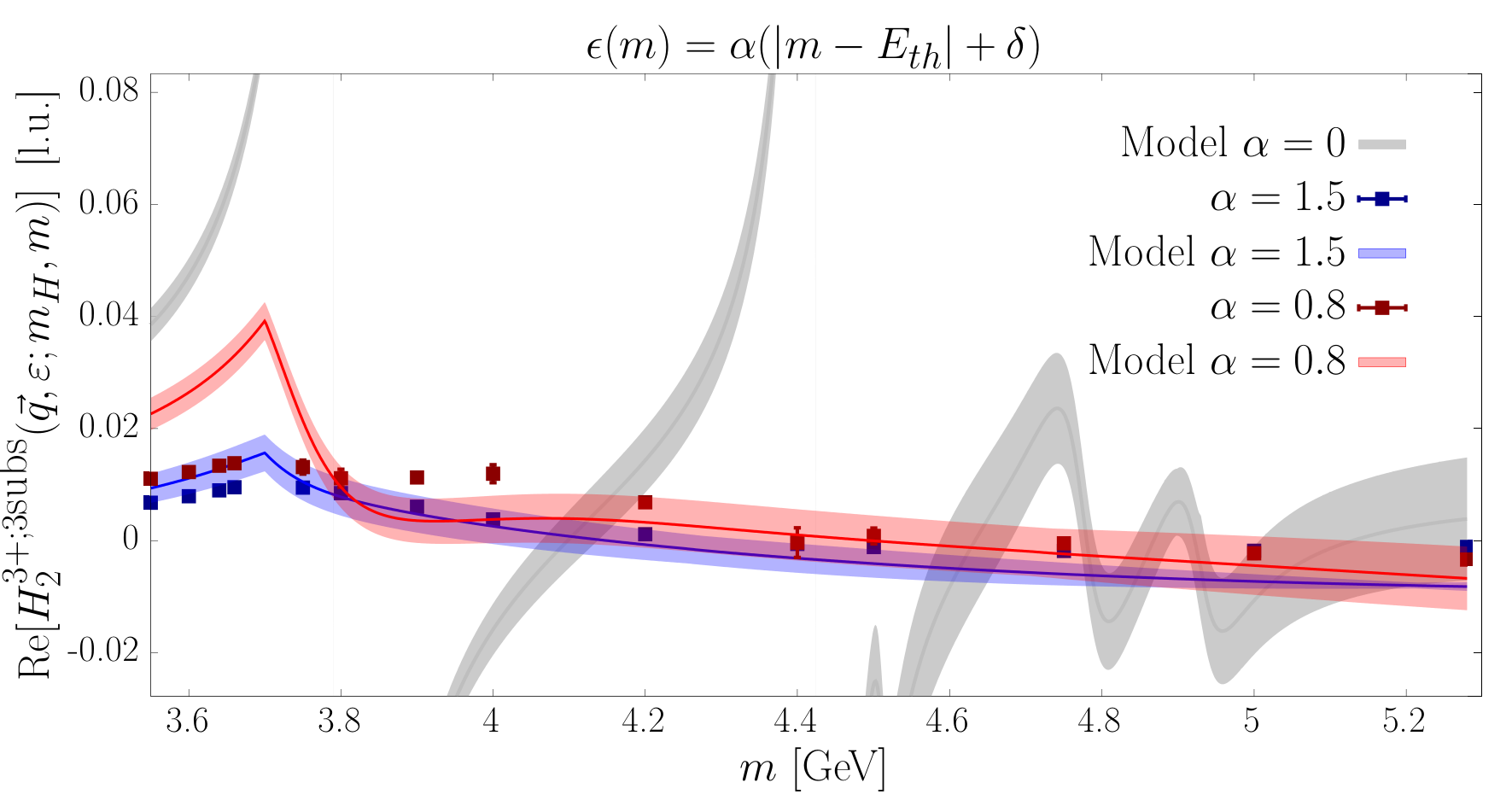}
    \caption{Same as in Fig.~\ref{fig:model_O1} for the $O_{2}^{(c)}$ contribution to $B_{s}\to\eta_{ss'}\ell^{+}\ell^{-}$.}
    \label{fig:model_O2}
\end{figure}

We find that in order to match the model predictions to the lattice data below threshold ($m<m_{J/\psi} + m_{\eta_{ss'}}$) for the $O_{1}^{(c)}$ contribution an additional multiplicative factor $2.3$ needs to be introduced. This is reasonable, as the typical magnitude of the missing renormalization is expected to be an $O(1)$ effect. In the figures the model predictions have been multiplied by this  factor. Overall, as the figure shows, we do observe a fairly good agreement between the model predictions and the lattice results for the contribution from the operator $O_{1}^{(c)}$. The analogous comparison for the $O_{2}^{(c)}$ contribution to $B_{s}\to\eta_{ss'}\ell^{+}\ell^{-}$ is shown in Fig.\,\ref{fig:model_O2} for $\alpha=1.5$ and $0.8$.
In this case the model prediction is larger in magnitude than our first-principles results, and moreover the model predictions seem to 
differ qualitatively from the lattice results. As already noted, this outcome is not unexpected: in our \emph{ab-initio} calculation the axial–axial (AA) contribution from the operator $O_{2}^{(c)}$, which is absent in the VSA, partially cancels the vector–vector (VV) term, so that the net $O_{2}^{(c)}$ contribution is smaller than the VSA estimate.\footnote{Clearly, one could adjust the overall normalization of the model by matching the model and lattice predictions below threshold for the $O_{2}^{(c)}$ contribution. However, since the model and lattice spectral function exhibit qualitatively different behaviour for this operator, the required normalization factor would vary with the choice of $\alpha$. The main conclusion to be drawn is that the VSA-based model fails to capture the essential features of the lattice data for the $O_{2}^{(c)}$ contribution.
}  Future investigations will be needed to establish whether this suppression is driven by the systematic uncertainties that presently affect our calculation—mainly lattice cut-off effects and the absence of non-perturbative renormalization—or whether it reflects a genuine physical effect.

In Figs.~\ref{fig:model_O1} and \ref{fig:model_O2}, we show in gray the model prediction in the physical limit $\varepsilon \to 0$. As discussed in the previous subsection, the lattice data computed at finite $\varepsilon$ can be reliably extrapolated to this limit using the Ansatz of Eq.~(\ref{eq:scaling_regime}), but only in the asymptotic regime where $\varepsilon \lesssim \Delta(m)$. This behaviour is clearly visible in Fig.~\ref{fig:model_O1}, where, in the vicinity of the dominant charmonium resonances, the $\varepsilon \to 0$ curve significantly deviates from the data points at nonzero $\varepsilon$. This indicates that naive polynomial extrapolation becomes unreliable in these regions, as $\Delta(m)$ is small and the spectral function varies too rapidly. In such cases, a more robust approach is to explicitly model the resonant part of the spectral density, for instance using Breit--Wigner functions with a small number of fit parameters, as proposed in the SFR work
~\cite{Frezzotti:2023nun}.

\begin{figure}
    \centering
    \includegraphics[width=0.7\linewidth]{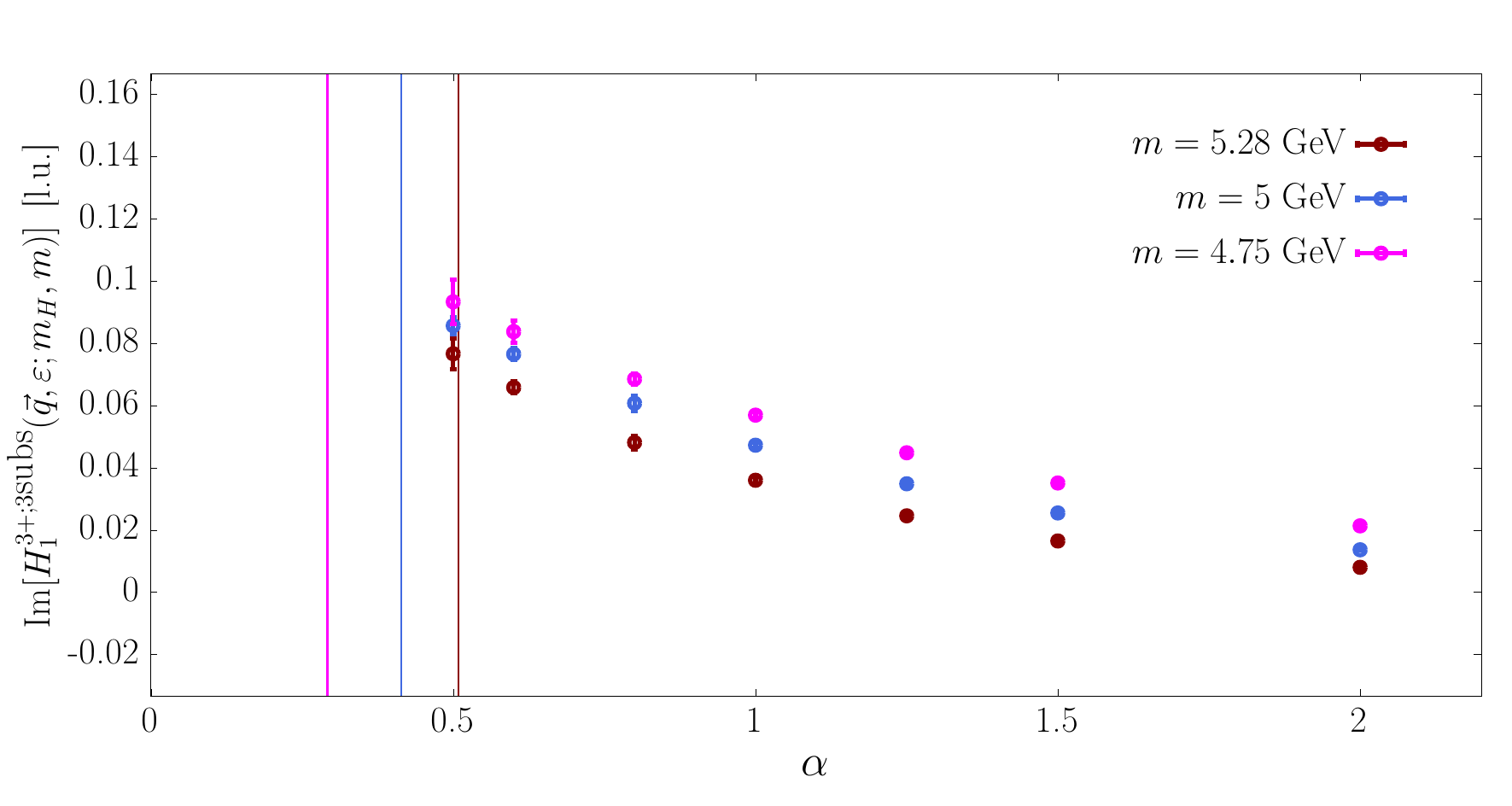}
    \includegraphics[width=0.7\linewidth]{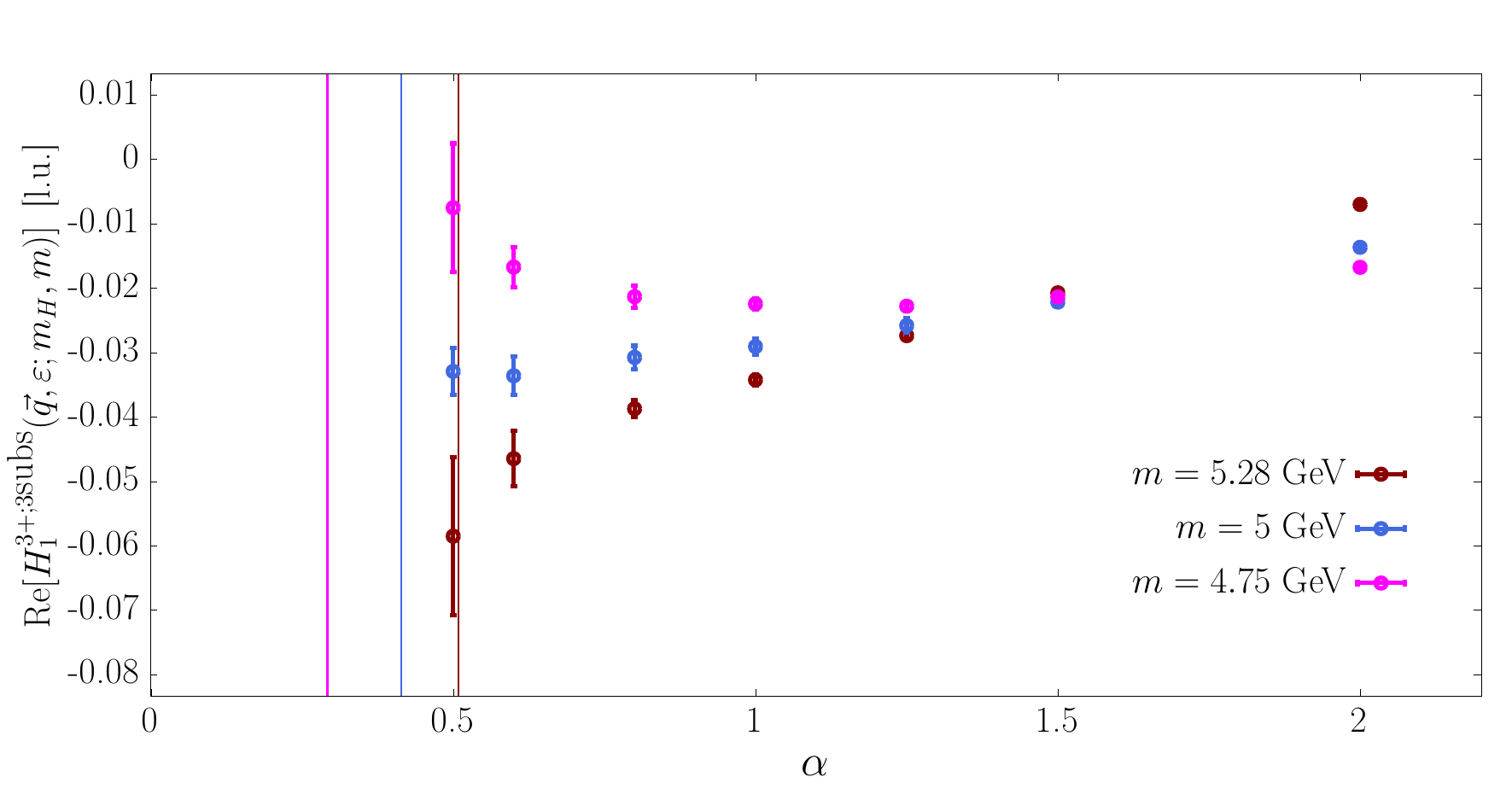}
    \caption{$\alpha$-dependence of our results for the contribution from $O_{1}^{(c)}$ contribution to the subtracted amplitude $B_{s}\to \eta_{ss'}\ell^{+}\ell^{-}$ in the region of large $m$. The top (bottom) panel show our result for the imaginary (real) part of the smeared amplitude. The different colors correspond to different values of $m$ in the high-$m$ region. The vertical lines correspond for each $m$ to the $\alpha$ value leading to $\epsilon(m) = \Delta(m)$, were $\Delta(m)$ is estimated according to Eq.~(\ref{eq:delta_m}). }
    \label{fig:alpha_Bs}
\end{figure}

\begin{figure}
    \centering
    \includegraphics[width=0.7\linewidth]{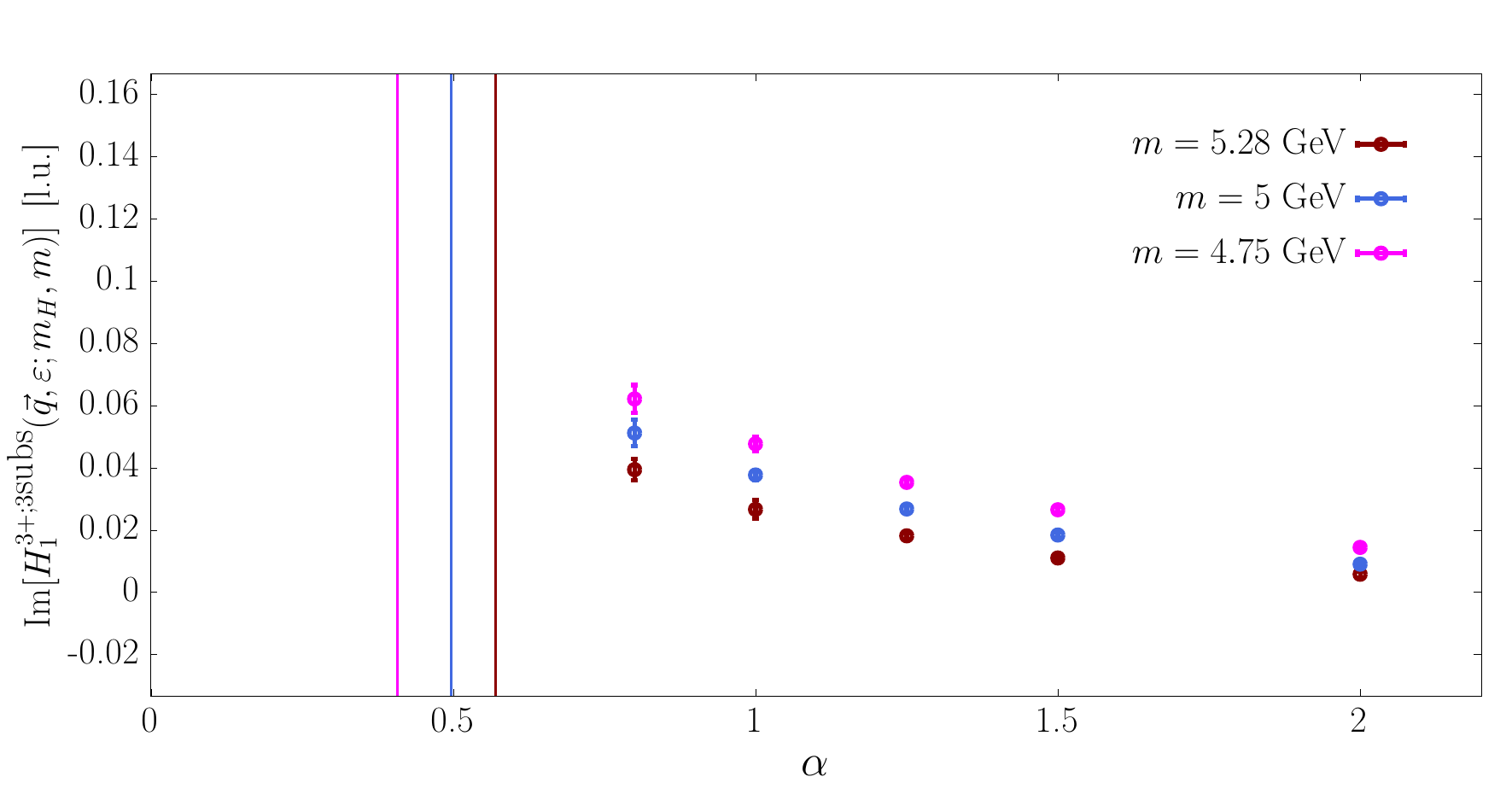}
    \includegraphics[width=0.7\linewidth]{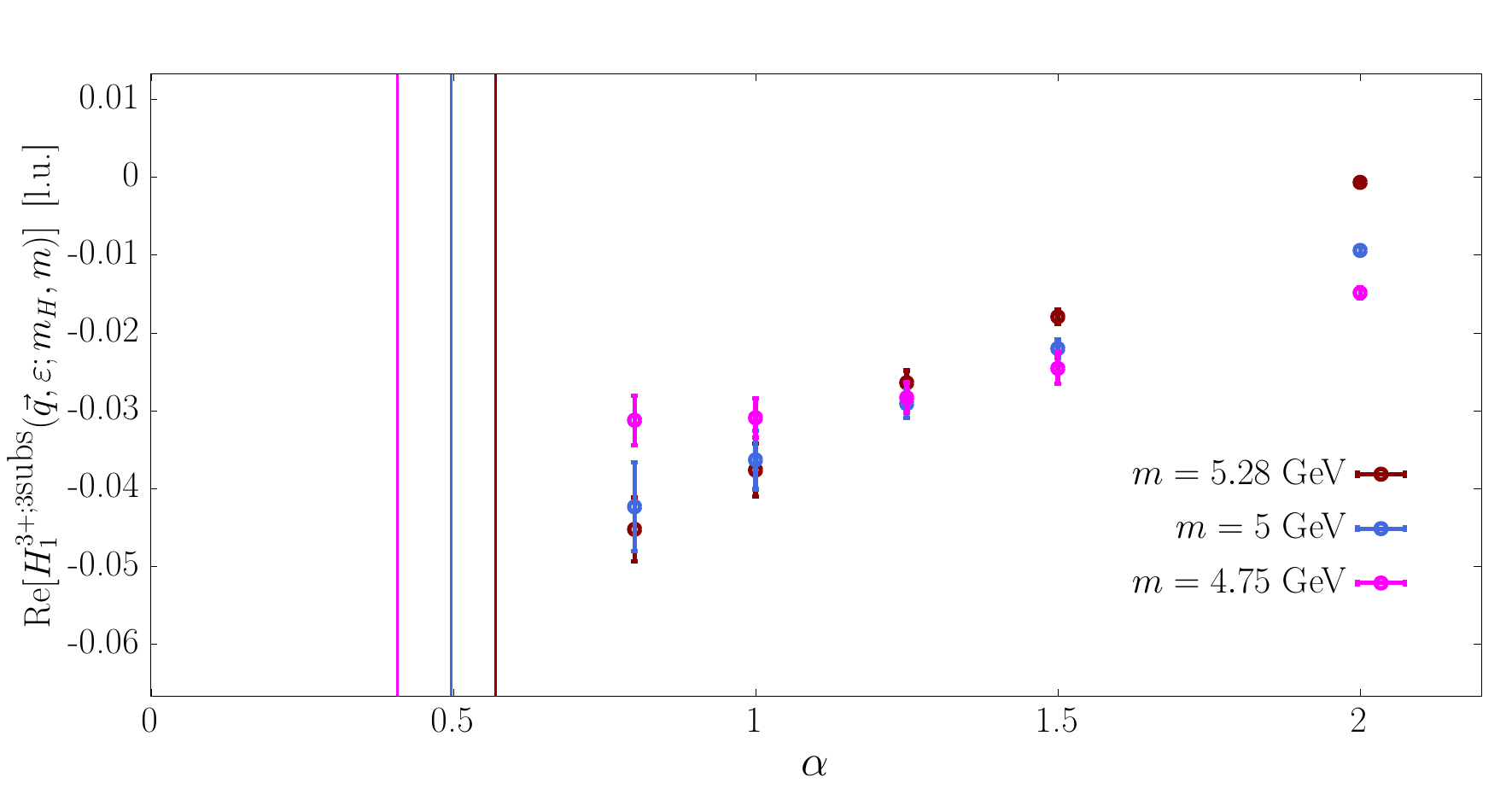}
    \caption{Same as in Fig.~\ref{fig:alpha_Bs} in the case of the $O_{1}^{(c)}$ contribution to $B\to K \ell^{+}\ell^{-}$.}
    \label{fig:alpha_B}
\end{figure}

In contrast, for mass values well above the resonance region, the dependence on $\varepsilon$ is much milder, and polynomial extrapolation becomes viable. However, even in this regime, a model-independent extrapolation still requires that $\varepsilon \lesssim \Delta(m)$ be satisfied. In Figs.~\ref{fig:alpha_Bs} and \ref{fig:alpha_B}, we show our results in this high-mass region, plotted as a function of $\alpha$, for the $B_s \to \eta_{ss'}$ and $B \to K$ transitions respectively. The vertical lines indicate the values of $\alpha$ corresponding to $\Delta(m)$, estimated as:
\begin{align}
\label{eq:delta_m}
\Delta(m) &= |m - m_{\psi(2S)} - m_{K}| \quad\quad (B \to K)~,\\[10pt]
\Delta(m) &= |m - m_{\psi(2S)} - m_{\eta_{ss'}}| \,\,\,\quad (B_s \to \eta_{ss'})~.
\end{align}
As the figures show, the current precision of the Euclidean correlation functions does not yet allow us to reach the regime $\varepsilon < \Delta(m)$, which is needed for a smooth polynomial extrapolation. Nonetheless, since this work represents a proof-of-principles study, we have not yet explored advanced noise-reduction strategies nor performed a high-statistics calculation of the correlation functions. Our results are therefore encouraging: the smallest $\varepsilon$ values we can currently achieve are only slightly larger than the estimated $\Delta(m)$, suggesting that reaching the required regime is within reach of a dedicated effort.

Since the primary limitation arises from the narrow charmonium resonances $J/\psi$ and $\psi(2S)$, a hybrid strategy remains viable: use a physically motivated model for the resonant contributions endowed with free parameters to be fitted to data as mentioned above, and describe the remaining smoother background with a polynomial in $\varepsilon$. This approach could enable a controlled extrapolation to the physical limit even with the current range of $\varepsilon$.  We plan to investigate all these aspects, including improved statistics, noise reduction techniques, and combined model–data fitting strategies, in future dedicated studies.

\section{Conclusions}\label{sec:concs}

A full computation of the decay amplitudes is very important for the phenomenological study of the FCNC decays $B\to K\ell^+\ell^-$ and $B\to\gamma\ell^+\ell^-$ and the analysis of the tensions with the Standard Model. In this paper we have presented a complete theoretical framework for the calculation of these amplitudes from lattice computations of Euclidean correlation functions.
The method is designed to overcome the difficulty in evaluating in Euclidean space the complex contributions to the amplitudes arising from diagrams which contain on-shell intermediate states propagating between the weak $b\to s$ transition operator and the electromagnetic current. 
Such contributions, which include those from ``charming penguins" and the chromomagnetic operator $O_8$, cannot be evaluated using standard lattice techniques, and we therefore rely on spectral density methods such as SFR\,\cite{Frezzotti:2023nun} and HLT\,\cite{Hansen:2019idp}.

Although our focus here is on the evaluation of $B\to K\ell^+\ell^-$ and $B\to\gamma\ell^+\ell^-$ decay amplitudes, we stress that the method is general and can be applied to other processes in which the hadronic factor is given in terms of matrix elements of multilocal operators with on-shell intermediate states propagating in one or more channels. 
This is explained in Sec.\,\ref{subsec:generalN}.

When evaluating the matrix elements of multilocal operators, in addition to the ultraviolet divergences present in the renormalization of the local operators, new ``contact-term" divergences can arise as two of the local operators approach each other. 
This happens, for example, with the weak $b\to s$ and electomagnetic current in $B\to K\ell^+\ell^-$ decays resulting in the appearance of the operators $O_7$ and $O_9$ in the effective Hamiltonian $\tilde{H}_{\mathrm{eff}}$ in Eq.\,(\ref{eq:Heff7910}).
Although, as a result of electromagnetic current conservation, these divergences are logarithmic, we show in Sec.\,\ref{subsec:contact} that each of the two time-orderings which contribute to the $B\to K\ell^+\ell^-$ amplitude is quadratically divergent and that the quadratic divergence cancels when the two contributions are summed. 
We present a method for the separation of the terms containing ultraviolet divergences (including the cancellation of the power divergences) which can be evaluated using standard lattice methods, from the long-distance contributions requiring spectral density techniques (see the discussion around Eq.\,(\ref{eq:3subs})).

In lattice QCD, the operators $O_{1,2}^{(c)}$ themselves contain power divergences, the details of their non-perturbative subtraction depend on the lattice discretization being used. In Sec.\,\ref{subsec:power} we show how such a subtraction can be achieved for the lattice action, based on twisted mass fermions, used in the exploratory numerical work reported in Sec.\,\ref{sec:exploratory}.

In Sec.\,\ref{sec:exploratory} we report on an exploratory numerical study of the charming penguin contribution to the $B\to K\ell^+\ell^-$ decay amplitudes. 
The computations are performed on a single gauge-field ensemble and with a lighter-than-physical $b$-quark ($m_b=2m_c$) to avoid large lattice artifacts. 
The lighter quarks have their physical masses.
In order to investigate the method more precisely, we have also applied it to the decay $B\to\eta_{ss^\prime}\ell^+\ell^-$, where $\eta_{ss^\prime}$ is a pseudoscalar meson with $s$ and $s^\prime$ degenerate quarks with masses equal to the physical strange quark mass.
The absence of $u$ and $d$ quarks in this decay results in significantly smaller statistical errors allowing for a more detailed investigation of the approach.
The results of this preliminary study are very encouraging; both the real and imaginary components of the three-times-subtracted hadronic factors $H_{1,2}^{3+;3\rm{subs}}$, defined in Eq.\,(\ref{eq:H3subs}), are determined for a range of values of the smearing parameter $\epsilon$. 
A comparison of the lattice results with those from a model based on the vacuum saturation approximation, which neglects non-factorizing effects, while naturally differing in detail was qualitatively satisfactory, providing further confidence in the method. 
The numerical study did however, underline the challenges which will have to be overcome in a full scale computation of the physical amplitude, particularly with regard to the $\epsilon\to0$ extrapolation. Within the available statistics, we were barely able, if at all, to reach the asymptotic regime where the extrapolation can be performed using the polynomial scaling ansatz in Eq.\,(\ref{eq:scaling_regime}) so that the correlation functions will need to be evaluated more precisely.

The approach developed in this paper provides the framework allowing us to embark on the long-term project of the complete evaluation of the amplitudes and the results will be reported in future publications. 
If the model expectations of several authors that the contributions from the charming penguins and of the operator $O_8$ are very small are true\,\cite{Gubernari:2022hxn,Khodjamirian:2010vf,Khodjamirian:2012rm,Isidori:2024lng}, then we will not require very small relative errors to establish this. 
Such a situation arose in the evaluation of the form factor $\bar{F}_T$ for $B\to\gamma\ell^+\ell^-$ decays, where the relative error was $O(100\%)$, but this was sufficient to establish that its contribution was negligible\,\cite{Frezzotti:2024kqk}.
If, on the other hand, the contributions from the charming penguins and/or $O_8$ are larger, then we will, of course, require a better relative precision to determine them. 
Much exciting flavor physics phenomenology is now becoming possible.  

\section*{Acknowledgements}
We thank all members of the ETM for the most enjoyable collaboration.
V.L., F.S., G.G., R.F., and N.T. are supported by the Italian Ministry of University and Research (MUR) under the grant PNRR-M4C2-I1.1-PRIN 2022-PE2 ``Non-perturbative aspects of fundamental interactions, in the Standard Model and beyond", F53D23001480006 funded by E.U.-NextGenerationEU. C.T.S. is partially supported by STFC consolidated grant ST/X000583/1. F.S. is supported by ICSC – Centro Nazionale di Ricerca in High Performance Computing, Big Data and Quantum Computing, funded by European Union -NextGenerationEU and by Italian Ministry of University and Research (MUR) project FIS 00001556. F.S. and S.S. are supported by MUR project PRIN 2022N4W8WR. The work of L.S. was supported in part by the European Union - Next Generation EU under Italian MUR grant PRIN-2022-RXEZCJ. 

We gratefully acknowledge the ICSC - Centro Nazionale di Ricerca in High Performance Computing for providing computing time under the allocations RAC 1916318. We gratefully acknowledge CINECA for the provision of GPU time on Leonardo supercomputing facilities under the specific initiative INFN-LQCD123, and under project IscrB VITO-QCD and project IscrB SemBD. We gratefully acknowledge EuroHPC Joint Undertaking for awarding the project ID EHPC-EXT-2023E02-052 access to MareNostrum5 hosted by at the Barcelona Supercomputing Center, Spain. We also thank the GENCI.fr for granting us access to the Jean Zay computers of the computing center IDRIS in Orsay.

\appendix
\section{Further remarks concerning the contact terms}\label{sec:appcontact}
In this appendix we add some further remarks concerning the contact terms discussed in Sec.\,\ref{subsec:contact}. Specifically, we describe the procedure for renormalizing the logarithmic divergence resulting in the appearance of the operators $O_9$ and $O_7$ in the effective Hamiltonian (\ref{eq:Heff7910}). We also explain how the transverse structure of the amplitude is recovered when the contributions from the two time-orderings are combined.

We start by explaining the appearance of the operator $O_9$ in the effective Hamiltonian and explain the procedure to renormalize the corresponding divergence into the $\MSbar$-scheme to correspond with the known Wilson coefficient function. The general procedure is analogous to that used to renormalize the contact terms in rare $K\to\pi\nu\bar\nu$ decays\,\cite{Christ:2016eae}.
\begin{enumerate} 
    \item   Let $\tilde{J}^\mu_{\mathrm{em}}(q)$ be the Fourier transform of  $J^\mu_{\mathrm{em}}(x)$. We compute the matrix elements\\ $\bra{s(\tilde{p}_s)}\big\{\mathrm{T}[\tilde{J}^\mu_{\mathrm{em}}(\tilde q)O_{1,2}^{\mathrm{SMom}}(0)]\big\}^{\mathrm{Lat}}_a\ket{b(\tilde{p}_b)}$ and, using the appropriate projectors, $\bra{s(\tilde{p}_s)}\bar{s}(0)\gamma^\nu P_L b(0)\ket{b(\tilde{p}_b)}$ ($\mu,\nu=1$\,-\,4) in the Landau gauge, at the kinematics chosen for the RI-SMom renormalization, $\tilde{p}_b^2=\tilde{p}_s^2=\tilde{q}^2=\mu^2$ in the limit of zero quark masses. 
Here the indices $\mathrm{Lat}$ and $a$ on the bilocal operator indicate that the divergence due to the contact term is regulated in the lattice regularization at a lattice spacing $a$. We have introduced the tildes over $\tilde{q}\,,\tilde{p}_s$ and $\tilde {p}_b$ to indicate that these are the momenta at which the renormalization conditions are applied. 
\item By electromagnetic gauge invariance $\bra{s(\tilde{p}_s)}T[\tilde{J}^\mu_{\mathrm{em}}(\tilde{q})O_{1,2}^{(c)}(0)]\ket{b(\tilde{p}_b)}$ is proportional to\\ $(\tilde{q}^2g^{\mu\nu}-\tilde{q}^\mu \tilde{q}^\nu)\bra{s(\tilde{p}_s)}\bar{s}(0)\gamma_\nu P_L b(0)\ket*{b(\tilde{p}_b)}$. 
    \item Next, at the renormalization point, we impose the condition 
\begin{equation}\label{eq:subcondition9}
    \bra{s(\tilde{p}_s)}\mathrm{T}[\tilde{J}^\mu_{\mathrm{em}}(\tilde{q})O_{1,2}^{\RS}(0)]-c_{1,2}(\mu,a)\,(\tilde{q}^2g^{\mu\nu}-\tilde{q}^\mu \tilde{q}^\nu)\bar{s}(0)\gamma^\nu P_L b(0)\ket{b(\tilde{p}_b)}=0
\end{equation}
and determine the coefficients $c_{1,2}(\mu,a)$.  We define the bilocal operator in the RI-SMom scheme by
\begin{equation}
    \big\{\mathrm{T}[\tilde{J}^\mu_{\mathrm{em}}(q)O_{1,2}^{\RS}(0)]\big\}^{\RS}_\mu\equiv \big\{\mathrm{T}[\tilde{J}^\mu_{\mathrm{em}}(q)O_{1,2}^{\RS}]-(q^2g^{\mu\nu}-q^\mu q^\nu)\,c_{1,2}(\mu,a)\,\bar{s}(0)\gamma^\mu P_L b(0)\big\}^{\mathrm{Lat}}_a.
\end{equation}
\item The relation between the local operators $O_{1,2}$ in the RI-SMom and \msbar schemes,  after subtraction of the power divergences, is obtained perturbatively in the standard way and here we assume that this has been done. Similarly the bilocal operator renormalized in the \msbar scheme is related to $\big\{\mathrm{T}[\tilde{J}^\mu_{\mathrm{em}}(q)O_{1,2}^{\RS}(0)]\big\}^{\RS}_\mu$ by 
\begin{equation}\label{eq:bilocalmatching}
    \big\{\mathrm{T}[\tilde{J}^\mu_{\mathrm{em}}(q)O_{1,2}^{\MSbar}(0)]\big\}^{\MSbar}_\mu
    =\big\{\mathrm{T}[\tilde{J}^\mu_{\mathrm{em}}(q)O_{1,2}^{\MSbar}(0)]\big\}^{\RS}_\mu + d_{1,2}(\mu)\,(q^2g^{\mu\nu}-q^\mu q^\nu)\,\bar{s}(0)\gamma_\nu P_L b(0)\,,
\end{equation}
where, for illustration we have chosen the scale $\mu$ to be the same as in Eq.\,(\ref{eq:subcondition9}). The coefficients $d_{1,2}(\mu)$ are obtained in perturbation theory by ensuring that Eq.\,(\ref{eq:bilocalmatching}) is satisfied between conveniently chosen external states. 
\item Using this procedure, the hadronic matrix elements of $\big\{\mathrm{T}[\tilde{J}^\mu_{\mathrm{em}}(q)O_{1,2}^{\MSbar}(0)]\big\}^{\MSbar}_\mu$, with all divergences renormalized in the \msbar scheme, can be obtained from lattice computations combined with, as usual and unavoidably, perturbation theory to match the RI-SMom scheme to \msbar.
\end{enumerate}
In higher orders of QCD perturbation theory, the operator $O_7$ appears and, with the use of tensor projection operators, the corresponding procedure can be followed to obtain $O_7^{\RS}$ non-perturbatively from the lattice operators and then to relate $O_7^{\RS}$ to $O_7^{\MSbar}$.

We now trace how the transverse structure of the matrix element is restored when the contributions from the two regions $t<0$ and $t>0$ are added together.
From the first line of Eq.\,(\ref{eq:Hmuplus}) we have
\begin{eqnarray}
    q_\nu\,H^{\nu+}_{1,2}(\vecp{q})&=&\int_0^\infty\!\!dt\int\!\dthree x~ (\partial_\nu e^{iq\cdot x}) \bra{K(-\vecpp{q})}J_\mathrm{em}^\nu(t,\vecp{x})\,O_{1,2}^{(c)}(0)\ket*{B(\vecpp{0})}\nn\\
    &&\hspace{-0.75in}=-\int_0^\infty\!\!dt\int\!\dthree x~ e^{iq\cdot x} \bra{K(-\vecpp{q})}(\partial_\nu J_\mathrm{em}^\nu(t,\vecp{x}))\,O_{1,2}^{(c)}(0)\ket*{B(\vecpp{0})}
    -\int\dthree x\,e^{-i\vec{q}\cdot\vec{x}}\,\bra{K(-\vecpp{q})}J_\mathrm{em}^0(0,\vecp{x})\,O_{1,2}^{(c)}(0)\ket*{B(\vecpp{0})}\nn\\
    &=&-\int\dthree x\,e^{-i\vec{q}\cdot\vec{x}}\,\bra{K(-\vecpp{q})}J_\mathrm{em}^0(0,\vecp{x})\,O_{1,2}^{(c)}(0)\ket*{B(\vecpp{0})}\,.
\end{eqnarray}
Similarly
\begin{equation}
 q_\nu\,H^{\nu-}_{1,2}(\vecp{q}) = \int\dthree x\,e^{-i\vec{q}\cdot\vec{x}}\,\bra{K(-\vecpp{q})}O_{1,2}^{(c)}(0)\,J_\mathrm{em}^0(0,\vecp{x})\ket*{B(\vecpp{0})}\,,
\end{equation}
so that 
\begin{eqnarray}
    q_\nu\,\big(H^{\nu+}_{1,2}(\vecp{q})+H^{\nu-}_{1,2}(\vecp{q})\big)&=&
    \int\dthree x\,e^{-i\vec{q}\cdot\vec{x}}\,\bra{K(-\vecpp{q})}\big[O_{1,2}^{(c)}(0),J_\mathrm{em}^0(0,\vecp{x})\big]\ket*{B(\vecpp{0})}\nn\\
    &=&\int\dthree x\,\bra{K(-\vecpp{q})}\big[O_{1,2}^{(c)}(0),J_\mathrm{em}^0(0,\vecp{x})\big]\ket*{B(\vecpp{0})}\nn\\
    &=&(Q_B-Q_K)\bra{K(-\vecpp{q})}O_{1,2}^{(c)}(0)\ket*{B(\vecpp{0})}=0\,,
\end{eqnarray}
where $Q_B$ and $Q_K$ are the electric charges of the $B$ and $K$ mesons respectively and we have used the fact that the equal-time commutator
$\big[O_{1,2}^{(c)}(0),J_\mathrm{em}^0(0,\vecp{x})\big]$ vanishes except at $\vec{x}=\vec{0}$. We have therefore demonstrated that the sum of the contributions from $t<0$ and $t>0$ is transverse, whereas the two contributions individually are not.

\section{Consequences of parity and time-reversal symmetries for the reality of the spectral density}
\label{sec:appreality}

In this appendix we exploit the parity and time reversal symmetries of QCD to show that the spectral density for the decay $B\to K\ell^+\ell^-$ is real. We denote the parity and time reversal operators by ${\mathbb{P}}$ and $\mathbb{T}$ respectively and define $\mathbb{X}\equiv\mathbb{PT}$. The vector and axial-vector currents transform under the combined $\mathbb{P}\mathbb{T}$ transformation as follows:
\begin{eqnarray}
  \mathbb{X}~\bar{\psi}_{f_1}(x)\gamma^\mu\psi_{f_2}(x)~( \mathbb{X})^{-1} &=&\bar{\psi}_{f_1}(-x)\gamma^\mu\psi_{f_2}(-x)\label{eq:PTV}\\
  \mathbb{X}~\bar{\psi}_{f_1}(x)\gamma^\mu\gamma^5\psi_{f_2}(x)~( \mathbb{X})^{-1} &=&-\bar{\psi}_{f_1}(-x)\gamma^\mu\gamma^5\psi_{f_2}(-x)\,.\label{PTA}
\end{eqnarray}
The subscripts $f_{1,2}$ denote the flavor quantum numbers of the quark fields. 

For the four quark operators we introduce the natural notation,
\begin{eqnarray}
 VV_{f_1f_2f_3f_4}(x)&=&\bar{\psi}_{f_1}(x)\gamma^\mu\psi_{f_2}(x)~
 \bar{\psi}_{f_3}(x)\gamma_\mu\psi_{f_4}(x)\,;\quad
 AA_{f_1f_2f_3f_4}(x)=\bar{\psi}_{f_1}(x)\gamma^\mu\gamma^5\psi_{f_2}(x)~
 \bar{\psi}_{f_3}(x)\gamma_\mu\gamma^5\psi_{f_4}(x)\,;\\
VA_{f_1f_2f_3f_4}(x)&=&\bar{\psi}_{f_1}(x)\gamma^\mu\psi_{f_2}(x)~
 \bar{\psi}_{f_3}(x)\gamma_\mu\gamma^5\psi_{f_4}(x)\,;\hspace{0.03in}
 AV_{f_1f_2f_3f_4}(x)=\bar{\psi}_{f_1}(x)\gamma^\mu\gamma^5\psi_{f_2}(x)~
 \bar{\psi}_{f_3}(x)\gamma_\mu\psi_{f_4}(x)\,. 
\end{eqnarray}
The transformation properties of the parity conserving $(VV,AA)$ and parity violating $(VA,AV))$ components of the four-quark weak operators differ in sign as follows:
\begin{eqnarray}
 \mathbb{X}~VV_{f_1f_2f_3f_4}(x)~
 \mathbb{X}^{-1} &=&+VV_{f_1f_2f_3f_4}(-x);\qquad
 \mathbb{X}~AA_{f_1f_2f_3f_4}(x)~
 \mathbb{X}^{-1} =+AA_{f_1f_2f_3f_4}(-x);
 \label{eq:PTVVAA}\\
\mathbb{X}~VA_{f_1f_2f_3f_4}(x)~
 \mathbb{X}^{-1} &=&-VA_{f_1f_2f_3f_4}(-x);\qquad
 \mathbb{X}~AV_{f_1f_2f_3f_4}(x)~
 \mathbb{X}^{-1} =-AV_{f_1f_2f_3f_4}(-x)
\,.\label{eq:PTVAAV}
\end{eqnarray}
Using the Eqs.\,(\ref{eq:PTV}), (\ref{eq:PTVVAA}) and (\ref{eq:PTVAAV}), together with the relations $\bra{\mathbb{T}f}\,\mathbb{T}O\,\mathbb{T}^{-1}\ket{\mathbb{T}i}
=\bra{f}\,O\,\ket{i}^\ast$ and $\mathbb{X}\ket{P}=-\ket{P}$ for a pseudoscalar state $\ket{P}$,
and noting that $\delta^{(4)}(\hat{P}-l)$ (where $\hat{P}=(\hat{H},\hat{\mathbf{P}})$ is the four-momentum operator and $l$ is a generic four-momentum) is invariant under the $\mathbb{PT}$ symmetry,
we find that the parity conserving contributions satisfy:
\begin{eqnarray}
\bra{K(\vec{p}_K)}J^\nu_\mathrm{em}(0)~\delta^{(4)}(\hat{P}-l) 
  ~VV_{f_1f_2f_3f_4}(0)~
\ket*{B(\vec{p}_B)}&=&+
\bra{K(\vec{p}_K)}J^\nu_\mathrm{em}(0)~\delta^{(4)}(\hat{P}-l) 
  ~VV_{f_1f_2f_3f_4}(0)\ket*{B(\vec{p}_B)}^\ast\,,\label{eq:VV2}\\
\bra{K(\vec{p}_K)}J^\nu_\mathrm{em}(0)~\delta^{(4)}(\hat{P}-l) 
  ~AA_{f_1f_2f_3f_4}(0)~
\ket*{B(\vec{p}_B)}&=&+
\bra{K(\vec{p}_K)}J^\nu_\mathrm{em}(0)~\delta^{(4)}(\hat{P}-l) 
  ~AA_{f_1f_2f_3f_4}(0)\ket*{B(\vec{p}_B)}^\ast\,,\label{eq:AA2}
\end{eqnarray}
whereas the parity odd terms satisfy
\begin{eqnarray}
\bra{K(\vec{p}_K)}J^\nu_\mathrm{em}(0)~\delta^{(4)}(\hat{P}-l) 
  ~VA_{f_1f_2f_3f_4}(0)~
\ket*{B(\vec{p}_B)}&=&-
\bra{K(\vec{p}_K)}J^\nu_\mathrm{em}(0)~\delta^{(4)}(\hat{P}-l) 
  ~VA_{f_1f_2f_3f_4}(0)\ket*{B(\vec{p}_B)}^\ast\,,\label{eq:VA2}\\
\bra{K(\vec{p}_K)}J^\nu_\mathrm{em}(0)~\delta^{(4)}(\hat{P}-l) 
  ~AV_{f_1f_2f_3f_4}(0)~
\ket*{B(\vec{p}_B)}&=&-
\bra{K(\vec{p}_K)}J^\nu_\mathrm{em}(0)~\delta^{(4)}(\hat{P}-l) 
  ~AV_{f_1f_2f_3f_4}(0)\ket*{B(\vec{p}_B)}^\ast\,.\label{eq:AV2}
\end{eqnarray}

We have already mentioned that the weak operators $VV$ and $AA$ are parity conserving and $VA$ and $AV$ are parity violating. In the present context this means that the matrix elements
\begin{equation}
    M^\nu(p_K,p_B,l)\equiv\bra{K(\vec{p}_K)}J^\nu_\mathrm{em}(0)~\delta^{(4)}(\hat{P}-l)\,O(0)\ket*{B(\vec{p}_B)}=\pm {\cal P}^\nu_\rho
    M^\rho({\cal P}p_k,{\cal P}p_B,{\cal P}l)\,,
\end{equation}
where the $+$ sign on the right-hand side corresponds to $O(0)=VV(0)$ or $AA(0)$ and the $-$ sign to $O(0)=VA(0)$ or $AV(0)$. 
The matrix ${\cal P}=\mathrm{diag}(1,-1,-1,-1)$ and, when acting on a four momentum $p=(p^0,\vecp{p})$ we have ${\cal P}p=(p^0,-\vecp{p})$. 
This implies that the matrix elements can be decomposed in terms of invariant form factors $F_B,\,F_K, F_l$ and $F_{\mathrm{odd}}$ as follows
\begin{eqnarray}
    M^\nu(p_K,p_B,l)&=&F_B\,p_B^\nu+F_K\,p_K^\nu+F_l\,l^\nu\,,\qquad O=VV~
    \mathrm{or}~AA\label{eq:FBKl}\\
    M^\nu(p_K,p_B,l)&=&i\epsilon^{\nu\alpha\beta\gamma}\,(p_B)_\alpha
    (p_K)_\beta\,l_\gamma\,F_{\mathrm{odd}}\,,\quad O=VA~\mathrm{or}~AV\,.
    \label{eq:Fodd}
\end{eqnarray}

We now deduce the consequences of the above discussion for the spectral density $\rho_{1,2}^{\nu+}(E,\vecpp{q})$ in Eq.\,(\ref{eq:rhoplus} which appears in the time-ordering requiring the SFR method for $B\to K\ell^+\ell^-$ decays. In this case $\vec{p}_B=\vec{l}=\vec{0}$ so that the coefficient of $F_{\mathrm{odd}}$ in Eq.\,(\ref{eq:Fodd}) is zero. Thus only the components $VV$ and $AA$ contribute to $\rho_{1,2}^{\nu+}(E,\vecpp{q})$ which is therefore real (see Eqs.\,(\ref{eq:VV2}) and (\ref{eq:AA2})). 

A similar analysis can be carried out for $\bar{B}_s\to\gamma\ell^+\ell^-$ decays. The three spectral densities corresponding to contributions requiring spectral density methods are given in Eqs.\,(\ref{eq:rho1munu}), (\ref{eq:rho2munu}) and (\ref{eq:rho3munu}).

\bibliography{hepbiblio}

\end{document}